\documentclass[acmsmall]{acmart}

\usepackage{xspace}
\usepackage{mathtools} %
\usepackage{enumitem}
\usepackage{changepage}

\usepackage{graphicx}
\usepackage{subcaption}
\usepackage{tikz}

\usepackage{tabularx}
\usepackage{multirow}
\usepackage{booktabs}
\usepackage{makecell}
\usepackage[figuresleft]{rotating}
\usepackage[flushleft]{threeparttable}

\usepackage[utf8]{inputenc} %
\usepackage{pifont}

\newcommand\encircle[1]{\tikz[baseline=(X.base)] 
    \node (X) [draw, shape=rectangle, inner sep=0pt, fill=black, text=white] {\strut #1};}

\renewcommand\footnotetextcopyrightpermission[1]{} %

\newcommand{\etal}{\textit{et al.}\xspace}

\begin{document}

\title{Deep Learning Workload Scheduling in GPU Datacenters: Taxonomy, Challenges and Vision}

\author{Wei Gao}
\authornote{Equal contribution. Alphabetical order of surname.}
\email{gaow0007@ntu.edu.sg}
\author{Qinghao Hu}
\authornotemark[1]
\affiliation{
  \institution{Nanyang Technological University}
  \country{Singapore}}
\email{qinghao.hu@ntu.edu.sg}
\author{Zhisheng Ye}
\authornotemark[1]
\affiliation{%
  \institution{Peking University}
  \country{China}}
\email{yezhisheng@pku.edu.cn}
\author{Peng Sun}
\affiliation{%
  \institution{SenseTime}
  \country{Singapore}}
\email{sunpeng1@sensetime.com}
\author{Xiaolin Wang}
\email{wxl@pku.edu.cn}
\author{Yingwei Luo}
\affiliation{
  \institution{Peking University}
  \country{China}}
\email{lyw@pku.edu.cn}
\author{Tianwei Zhang}
\email{tianwei.zhang@ntu.edu.sg}
\author{Yonggang Wen}
\affiliation{
  \institution{Nanyang Technological University}
  \country{Singapore}}
\email{ygwen@e.ntu.edu.sg}

\renewcommand{\shortauthors}{Gao, Hu and Ye, et al.}

\begin{abstract}
    Deep learning (DL) shows its prosperity in a wide variety of fields. The development of a DL model is a time-consuming and resource-intensive procedure. Hence, dedicated GPU accelerators have been collectively constructed into a GPU datacenter. An efficient scheduler design for such GPU datacenter is crucially important to reduce the operational cost and improve resource utilization. However, traditional approaches designed for big data or high performance computing workloads can not support DL workloads to fully utilize the GPU resources. Recently, substantial schedulers are proposed to tailor for DL workloads in GPU datacenters. This paper surveys existing research efforts for both training and inference workloads. We primarily present how existing schedulers facilitate the respective workloads from the \textit{scheduling objectives} and \textit{resource consumption features}. Finally, we prospect several promising future research directions. More detailed summary with the surveyed paper and code links can be found at our project website: \url{https://github.com/S-Lab-System-Group/Awesome-DL-Scheduling-Papers}.

\end{abstract}



\maketitle

\section{Introduction}
\label{sec_intro}

Recent decades have witnessed a dramatic increase in deep learning (DL) research, development, and application in many fields, including Go~\cite{silver2017mastering}, medical analysis~\cite{shen2017deep}, robotics~\cite{gu2017deep}, etc. A standard DL development pipeline consists of \textit{model training} and \textit{model inference}. Each stage requires high-grade hardware resources (GPU and other compute systems) to produce and serve production-level DL models~\cite{Philly,MLaaS,Gavel,Helios}. Therefore it becomes prevalent for IT industries~\cite{Helios,MLaaS} and research institutes~\cite{Philly,RIFLING,DLQoSSched} to set up \textit{GPU datacenters} to meet their ever-growing DL development demands. A GPU datacenter possesses large amounts of heterogeneous compute resources to host large amounts of DL workloads.  An effective scheduler system is urgently required to orchestrate these resources and workloads to guarantee the efficiency of DL workload execution, hardware utilization, and other scheduling objectives.

The scheduler is responsible for determining the resource utilization of the entire datacenter and the performance of each job, which further affects the operation cost and user experience~\cite{Chronus}. Specifically, (1) for model training, the scheduler allocates resources requested by the users to support the long-running offline training workloads. The scheduler needs to achieve high performance for each individual workload, high resource utilization for the entire datacenter, and high fairness among different users. Due to the unique and complicated features of DL training jobs, conventional scheduling algorithms for high performance computing (HPC) and big data workloads could cause unbalanced resource utilization and exorbitant infrastructure expense~\cite{ASTRAEA}, and new solutions tailored for GPU datacenters are required. (2) For model inference, DL applications often serve as online services to answer users' requests. They often have a higher expectation on the response latency and inference accuracy~\cite{Daniel2017clipper,zhang2019mark}. Applications that fail to be completed within the specified time (Service Level Agreement) or have lower accuracy than expected may have little or no commercial values. Therefore, it is critical for the scheduler to balance the inference latency, accuracy and cost.

Over the years a variety of DL schedulers have been proposed for GPU datacenters \cite{Gandiva, Pollux, Gavel, Tiresias, Daniel2017clipper, zhang2019mark, INFaaS}. However, most of these systems are designed in an ad-hoc way for some specific objectives. There is still a lack of comprehensive exploration towards efficient scheduling of DL workloads. We are interested in the following questions: (1) what are the main challenges for designing a satisfactory scheduler to manage DL workloads and resources? (2) Do existing solutions share common strategies to achieve their scheduling objectives? (3) How do we need to refine the schedulers to adapt to the rapid development of DL technology? Those questions are important for system researchers and practitioners to understand the fundamental principles of DL workload scheduling and management, and design innovative schedulers for more complex scenarios and objectives. Unfortunately, there are currently no such works to summarize and answer these questions from a systematic point of view.

To the best of our knowledge, this paper presents the \textit{first} survey for scheduling both DL training and inference workloads in research and production GPU datacenters. We make the following contributions. First, we perform an in-depth analysis about the characteristics of DL workloads and identify the inherent challenges to manage various DL workloads in GPU datacenters. Second, we comprehensively review and summarize existing DL scheduling works. We categorize these solutions based on the scheduling objectives and resource consumption features. We also analyze their mechanisms to address the scheduling challenges. Such summary can disclose the common and important considerations for existing DL scheduler designs. Third, we conclude the limitations and implications from existing designs, which can shed new light on possible directions of scheduler designs in GPU datacenters. We expect this survey can help the community understand the development of DL schedulers and facilitate future designs.

\noindent\textbf{Existing surveys.} Past works also presented some surveys, which are relevant to but distinct from ours. (1) Some works summarized the optimization techniques for DL applications, such as distributed training acceleration \cite{SurveyDistDL, SurveyDML}, efficient model inference \cite{SurveyPruneQuant, SurveyInfer21}, etc. These surveys primarily focused on the acceleration of individual jobs, while we consider the global optimization of the entire datacenter with plenty of workloads for various objectives. (2) Some works surveyed the scheduler designs for conventional cloud big data \cite{SurveyCloudSched, SurveyCloudSched17} and HPC \cite{SurvHPC, SurveyHPCBigData} workloads. As discussed in Sec. \ref{sec_characteristics}, DL workloads have significantly distinct characteristics from these traditional jobs, and their scheduling mechanisms are not quite adaptable for DL training or inference. (3) Very few surveys conducted investigations on DL workload scheduling. Mayer and Jacobsen \cite{SurvScalableDL} summarized early designs of DL training job schedulers before 2019. This summary is outdated due to the emerging scheduling algorithms in recent years. Yu \etal \cite{SurveyDLServing} proposed a taxonomy for DL inference system optimization based on the computing paradigm. However, it mainly investigated the single node scenario instead of the datacenter scale.
A recent work~\cite{yu2022survey} considered the inference scheduling by colocating multiple workloads on the same GPU from both the cluster level and workload level. 
Different from those works, we provide a very comprehensive and up-to-date survey for scheduling techniques of both DL training and inference in the entire GPU datacenters.

\begin{figure*}[t]
    \centering
    \includegraphics[width=0.95\textwidth]{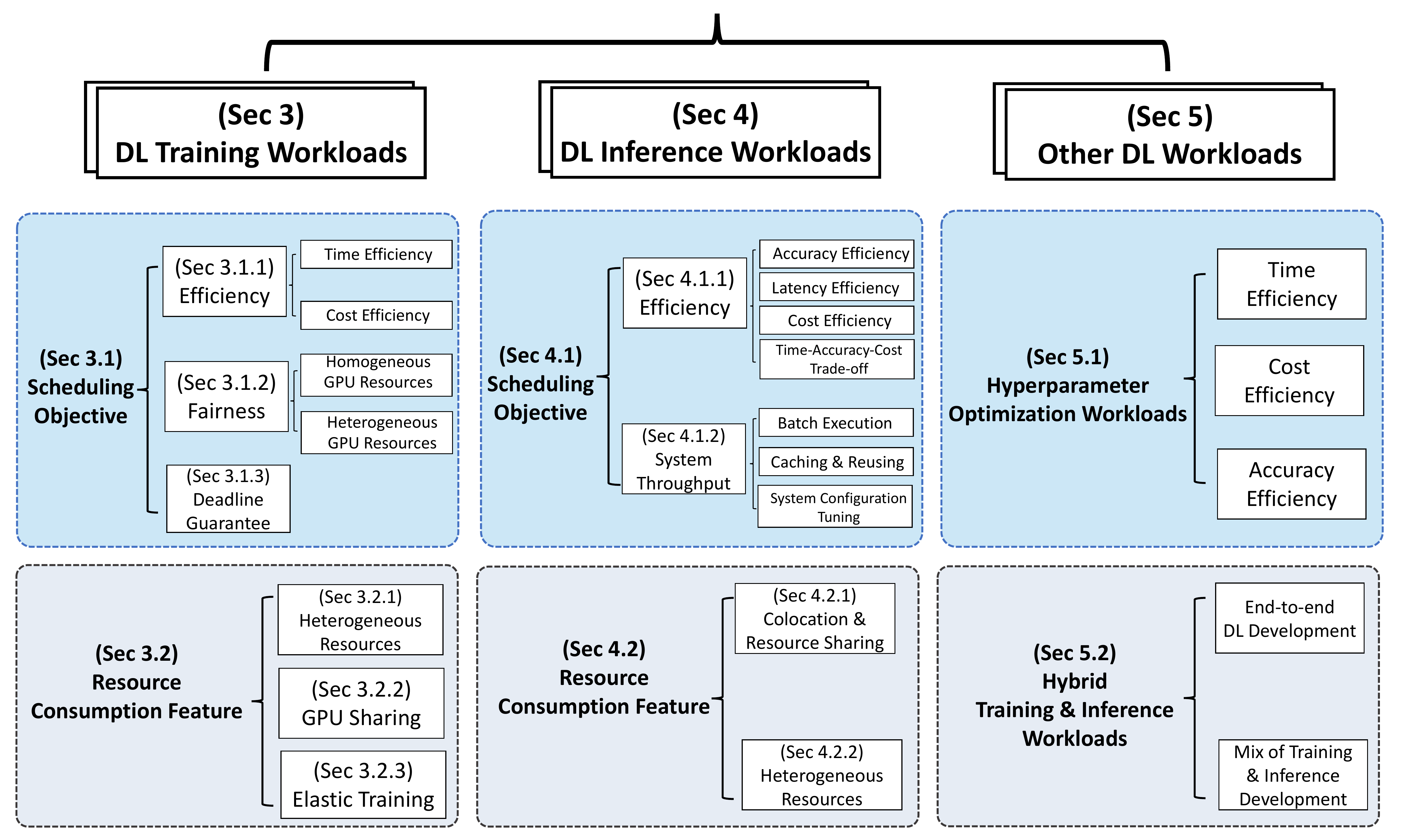}
    \vspace{-7pt}
    \caption{\textbf{The overall structure of this survey.}}
    \vspace{-15pt}
    \label{figure_structure}
\end{figure*}

\noindent\textbf{Paper organization.} The paper is organized as follows: %
Sec. \ref{sec_background} describes the unique characteristics of DL workloads and challenges for scheduling in GPU datacenters. It also illustrates the scope of this survey. The main body of this survey is presented in Fig \ref{figure_structure}. Concretely, Sec. \ref{sec_training} and Sec. \ref{sec_inference} present detailed categorizations of training and inference workloads based on the scheduling objectives and resource consumption features, respectively. Sec. \ref{sec_other_workload} discusses the other workloads, e.g., hyperparameter optimization, mixed training and inference workloads. Implications from these works are also given at the end of each section. Sec. \ref{sec_conclusion} concludes this survey paper and identifies the future directions of scheduler designs.
\section{Background}
\label{sec_background}

\subsection{DL Workloads and Their Characteristics}
\label{sec_characteristics}

A DL development pipeline typically consists of three stages: data processing, model training and model inference. In this survey, we narrow down our focus to training and inference workloads which account for the most computation and consume the majority of resources in the datacenter.

\begin{figure*}[t]
    \centering
    \includegraphics[width=0.95\textwidth]{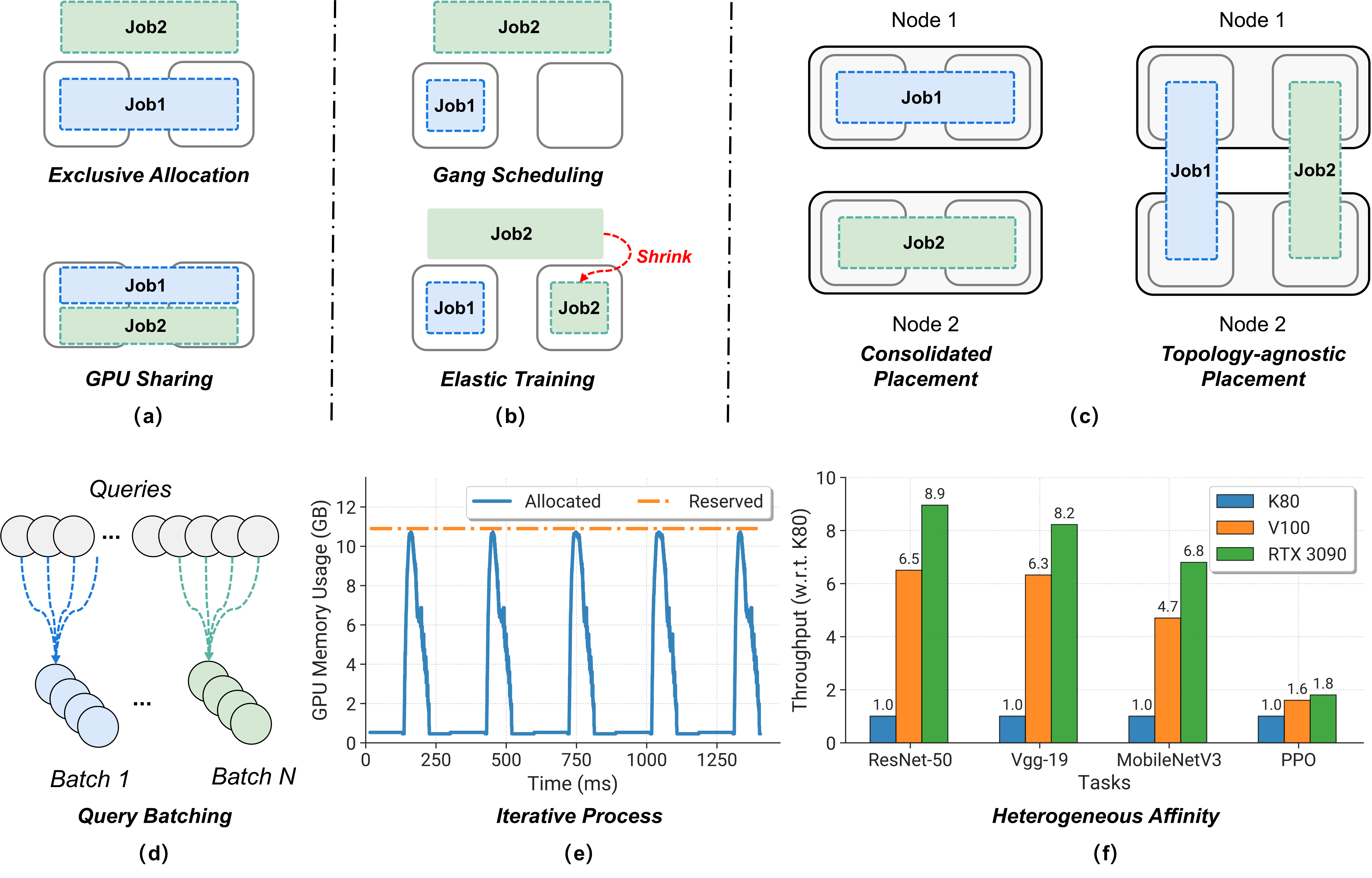}
    \vspace{-5pt}
    \caption{\textbf{Characteristics of training and inference workloads.} (\textbf{a}) Exclusive Allocation versus GPU Sharing. (\textbf{b}) Gang Scheduling versus Elastic Training. (\textbf{c}) Consolidated Placement versus Topology-agnostic Placement. (\textbf{d}) Query Batching mechanism in inference. (\textbf{e}) Iterative Process: allocated and reserved GPU memory trace profiled through \texttt{torch.profiler} (ResNet-50 ImageNet classification task). (\textbf{f}) Heterogeneous Affinity: the magnitude of speedup across GPU generations varies significantly across different tasks. }
    \vspace{-15pt}
    \label{figure_illustrate}
\end{figure*}

\subsubsection{DL Training.}
\label{sec:training-feature}
A DL training workload builds models by extracting features from existing data. A DL framework (e.g., PyTorch \cite{PyTorch}, TensorFlow \cite{TensorFlow}) is commonly adopted to fully utilize heterogeneous compute resources to accelerate the training process. To further reduce the training time, the workload is deployed across multiple GPUs with a data-parallel training scheme, which is implemented via distributed training libraries (e.g., Horovod \cite{Horovod}, \texttt{DistributedDataParallel} in Pytorch, \texttt{MultiWorkerMirroredStrategy} in Tensorflow).

DL training workloads exhibit some unique features compared to traditional big data or HPC jobs, which need to be particularly considered for GPU datacenter scheduling. A series of studies have characterized training workloads from the production GPU datacenters, including  Microsoft \cite{Philly}, SenseTime \cite{Helios} and Alibaba \cite{AliPAI, MLaaS}. The characteristics are summarized as below.

\textbf{T1: Inherent heterogeneity} \cite{Philly, Gandiva}. GPU resources play a dominant role in DL training. However, CPUs and memory might interfere with the input processing and then delay the training execution. A GPU datacenter generally offers an ample pool of CPU and memory resources compared to GPUs. Arbitrary selection of heterogeneous resource combinations by users may lead to imperfect training progress. Figure \ref{figure_illustrate} (f) shows the training performance speedups of common DL models with various generations of GPUs. Different models have diverse affinities to GPU types.

\textbf{T2: Placement sensitivity} \cite{Gandiva,Themis}. Distributed DL jobs are sensitive to the locality of allocated GPU resources. Specifically, the runtime speed of some distributed DL jobs are bounded by device-to-device communication. Figure \ref{figure_illustrate} (c) shows two types of placement, where a consolidated placement can efficiently reduce the communication overhead compared with topology-agnostic placement. The communication sensitivity of training jobs depends on the inherent property of the model structure. Advanced interconnect link (e.g., NVlink) can offer an order of magnitude higher bandwidth than PCIe. Therefore, distributed training jobs tend to request advanced interconnect to further obtain communication time reduction. Besides, jobs colocated in one server may suffer from PCIe bandwidth contention.

\textbf{T3: Iterative process} \cite{Optimus, Chronus}. DL training repeats a similar iterative pattern for up to thousands of times, as shown in Figure \ref{figure_illustrate} (e). Each iteration consists of forward propagation, backward propagation and parameter update. It motivates that profiling a small number of iterations suffices to predict the pattern of future GPU memory usage and job completion time.

\textbf{T4: Feedback-driven exploration} \cite{Gandiva, JPAS}. Training a DL model is a typical trial-and-error process. Users may explore a number of trial configurations and terminate unpromising trials by the early feedback. Such early feedback can further motivate to launch new trial configurations. Hence, a GPU datacenter hosts abundant repetitive training trials and short duration trials.

\textbf{T5: Exclusive allocation} \cite{Helios} \textbf{versus GPU sharing} \cite{MLaaS}. Figure \ref{figure_illustrate} (a) depicts the difference between exclusive allocation and GPU. Exclusive allocation refers to that a DL job exclusively has the resource usage ownership. On the contray, GPU sharing allows multiple jobs to co-locate in the same GPU device and take advantage of resources in a time-/space- sharing manner. Unlike CPUs, GPUs basically do not have the intrinsic hardware-level support for fine-grained sharing across users and thus they are allocated to DL training jobs exclusively. Due to the increasing hardware compute capability, plenty of DL training jobs can not fully utilize recent generations of GPU chips. To address this issue, datacenters enable GPU sharing through various technologies, e.g., NVIDIA Multi-Instance GPU (MIG) \cite{mig}, Multi-Process Service (MPS) \cite{mps}, GPU virtualization \cite{SurvGPUVirtual}.

\textbf{T6: Gang scheduling} \cite{Helios} \textbf{versus elastic training} \cite{Pollux}. Figure \ref{figure_illustrate} (b) illustrates two scheduling mechanisms for data-parallel DL jobs. In particular, gang scheduling is that DL training requires all the GPUs to be allocated simultaneously in an all-or-nothing manner \cite{gang}. The requirement of gang scheduling results from the native support of DL frameworks and runtime speed performance guarantee. In contrast, elastic training removes the strict GPU request constraint, and allows a dynamic number of GPUs to run training jobs. Many scheduling systems support elastic training in order to improve GPU utilization and accelerate the training process. They take advantage of the elasticity of DL training workloads: a DL training job can adapt to a wide range of GPU counts and the training processes can be suspended and resumed via checkpoints \cite{Helios}.

\subsubsection{DL inference.}
\label{sec_background-inference}Model inference is the process of making predictions to users' inputs. It is commonly applied as online services (e.g., personalized recommendations, face recognition, language translation). DL frameworks also make efforts to support inference workloads, like TensorFlow Serving \cite{TensorFlow-Serving}, MXNet Model Server \cite{MXNET-mms}, etc. The inference jobs must be performed in a real-time manner, facing dynamic queries with strict latency requirements \cite{zhang2019mark}. They may process each inference request individually, or batch multiple requests concurrently to balance the resource usage and latency. Since many inference systems are deployed in the public cloud alternative to on-premise clusters, there exist many works emphasizing how to exploit cloud resources at scale to handle inference requests.
According to the report from AWS \cite{aws2019}, the cost of DL inference has already taken up the majority (more than 90\%) of the total infrastructure cost for machine learning as a service. A DL inference workload also gives unique characteristics that can affect the scheduling system designs. They are summarized as follows.

\textbf{I1: Deterministic online execution}~\cite{Arpan2020clockwork,cuiEnable2021}. Different from offline training which could be resource-intensive and last for days or weeks, the inference for each query is often completed with sub-second response time and consumes much less resources.  Moreover, many inference jobs reveal deterministic execution flows and duration under fixed-size query input. This gives predictable resource usage and execution speed, offering the opportunities of fine-grained optimization.

\textbf{I2: High demands on latency and accuracy}~\cite{Daniel2017clipper,zhang2019mark}. First, the inference service is expected to respond to the incoming queries promptly. Delays of inference responses can cause bad user experience. For example, an online recommend service is required to provide recommendations at interactive latencies (<100ms) to prevent user losses~\cite{Daniel2017clipper}. Other kinds of inference services also have strong latency requirements (e.g., <200ms~\cite{zhang2019mark}). Second, the prediction accuracy is also critical for building a reliable inference service.
Inference workloads in some critical domains, e.g., healthcare and finance, may have stronger accuracy requirements~\cite{Halpern2019onesize}. The tight latency and accuracy demands pose great difficulty in managing inference jobs on GPUs, and there exist a trade-off between high accuracy and low latency. The datacenter managers need to carefully balance the latency overhead and prediction performance of the inference workloads.

\begin{figure*}[t]
    \centering
    \includegraphics[width=0.95\textwidth]{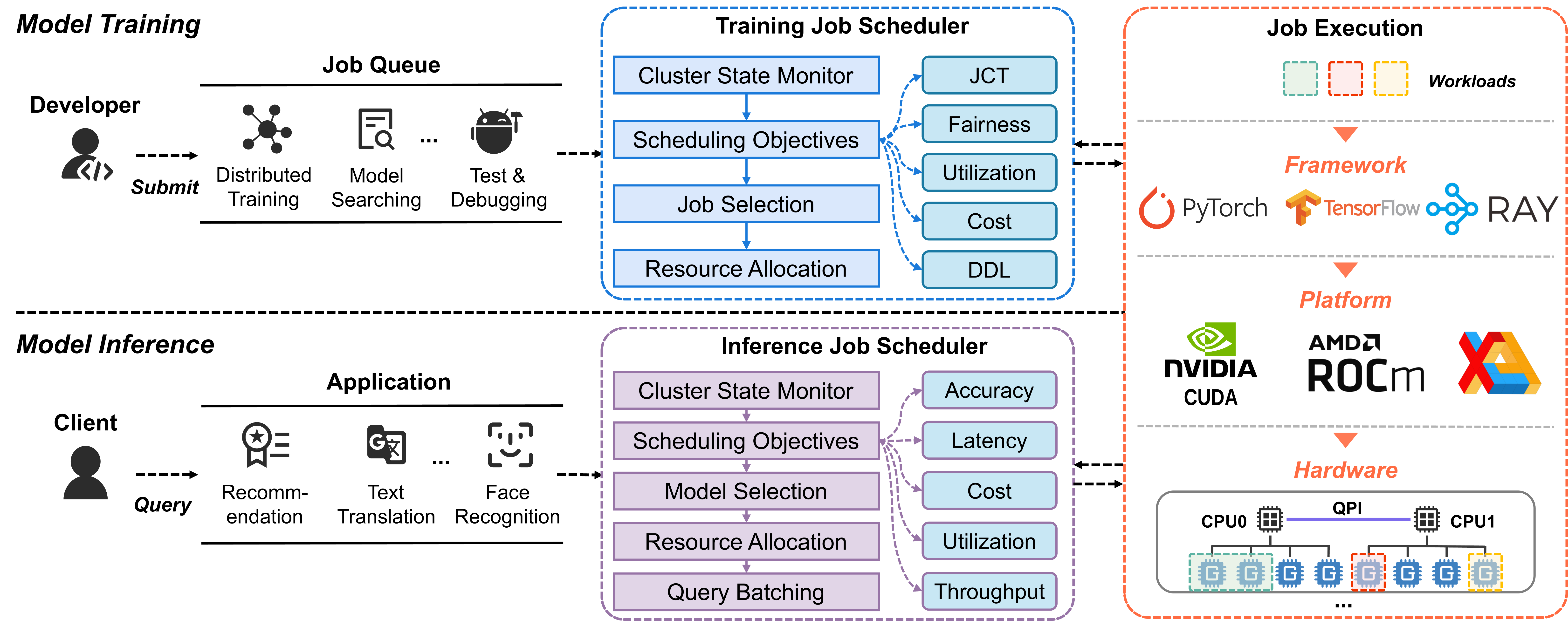}
    \vspace{-5pt}
    \caption{\textbf{Scheduling workflow for model training and inference workloads.}}
    \vspace{-15pt}
    \label{figure_a}
\end{figure*}

\subsection{Scheduler in GPU Datacenters and Design Challenges}

Scheduling has continuously drawn public attention for several decades~\cite{epema1995analysis,feitelson1995parallel,feitelson1997job}. Similar to scheduling at the level of the operating system, networking system or applications, parallel job scheduling at the datacenter level makes decisions about the allocation of computing resources to competing jobs for specific scheduling objectives~\cite{feitelson1997job}, which forms an NP-hard problem. In particular, it matches available resources with pending jobs and decides the optimal moment and amount of resources to be allocated to each job. Modern datacenters have introduced a number of schedulers to manage conventional workloads. For instance, HPC schedulers (e.g., Slurm \cite{SLURM}, OpenPBS \cite{OpenPBS}) are used to support HPC applications and scientific computing; cloud schedulers (e.g., Mesos~\cite{Mesos}, Kubernetes~\cite{BorgK8S}, Yarn~\cite{YARN}) help allocate heterogeneous compute resources for big data applications at scale.

As a special case, DL workload scheduling in GPU datacenters shares many similar features as conventional parallel job scheduling. Figure \ref{figure_a} shows the general workflow of DL schedulers in a GPU datacenter. The scheduler works on top of the DN frameworks, and assigns appropriate resources to satisfy a variety of DL workloads. It receives different types of workloads from the users. By monitoring the usages of existing compute resources in the datacenter, it delivers an efficient scheduling solution for these workloads to optimize the predetermined scheduling objective, e.g, JCT, fairness. Then it allocates the jobs to a set of hardware resources for execution. The schedulers for model training and model inference share similar logic flows but have totally different scheduling objectives, workload types, and target users. So our survey will investigate them separately (Sec. \ref{sec_training} and \ref{sec_inference}), and consider the mix of them in Sec. \ref{sec_other_workload}.

\subsubsection{Scheduling Techniques}

Some techniques and mechanisms of conventional parallel job scheduling may also apply to DL workloads scheduling in GPU datacenters. For example, to manage computing resources more efficiently and provide guaranteed service for users, it is common to divide computing resources into separate partitions and set up different queues for different users or jobs with different characteristics~\cite{feitelson1997theory, Tiresias, ASTRAEA}. Queues may also have different priorities and be equipped with different queuing policies, e.g., First-Come-First-Served and Shortest-Remaining-Time-First. Schedulers also pursue a better comprehension of affinities between workloads and resources to make wiser decisions. Therefore, mechanisms like performance modeling of workloads (e.g., online profiling~\cite{Chronus} and performance prediction~\cite{Tiresias}) and trace analysis for characterizing the cluster-level workload distribution~\cite{Helios, MLaaS} are widely adopted. Other traditional scheduling techniques (e.g., backfilling~\cite{mu2001backfilling, Tiresias}) and mechanisms (e.g., time-slicing~\cite{Gandiva}, checkpointing~\cite{ASTRAEA}, and migration~\cite{Gandiva, Gandivafair}) are also adopted for more flexible job arrangements and better resource utilization in DL workloads scheduling.

However, due to the distinct characteristics of DL jobs (Sec. \ref{sec_characteristics}), simply adopting these techniques can cause a series of issues, e.g., serving job blocking, resource under-utilization, high operation cost. Below we summarize the challenges of scheduler designs caused by DL workload features.

\begin{figure*}[t]
    \centering
    \includegraphics[width=\textwidth]{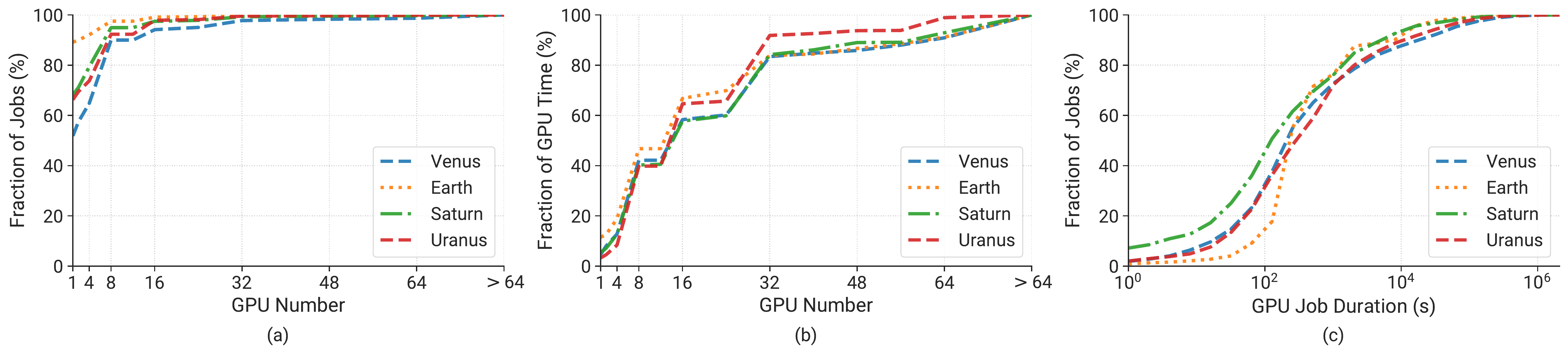}
    \vspace{-20pt}
    \caption{\textbf{Characterization of four clusters in SenseTime GPU datacenter Helios}. (a) CDF of job size with job distributions. (b) CDF of job size with GPU time distribution.  (c) CDF of job duration.}
    \vspace{-15pt}
    \label{figure_character}
\end{figure*}

\subsubsection{Challenges for Scheduling Training Jobs}
\label{subsec_challenge_training}
As discussed in Sec. \ref{sec:training-feature}, DL training workloads have some unique requirements compared to HPC or cloud jobs, which raises some challenges for scheduling them. We discuss these challenges with a workload trace analysis from four private clusters (\emph{Venus}, \emph{Earth}, \emph{Saturn} and \emph{Uranus}) in SenseTime GPU datacenter \texttt{Helios} \cite{Helios}. These clusters contain over 6000 GPUs and 1.5 million GPU jobs in total, spanning 6 months in 2020.

\textbf{C1: Intensive resource consumption}. The adoption of distributed training aims to reduce the training time yet prompts users to overclaim GPU resources for their jobs.
Figures \ref{figure_character} (a) and (b) depict the distributions of requested GPUs pertaining to job and GPU resource occupation respectively. We observe that large-size jobs ($\leq$8 GPUs) account for 10\% of the entire trace, and they consume over half of computing resources. Such intensive resource requests can aggravate the job pending issue due to the shortage of GPU resources. If the scheduler prioritizes those large-scale jobs, the situation becomes worse as subsequent jobs have to compete for much less resources. Existing solutions often favor small jobs or treat large and small jobs equally. How to balance the trade-off between intensive and light-weight resource consumption remains a challenging problem.

\textbf{C2: Unbalanced runtime distribution}. Recent trace analysis works \cite{Philly, Helios, AliPAI, MLaaS} presented the long-tail runtime distribution of DL training workloads in production GPU datacenters. Figure \ref{figure_character} (c) compares the GPU job duration distribution of each cluster. We observe it is common that job runtime varies from seconds to weeks even months among different production-level GPU clusters. The majority of workloads only finish within a short period of time, while the minority part consume many orders magnitudes of GPU time. Prioritizing short jobs is an effective way to reduce average job completion time but incurs low GPU utilization. More research efforts should be devoted to balance between short jobs and time-consuming jobs.

\textbf{C3: Heterogeneous resource affinity}. The runtime speed of a DL training job is affected by a variety of hardware factors, among which GPU heterogeneity and network link heterogeneity are the most important ones. For the impact of GPUs, DL training can benefit from newer generations of GPUs. However, the marginal benefit brought by new GPU versions varies significantly (Figure \ref{figure_illustrate} (f)). Also, the speedup ratio is unpredictable, which complicates the heterogeneous GPU resource allocation. For the impact of network links, the recently released high-end GPU interconnect including NVlink and NVswitch can significantly reduce the communication overhead across GPUs in the same sever. Along with PCI Express, InfiniBand, Ethernet and QPI, distributed training has several alternatives for cross-GPU communications. As these links differ considerably in bandwidths, and different jobs have different data sizes for exchange, it is non-trivial to allocate these network resources to the jobs to maximize the benefits and minimize the bandwidth contention.

\textbf{C4: Preemption overhead}. DL frameworks usually provide functions to pause/resume the training jobs at any time for better fault-tolerance. The overhead of such processes primarily depends upon the job scale, which ranges from seconds to minutes. In this paper, the preemption overhead is considered as the addition of the costs of pausing and resuming the job. For time-consuming jobs, the preemption overhead is relatively small with the benefit of higher scheduling flexibility. But for short jobs, the preemption overhead is non-negligible, and frequent preemption will delay their progress. Designing an appropriate preemptive mechanism requires meticulous considerations of both short and time-consuming jobs as well as their preemption overheads.

\subsubsection{Challenges for Scheduling Inference Jobs.}
\label{sec_inference_challenge}The online execution fashion and high latency requirement of inference workloads also give the following challenges for designing a scheduler.

\textbf{C5: Low GPU utilization for each request.}
Compared to training jobs, the inference service mainly involves small convolutional operations (e.g., 1x1, 3x3), and consumes small amounts of GPU resources. Besides, the peak performance of new GPUs are increasing rapidly~\cite{IOS}. This often leads to low GPU utilization for inference workloads \cite{lemay2020Perseus,zhang2020mark}. A common practice to improve the GPU utilization is to batch multiple inference requests and execute them at the same time \cite{Daniel2017clipper}.

\textbf{C6: Latency-accuracy-cost tradeoff.}
The inference jobs are relatively malleable in terms of latency, accuracy and cost. To improve the resource utilization and cluster-wide job throughput, we can colocate multiple inference jobs or increase the batch size. However, this can increase the inference latency. To increase the accuracy, effective ways include model ensemble or augmentation evaluation, which can also incur latency delay \cite{Jashwant2021Cocktail}.
The adoption of high-class hardware resources can accelerate the inference execution, but charges more for online services. Different users or inference jobs may have different demands towards latency, accuracy and cost. The scheduler needs to figure out a sweet spot for each job over an assortment of algorithms and hardware units.

\textbf{C7: Bursty and fluctuating requests.}
As an online service, it is common for the inference application to receive bursty and fluctuating requests, which are unpredictable. This must be considered when determining the resources for the workload. How to guarantee the latency with the minimal operational cost even in extremely overloading scenarios raises a new challenge. In practice, resources are often over-provisioned for inference workloads to guarantee their latency during the rush hours. Then an efficient scheduler needs to consider how to exploit the unused resources of these workloads when there are less queries.

\subsection{Relevant Studies Not Included in This Survey}

This survey mainly focuses on the scheduling of DL training and inference workloads in GPU datacenters. Other relevant works beyond the scope of this paper will not be summarized in the following sections. Here we briefly discuss these directions. Readers who are interested in these works can refer to relevant surveys \cite{SurveyDistDL, SurveyDML,SurveyPruneQuant, SurveyInfer21,SurveyCloudSched, SurveyCloudSched17,SurvHPC, SurveyHPCBigData}.

First, we do not consider the \textit{optimization solutions} for \textit{individual} training or inference jobs. Training job optimization mainly contains distributed training acceleration \cite{BytePS, Poseidon, LB-BSP} and job placement optimization \cite{HeteroG, Pack, OJPP}. Inference job optimization techniques include workload characterization~\cite{chahal2021performance}, pipeline execution~\cite{kang2020jointly}, etc. Their objectives are to achieve high performance for a single job instead of an entire datacenter.  It is worth noting that scheduling hyperparameter optimization jobs will not be considered as single job optimization, because it involves a collection of training tasks (e.g., RubberBand \cite{RubberBand}, HyperSched \cite{HyperSched}). They will be summarized in Sec. \ref{sec_other_workload}.

Second, we consider the scheduling at the job level, and do not cover the scheduling approaches at the hardware resource level (e.g., network I/O, power). For instance, HIRE \cite{HIRE} proposed a novel in-network computing scheduling algorithm for datacenter switches. A number of works \cite{DVFS-AWARE, mei2021energy, DVFS2020} utilized the DVFS mechanism on CPU and GPU chips to achieve cluster energy conservation. These works are not included in this survey.

Third, we focus on the GPU datacenters where GPUs are the primary resources for the DL workloads. Those datacenters can also exploit other resources (e.g., CPU, FPGA, ASIC) as subsidiary. This can reflect the current status of mainstream DL infrastructures. Some scheduling systems mainly utilize the CPU~\cite{yan2016SERF,Ishakian2018Serving,Bhattacharjee2019BARISTA,Gunasekaran2020Fifer}, FPGA~\cite{jiang2021MicroRec,hwang2020Centaur}, or hybrid resources~\cite{jiang2021FleetRec} where GPUs are not dominant. Some papers consider the DL services on mobile devices~\cite{Ogden2021PieSlicer} or edge computing~\cite{seo2021SLO_Aware,zhang2020QoS} other than datacenters. Those works are also out of the scope of this survey.

Fourth, we target the scheduling of general DL training and inference workloads. Some works studied other types of DL jobs, e.g., data processing, model re-training, model validation. Some papers considered the optimization of specific DL applications based on their unique behaviors, including RNN-based service~\cite{gao2018low,holmes2019grnn}, recommendation systems~\cite{gupta2020deeprecsys, liu2021jizhi,jiang2021MicroRec,jiang2021FleetRec} and video analytics~\cite{shen2019nexus,zhang2020hysia}. These works are not summarized in this paper. Besides, our aim is to enhance the system and workloads in terms of performance, efficiency and user experience. Other objectives like privacy protection~\cite{DPF,Sage} is not considered either.

\section{Scheduling DL Training Workloads}
\label{sec_training}

DL training jobs consume a majority of compute resources in GPU datacenters. Therefore an effective and efficient scheduler for training workloads is of critical importance. Existing scheduling systems can be generally categorized from two dimensions: scheduling objective and resource consumption feature. Table \ref{tab_training} summarizes the past works for DL training scheduling. We detail them in the rest of this section.

\begin{table*}[t]
    \centering
    \caption{\textbf{Summary of schedulers for DL training workloads in GPU Datacenters}.
    }
    \vspace{-10pt}
    \resizebox{\textwidth}{!}{

        \begin{tabular}{@{}cccccccccc@{}}
            \toprule
            \textbf{Year}          & \textbf{Scheduler}                   & \textbf{Objectives}           & \textbf{Approaches}                                & \textbf{Advantages}                    & \textbf{Heter.} & \textbf{Elastic} & \textbf{AutoML} & \textbf{\makecell[c]{Exp.             \\ Scale}} & \textbf{\makecell[c]{Source \\ Code}} \\ \midrule
            \multirow{2}{*}{2017}  & Dorm \cite{Dorm}                     & \ding{170}                    & Linear Programming                                 & Fairness Guarantee; JCT Reduction      & -               & -                & -               & S                         & -         \\
                                   & Topology-Aware \cite{Topology-Aware} & \ding{168}                    & Best-effort topology-aware placement               & Lower interference                     & -               & -                & -               & S                         & \ding{52} \\ \midrule
            \multirow{3}{*}{2018}  & Gandiva \cite{Gandiva}               & \ding{95}\ding{168}           & Time-slicing; Migration; Grow-shrink               & Better GPU Utilization                 & -               & \ding{52}        & \ding{52}       & L                         & -         \\
                                   & OASiS \cite{OASiS}                   & \ding{95}\ding{168}           & Primal-dual framework                              & Better GPU utilization                 & -               & \ding{52}        & -               & S                         & -         \\
                                   & Optimus \cite{Optimus}               & \ding{168}                    & Performance Modelling                              & JCT Reduction                          & -               & \ding{52}        & -               & S                         & \ding{52} \\ \midrule
            \multirow{9}{*}{2019}  & $FC^2$ \cite{FC2}                    & \ding{169}                    & Automatic Resource Configuration                   & Cost-effectiveness                     & \ding{52}*      & \ding{52}        & -               & S                         & -         \\
                                   & $Sched^2$ \cite{Sched2}              & \ding{168}                    & Q-Network based Scheduler                          & Reduce Cluster Fragmentation           & -               & -                & -               & L                         & -         \\
                                   & Cynthia \cite{Cynthia}               & \ding{169}                    & Performance Modelling                              & Monetary Cost Reduction                & -               & \ding{52}        & -               & S                         & -         \\
                                   & Dragon \cite{Dragon}                 & \ding{95}\ding{168}           & GPU time-sharing; Autoscaling                      & Better GPU utilization                 & -               & \ding{52}        & -               & S                         & -         \\
                                   & FfDL \cite{FfDL}                     & \ding{168}                    & Lesson-motivated Design                            & production DL platform                 & -               & -                & -               & S                         & \ding{52} \\
                                   & Harmony \cite{Harmony}               & \ding{168}                    & RL Scheduler; Bin Packing                          & JCT Reduction; Better GPU Utilization  & -               & -                & -               & S                         & -         \\
                                   & JPAS \cite{JPAS}                     & \ding{169}                    & Accuracy Curve Modelling                           & JCT Reduction                          & -               & -                & \ding{52}       & S                         & -         \\
                                   & Philly \cite{Philly}                 & \ding{168}                    & Locality-Relaxity                                  & Workload Analysis; JCT Reduction       & -               & -                & -               & -                         & \ding{52} \\
                                   & Tiresias \cite{Tiresias}             & \ding{168}                    & Gittins index;  Least-Attained Service (LAS)       & Information-agnostic                   & -               & -                & -               & M                         & \ding{52} \\ \midrule
            \multirow{20}{*}{2020} & $Gandiva_{fair}$ \cite{Gandivafair}  & \ding{168}                    & Gang-Aware Lottery; Automatic Trading              & Inter-user fairness guarantee          & \ding{52}       & -                & -               & XL                        & -         \\
                                   & Ada-SRSF \cite{Ada-SRSF}             & \ding{170}\ding{168}          & AdaDUAL; Least workload first                      & Less communication contention          & \ding{52}*      & -                & -               & S                         & -         \\
                                   & Antman \cite{Antman}                 & \ding{168}                    & Framework-Cluster Co-Design                        & Better GPU utilization                 & -               & \ding{52}        & -               & XL                        & \ding{52} \\
                                   & CODA \cite{Effective20}              & \ding{95}\ding{168}           & Adaptive CPU allocator; Contention eliminator; etc & Non-dominant resource aware            & \ding{52}*      & -                & -               & -                         & -         \\
                                   & Co-scheML \cite{Co-scheML}           & \ding{95}\ding{168}           & Interference-Aware Scheduler; Random Forest        & Better GPU Utilization; JCT Reduction  & -               & -                & -               & S                         & -         \\
                                   & Elan \cite{ELAN}                     & \ding{168}                    & Hybrid scaling; IO-free replication;etc            & Better GPU utilization; Less IO        & -               & \ding{52}        & -               & L                         & -         \\
                                   & E-LAS \cite{E-LAS}                   & \ding{95}\ding{168}           & Real-time Epoch Progress Rate; LAS                 & Information-agnostic                   & -               & -                & -               & -                         & -         \\
                                   & Gavel \cite{Gavel}                   & \ding{168}                    & Linear Programming; Round-based Scheduling         & Heterogeneity-Aware                    & \ding{52}       & -                & -               & M                         & \ding{52} \\
                                   & GENIE \cite{GENIE}                   & \ding{168}\ding{170}          & Light Profiler                                     & QoS Guarantee                          & -               & \ding{52}        & -               & S                         & -         \\
                                   & HiveD \cite{HiveD}                   & \ding{171}                    & Buddy Cell Allocation                              & Better Resource Utilization            & -               & -                & -               & L                         & \ding{52} \\
                                   & MARBLE \cite{Marble}                 & \ding{168}                    & Offline profiling based scaling                    & Better GPU utilization                 & -               & \ding{52}        & -               & S                         & -         \\
                                   & MLCloudPrice \cite{MLCloudPrice}     & \ding{95}\ding{168}           & Linear programming; Spot-Instance Training         & Cloud Cost Reduction                   & -               & -                & -               & -                         & \ding{52} \\
                                   & MLFS \cite{MLFS}                     & \ding{168}\ding{169}          & RL Scheduler                                       & Optimize Multiple Objectives           & -               & -                & -               & -                         & \ding{52} \\
                                   & Non-Intrusive \cite{NonIntrusive}    & \ding{168}\ding{171}          & SideCar; Early initialization                      & Framework non-intrusive                & -               & \ding{52}        & -               & S                         & -         \\
                                   & Parrot \cite{Parrot}                 & \ding{95}\ding{168}           & Linear Programming                                 & Better Bandwidth Utilization           & -               & -                & -               & -                         & -         \\
                                   & Salus \cite{Salus}                   & \ding{168}                    & Fast job switching; Memory sharing                 & Better GPU utilization                 & -               & -                & -               & S                         & \ding{52} \\
                                   & SPIN \cite{SPIN}                     & \ding{95}\ding{168}           & Rounding-based Randomized Approximation            & Robust Time Misestimation              & -               & -                & -               & -                         & -         \\
                                   & Themis \cite{Themis}                 & \ding{168}                    & Finish-Time Fairness; Auction Bid                  & Better Fairness; GPU Utilization       & -               & -                & -               & L                         & -         \\
                                   & Vaibhav et al. \cite{Effective20}    & \ding{170}                    & Dynamic programming optimization                   & Better GPU utilization                 & -               & \ding{52}        & -               & M                         & -         \\
                                   & Yeung \cite{Yeung}                   & \ding{95}\ding{168}           & GPU Utilization Prediction                         & Better GPU utilization                 & -               & -                & -               & -                         & -         \\ \midrule
            \multirow{14}{*}{2021} & $DL^2$ \cite{DL2}                    & \ding{95}                     & RL Scheduler                                       & JCT Reduction                          & -               & \ding{52}        & -               & S                         & \ding{52} \\
                                   & AFS \cite{AFS}                       & \ding{168}                    & Apathetic Future Share; CoDDL                      & Better GPU utilization                 & -               & \ding{52}        & -               & L                         & -         \\
                                   & ANDREAS \cite{ANDREAS}               & \ding{95}\ding{168}           & Randomized Greedy Algorithm                        & Energy Cost Reduction                  & -               & -                & -               & S                         & -         \\
                                   & Astraea \cite{ASTRAEA}               & \ding{169}                    & Long-Term GPU-time Fairness                        & Fairness Guarantee                     & -               & -                & -               & -                         & -         \\
                                   & Chronus \cite{Chronus}               & \ding{170}                    & Linear Programming; Local Search Allocation        & SLO Guarantee                          & -               & -                & -               & S                         & \ding{52} \\
                                   & DynamoML \cite{DynamoML}             & \ding{171}                    & Combine KubeShare and Dragon                       & Better GPU utilization                 & -               & \ding{52}        & -               & S                         & -         \\
                                   & Helios \cite{Helios}                 & \ding{95}\ding{168}           & Data Driven Prediction; QSSF; CES                  & Workload Analysis; Energy Conservation & -               & -                & -               & -                         & \ding{52} \\
                                   & Horus \cite{Horus}                   & \ding{168}                    & XGBoost-based interference prediction              & No need to online profiling            & -               & -                & -               & S                         & -         \\
                                   & Jigsaw \cite{JIGSAW}                 & \ding{95}\ding{168}           & Structured Partial Training                        & Algorithm-System Co-Design             & -               & -                & -               & M                         & -         \\
                                   & Liquid \cite{Liquid}                 & \ding{168}                    & Best-fit; Grouping genetic                         & Accelerate  job execution              & -               & -                & -               & M                         & \ding{52} \\
                                   & ONES \cite{ONES}                     & \ding{168}                    & Online evolutionary search                         & Better GPU utilization                 & -               & \ding{52}        & -               & S                         & \ding{52} \\
                                   & Pollux \cite{Pollux}                 & \ding{95}\ding{168}           & Goodput; Dynamic batch size and learning rate      & Better GPU Utilization                 & -               & \ding{52}        & \ding{52}       & L                         & \ding{52} \\
                                   & POP \cite{POP}                       & \ding{95}\ding{168}\ding{170} & Partitioned Optimization                           & Reduce Scheduling Overhead             & \ding{52}       & -                & -               & L                         & \ding{52} \\
                                   & SMD \cite{SMD}                       & \ding{169}                    & Multi-dimensional-knapsack Decomposition           & JCT Reduction                          & -               & -                & -               & -                         & -         \\ \midrule
            \multirow{6}{*}{2022}  & Aonline \cite{AOnline, AOnlineTPDS}  & \ding{168}                    & Integer Linear Programming                         & JCT Reduction                          & -               & \ding{52}        & -               & -                         & -         \\
                                   & EDL \cite{EDL}                       & \ding{95}\ding{168}           & Stop-free scaling; Graceful exit;etc               & Better GPU utilization                 & -               & \ding{52}        & -               & L                         & -         \\
                                   & GADGET \cite{GADGET}                 & \ding{95}\ding{168}           & Greedy; G-VNE                                      & JCT Reduction                          & -               & \ding{52}        & -               & -                         & \ding{52} \\
                                   & Ali-MLaaS \cite{MLaaS}               & \ding{95}\ding{168}           & GPU Sharing; Predictable Duration                  & Fine-grained Workload Analysis         & -               & -                & -               & -                         & \ding{52} \\
                                   & Singularity \cite{Singularity}       & \ding{95}\ding{168}           & Device Proxy; Replica splicing; etc                & User code non-intrusive; Efficient     & -               & \ding{52}        & -               & M                         & -         \\
                                   & Synergy \cite{Synergy}               & \ding{95}\ding{168}           & Optimistic Profiling; Greedy Scheduling            & Non-dominant resource aware            & -               & -                & -               & M                         & -         \\ \bottomrule
        \end{tabular}
    }

    \begin{flushleft}
        \begin{tablenotes}[para,flushleft]
            \scriptsize
            \textbf{Objectives:} \ding{95} Utilization \ding{168} JCT \ding{169} Cost \ding{170} Fairness \ding{171} DDL; \hspace{10pt} \textbf{Heterogeneous:}  \ding{52} heterogeneous GPUs of different generations. * heterogeneous resources (e.g., CPU, networking); \hspace{10pt}  \textbf{Experiment GPU Scales:} the scale of physical testbed. S $(0, 30]$ M $(30, 60]$ L $(60, 120]$ XL $(120, \infty]$ -: no evaluation on a physical cluster or not clearly specified.
        \end{tablenotes}
    \end{flushleft}
    \vspace{-15pt}
    \label{tab_training}
\end{table*}

\subsection{Scheduling Objectives}
Different schedulers are designed to achieve different objectives, including efficiency, fairness and deadline guarantee. We first review past works from this perspective.

\subsubsection{Efficiency}
\label{sec_training_efficiency}
Efficiency is a main objective to pursue when designing the workload schedulers. The GPU datacenter manager can consider different types of efficiency. We classify the efficiency-aware schedulers into three categories, as discussed below.

\textbf{1) Timing efficiency.} This scheduling goal is to reduce the average queuing and execution time of training workloads in a datacenter. Some advanced strategies with special training configurations (e.g., sharing training, elastic training, heterogeneous training) can help improve the timing efficiency~\cite{Gandiva,Antman,JIGSAW,ELAN,EDL,AFS,Pollux,Aryl}, which will be elaborated in Sec. \ref{sec_training_resource}. Here we mainly discuss the techniques over common training configurations that support gang scheduling, resource exclusive usage and preemptive operations.

One of the most common and effective ways for guaranteeing timing efficiency is to adopt some heuristic functions to determine the job scheduling priority {\tiny\encircle{\normalsize C1}}\footnote{{\tiny\encircle{\normalsize CX}} indicates the challenge (Sec. \ref{subsec_challenge_training} and \ref{sec_inference_challenge}) to be addressed.  {\tiny\encircle{\normalsize TX}} and {\tiny\encircle{\normalsize IX}} indicate the training and inference job characteristic (Sec. \ref{sec:training-feature} and \ref{sec_background-inference}) to be considered in the scheduler design.}. For instance, Tiresias~\cite{Tiresias} designs the \emph{Least Attained Service} (LAS) algorithm to prioritize jobs based on their \textit{service}, a metric defined as the multiplication of requested GPU resources and execution time. It devises the priority discretization to mitigate the frequent preemption issue {\tiny\encircle{\normalsize C4}}, which is inspired by the classic Multi-Level Feedback Queue (MLFQ) algorithm~\cite{corbato1962experimental,arpaci2018operating,chowdhury2015efficient}. These techniques enable Tiresias to beat the classical YARN-CS~\cite{YARN} significantly. E-LAS~\cite{E-LAS} improves over Tiresias by prioritizing jobs with the  \emph{real-time epoch progress rate}, which is computed as the proportion of the current training epoch over the total number of training epochs. With such improvement, E-LAS outperforms Tiresias in terms of average job timing efficiency. 
FfDL~\cite{FfDL} is an open-sourced scheduler platform developed by IBM. It uses the operating lessons from the industry practice to guide the management of DL training workloads in the cloud.

An alternative strategy is to use machine learning techniques for job scheduling.
$Sched^2$~\cite{Sched2} is a scheduler based on reinforcement learning (RL). It utilizes a Q-network which takes the job state and GPU datacenter state as input, and outputs the optimal job to be scheduled. MLFS~\cite{MLFS} also leverages RL to determine the job priority and resource allocation. The RL model takes as input the job time information, resource demand, and accuracy requirements. It can effectively improve the average latency of a mix of data-parallel and model-parallel training jobs. Helios~\cite{Helios} characterizes the production training jobs from a shared GPU datacenter in SenseTime, and then adopts a variety of machine learning algorithms to predict the job priority from the history job information. The prediction result suffices to minimize the cluster-wide job latency. JPAS~\cite{JPAS} is a scheduler based on the accuracy curve fitting technique to expedite the feedback-driven exploration of general training workloads {\tiny\encircle{\normalsize T4}}. The feedback-driven exploration readily expects the scheduler to allocate more resources for more accurate models. JPAS leverages the accuracy curve fitting to predict the potential maximal accuracy improvement of each job, and then prioritize the jobs in a time interval. With this technique, JPAS can facilitate the early-stage progress of the training workloads and satisfy the needs for the feedback-driven exploration.

The timing efficiency of DL training jobs is highly dependent on the job placement {\tiny\encircle{\normalsize C3}}, where different placement policies can lead to different communication overheads. Users prefer the strict placement locality to maintain the DL training speed {\tiny\encircle{\normalsize T2}}. Amaral \etal \cite{Topology-Aware} found that packing jobs on the same CPU socket could bring up to 1.3$\times$ speedup compared to spreading jobs across different sockets. Then they designed the Topology-Aware scheduler, which uses a profiler to measure the placement sensitivity of each job, and thus performs a best-effort approach to schedule locality-sensitive jobs in a packing manner.
Similarly, Tiresias~\cite{Tiresias} and E-LAS~\cite{E-LAS} also adopt the profiling strategy to identify the optimal job placement solutions. SMD~\cite{SMD} is a scheduler for parameter-server (PS) training jobs, which allows multiple jobs to contend the communication bandwidth. It models the scheduling problem as a non-convex integer non-linear program with the bin-packing constraints, and then develops an $\epsilon$-approximation algorithm called sum-of-ratio multi-dimensional-knapsack decomposition to solve it. The effectiveness of the SMD scheduler is validated both theoretically and empirically. Philly~\cite{Philly} investigates a production workload trace from Microsoft and conducts a thorough analysis about the impact of gang scheduling and locality constraints on the queuing delay and job runtime. Motivated by this, it proposes to relax locality constraints to improve the job timing efficiency.

Sometimes the scheduler can satisfy the GPU capacity request but fail to meet the placement locality. This will lead to the cluster fragmentation issue, which is often caused by the scattered GPU resource allocation. HiveD~\cite{HiveD} emphasizes that sharing the GPU datacenter without the consideration of cluster fragmentation will cause significant job queuing delay. Therefore it develops a buddy cell allocation mechanism to ensure \emph{sharing safety}. HiveD can be easily incorporated with Tiresias~\cite{Tiresias} to reduce the queuing delay and further improve the job latency. $Sched^2$~\cite{Sched2} addresses the cluster fragmentation problem with an RL model, which is able to satisfy the locality constraint as much as possible. SPIN~\cite{SPIN} observes that delay scheduling~\cite{zaharia2010delay} can bring reward to the GPU datacenter in the long term for satisfying the placement locality in the near future. It requires the job runtime information to determine the delay scheduling policy. SPIN proposes a rounding-based randomized approximation method to achieve this goal, which has strong robustness even with inaccurate job runtime estimation.

\textbf{2) Cost efficiency.} This refers to the reduction of power consumption or financial cost for renting cloud services. This is another significant objective for training workload scheduling.

Existing GPU datacenters have considerable power waste as not all the GPUs are actively used all the time, while the datacenter managers prefer to keep all the devices on. To reduce the energy cost, ANDREAS~\cite{ANDREAS} considers a scenario where the execution of each job can be postponed within a certain period. Then it judiciously schedules jobs at appropriate moments to keep all the GPUs busy in the datacenter. It formulates the power consumption as a Mixed Integer Non-Linear Programming problem, and proposes an effective greedy heuristic algorithm to achieve a significant cost reduction. Different from ANDREAS, the \emph{Cluster Saving Service} (CES) in Helios~\cite{Helios} has no assumption about postponing the execution of DL training jobs. It leverages a prediction model to estimate the future resource utilization from the history logs. Then the scheduler can decide how many GPU nodes should be turned on/off. CES can save the electricity by up to 1.65 million kilowatt-hours per year in four production clusters from SenseTime. Additionally, recent energy optimization frameworks such as GPOEO~\cite{GPOEO} can significantly save the power consumption of training workloads. Although they are not tailored for GPU datacenters, they can be easily transplanted into the GPU datacenter with a customized scheduler to orchestrate between datacenters and jobs.

Cloud GPU resources are billed based on the amount and duration of usage. Training a model can be very time-consuming and resource-intensive {\tiny\encircle{\normalsize C1}}. As such, the cost of a training workload could be considerably expensive. It is critical to reduce such financial cost to produce the model with the same quality. Particularly, PS training is a common method for distributed data-parallel model training in the cloud.
Cynthia~\cite{Cynthia} is a scheduler to guarantee the cost-effectiveness of cloud resource provision for PS training. It introduces an analytical performance model to characterize the relationship between throughput and resource provision. Through this performance model, this scheduler can identify an optimal resource type and PS configurations to maintain the training throughput while minimizing the monetary cost.
Analogously, $FC^2$~\cite{FC2} is a scheduler, which recommends cost-effective and high-performing cloud resource configurations for PS training jobs. It selects the instances with the largest network bandwidth within the budget for the parameter server in order to avoid the communication bottleneck. It also proposes a heuristic method named \emph{Scala-Opt} to decide the work instances which can guarantee the job throughput while maximizing the cost savings.
Jahani~\cite{Jahani} treats the compute node with different numbers of GPUs as different virtual machines (VMs). The renting cost and job throughput vary with different VM types.
Then it models the scheduling process as a Mixed Integer Linear Programming (MILP) problem, and reduces the renting cost in a global manner while maintaining the job latency.
MLCloudPrice \cite{MLCloudPrice} makes a quantitative analysis on the price difference among different GPU specifications and dynamic prices of the public cloud. It moves the workloads between spot and on-demand instances, which opportunistically utilizes the low-pricing spot instance to push forward the training progress.

\subsubsection{Fairness}
Fairness indicates how fairly the compute resources are allocated among different entities, including user groups (i.e., tenants) and workloads. Fairness schedulers aim to guarantee that each entity can achieve better performance with the resource sharing mechanism than exclusively using the same portion of resources. For conventional workloads, the design of fairness schedulers follows some typical fairness principles, such as sharing incentive, strategy-proofness, envy-freeness and pareto efficiency~\cite{DRF}.
It is more challenging to maintain fairness for DL training workloads for two reasons: (1) A GPU is an indivisible resource in common settings (gang scheduling) {\tiny\encircle{\normalsize T6}}; (2) DL training exhibits resource heterogeneity preference {\tiny\encircle{\normalsize T1}} {\tiny\encircle{\normalsize C3}}. Below we discuss the new works that can address these two challenges for fairness scheduling of training workloads.

\textbf{1) Homogeneous GPU resources.} A datacenter with only one generation of GPU devices can be considered as a homogeneous GPU environment. The scheduler in this system achieves fairness sharing of indivisible GPU resources from the timing dimension.
For instance, Themis~\cite{Themis} maintains the job-level fairness by introducing a new metric called ~\emph{finish-time fairness}. This metric inspires the scheduler to allocate more resources to the jobs whose attained service is less than the deserved amount. Moreover, in existing fairness schedulers (e.g., DRF~\cite{DRF}), the placement preferences of DL training workloads can result in severe fairness sharing loss. To address this problem, Themis builds a two-level scheduling architecture for biding resource allocation among jobs and uses the game theory to guarantee the performance.
Astraea~\cite{ASTRAEA} concentrates on the fairness across workloads and tenants.
It introduces the Long-Term GPU-time Fairness (LTGF) metric to measure the sharing benefit of each tenant and job, and proposes a two-level max-min scheduling discipline to enforce job-level and tenant-level LTGF in a shared GPU datacenter.

\textbf{2) Heterogeneous compute resources.} It is relatively easy to maintain fairness over one type of GPUs. However, the existence of multiple generations of GPUs and other compute resources (e.g., CPUs, network links) can also exacerbate the fairness of workloads or user groups {\tiny\encircle{\normalsize T1}} {\tiny\encircle{\normalsize C3}}. A couple of works have introduced solutions to achieve fairness in the heterogeneous environment\footnote{Note here we focus on how to fairly allocate heterogeneous resources. The consumption optimization of specific heterogeneous resources will be discussed in Sec \ref{sec:train-resource-heter}.}.

To achieve the fairness over GPUs and other compute resources, Allox~\cite{Allox} is a fairness scheduler, which
assumes that both GPUs and CPUs are interchangeable resources, and takes into account the affinity of workloads towards different compute resources. It models the resource allocation as a min-cost bipartite matching problem with a theoretically optimal property. Then it proposes a greedy heuristic solution to solve this problem in an effective and scalable way. Dorm~\cite{Dorm} is another fairness scheduler for the fair sharing of GPUs, CPUs and memory resources. It assumes
that GPUs, CPUs and memory are complementary resources and the capacity of each one can influence the training job throughput.
It dynamically partitions different types of compute resources for each DL training job. It formulates the resource allocation as an MILP problem with the resource utilization fairness as the optimization objective. The scheduling decision in each round is made by calling the MILP solver to optimize the utilization fairness.

It is also challenging to achieve fairness over different generations of GPUs. Datacenter users prefer to request the most powerful GPU resources for their training jobs. However, many jobs can not saturate the peak performance of these high-end GPUs. Besides, different DL training jobs have different sensitivities of runtime speed to the compute capability of GPUs. $Gandiva_{fair}$~\cite{Gandivafair} is an early fairness scheduler dedicated for the heterogeneous GPU resource environment. It targets the inter-user fairness in the GPU heterogeneity.
To maintain such fairness while maximizing the cluster-wide job efficiency, $Gandiva_{fair}$ allows users to transparently trade heterogeneous GPU-time by a couple of techniques including profiling and automatic trade pricing.
Gavel~\cite{Gavel} is another heterogeneity-aware fairness scheduler. It profiles the performance heterogeneity between different types of GPUs and DL model architectures. A round-based scheduling technique is adopted to improve the scheduling flexibility and ensure timely GPU-time re-allocation. This scheduler can satisfy different types of fairness defnitions, e.g., max-min fairness, makespan minimization, finish-time fairness minimization. However, it is prohibitive to scale up Gavel to a large datacenter due to the time-consuming mathematical solving process. To this end, POP~\cite{POP} proposes to partition a large datacenter into several smaller ones. Then the original complex optimization formulation is decomposed into multiple smaller problems and can be solved in parallel. It provides a theoretical proof and several empirical evidences to demonstrate the effectiveness of this optimization technique.

\subsubsection{Deadline Guarantee}
Different from the efficiency goal which aims to complete the job as soon as possible, this objective is to ensure the job can be done before the specified deadline. It is relatively less studied due to the lack of comprehensive analysis about the deadline requirement in DL workloads. An early deadline-aware scheduler for DL training workloads is GENIE~\cite{GENIE}.
It develops a performance model to predict the job throughput on different resource placement policies. The performance model only requires a small number of training iterations to profile without any significant degradation of job execution {\tiny\encircle{\normalsize T3}}. With this performance model, GENIE can identify the best placement policy for each job to satisfy the corresponding deadline requirement. However, GENIE~\cite{GENIE} does not investigate the deadline requirement from users and cannot support a mixed workload of deadline and best-effort jobs. In \cite{Chronus}, a user survey is conducted to uncover users' latent needs about the deadline guarantee, and comprehensively discuss the deadline requirement from GPU datacenter users. Motivated by this survey, it introduces Chronus, a scheduler to improve the deadline guarantee for Service-Level-Objective (SLO) jobs and latency of best-effort jobs at the same time. It formulates the deadline-guarantee scheduling task as an MILP problem with the resource and time constraints. The MILP solver can make effective scheduling decisions for a collection of jobs. Moreover, in consideration of the placement sensitivity of different training jobs, it proposes round-up and local-search techniques to make placement decisions. These designs successfully enable Chronus to outperform existing deadline schedulers in reducing deadline miss rates and improving the latency of best effort jobs.

\subsection{Resource Consumption Feature}
\label{sec_training_resource}
In addition to the scheduling objective, another orthogonal view to categorize training workloads is their resource consumption features. We discuss prior works based on whether they adopt heterogeneous resources, GPU sharing and elastic training.

\subsubsection{Heterogeneous Resources}
\label{sec:train-resource-heter}
Most schedulers focus on the allocation of GPU resources, as they dominate the DL training. However, the consumption of CPUs and memory can also affect the training performance {\tiny\encircle{\normalsize C3}}. Synergy \cite{Synergy} observes that different DL training jobs exhibit different levels of sensitivity to the CPU and memory allocation. An optimal allocation can improve the overall cluster utilization and efficiency. Therefore, it introduces \emph{optimistic profiling} to empirically profile the job throughput for various CPU allocations and analytically estimate all the combinations of CPUs and memory along with the respective storage bandwidth requirement. Based on the profiling results, it performs round-based scheduling and greedily packs runnable jobs along multiple resource dimensions with the objective of minimizing the fragmentation in each round {\tiny\encircle{\normalsize T1}}. CODA \cite{CODA} observes that CPU jobs colocating within the same compute node can interfere with the training jobs due to the CPU resource contention. It then designs three components to optimize system-wide performance: an \emph{adaptive CPU allocator} identifies the optimal CPU cores for each DL training job; a \emph{real-time contention eliminator} monitors and throttles the memory bandwidth of each CPU job to reduce its interference with the GPU training jobs; a \emph{multi-array job scheduler} allows CPU jobs to preempt the CPU cores reserved by the GPU jobs accordingly, and vice versa. Experimental results demonstrate CODA can efficiently improve the GPU utilization without sacrificing the performance of CPU jobs.

Beyond the CPU and memory resources, network bandwidth is another bottleneck for efficient DL training. Ada-SRSF \cite{Ada-SRSF} is a two-stage framework for mitigating the communication contention among DLT jobs. In the job scheduling stage, it is combined with the classical SRSF algorithm to relax the contention of two jobs if it can reduce the job completion time. In the job placement stage, it strives to balance the resource utilization and communication overhead. Liquid \cite{Liquid} proposes a cluster network-efficient scheduling solution to achieve better placement for PS-based distributed workloads. Specifically, it adopts a random forest model to predict job resource requirements and then uses the best-fit algorithm and grouping genetic algorithm to optimize the execution performance of DL jobs. Parrot~\cite{Parrot} is a framework to manage network bandwidth contention among training jobs using the PS architecture. The communication scheme in a PS workload exhibits a coflow chain dependency where the event of parameter-pull happens after the event of parameter-push. Parrot tries to assign the bandwidth of each physical link to coflows while satisfying the dependency constraints in order to minimize the JCT. It adopts a least per-coflow attained service policy to prioritize jobs. Then it uses a linear program (LP) solution to derive a weighted bandwidth scaling strategy to minimize the time cost in the communication stage.

\subsubsection{GPU Sharing}

With the increased compute capability and memory capacity of GPUs, the conventional placement approach which makes each DL job exclusively use the GPU can lead to severe resource underutilization. It is now more promising to perform GPU sharing to fully exploit GPU resources and improve the system throughput {\tiny\encircle{\normalsize T5}}. In this context, \emph{utilization} is more inclined to the usage of every single GPU instead of the occupied GPU quantity at the datacenter scale.

Some works profile and revoke unsuitable jobs to achieve efficient GPU sharing. Salus \cite{Salus} focuses on fine-grained GPU sharing with two primitives: \emph{fast job switching} enables efficient time sharing and rapid preemption for active DL jobs on a GPU; \emph{memory sharing} addresses the memory management issues to ensure high utilization by packing more small DL jobs on the same device. Gandiva \cite{Gandiva} designs a \emph{packing} mechanism to pack multiple jobs on one GPU under the constraints of GPU memory and job performance. It utilizes a profiling mechanism to monitor and unpack jobs that could affect jobs’ performance. Jigsaw~\cite{JIGSAW} is designed upon a novel distributed training scheme named \emph{Structured Partial Backpropagation} (SPB). SPB allows each worker not to perform the entire backward pass in the distributed training. This can save lots of compute resources, and enable efficient time- and space-multiplexing across jobs in a single GPU.
Although SPB can reduce the cluster-wide JCT, it might lead to accuracy loss to some extent. Recently, Antman \cite{Antman} is introduced, which co-designs the infrastructure between the cluster scheduler and DL framework engine to efficiently manage GPU resources in a fine-grained manner. It supports the co-execution of multiple jobs on a GPU device, and thus largely improves the overall compute resource utilization. Ali-MLaaS \cite{MLaaS} provides a comprehensive analysis of large-scale workload traces in Alibaba, and discloses the benefit of GPU sharing in production GPU datacenters.

Alternatively, some works use data-driven approaches to make the GPU sharing decision. Horus \cite{Horus, Yeung} designs a prediction-based interference-aware mechanism that can be integrated with existing DL training scheduling frameworks. The \emph{prediction engine} in Horus is in charge of estimating the GPU usage of each DL job by accessing its graph and dry running the model upon the job submission. Based on the prediction results, Horus allocates GPU resources to DL jobs via de-prioritizing co-location placement decisions that would result in JCT slowdown from the severe interference and communication delays. Co-scheML \cite{Co-scheML} also measures some metrics for each DL job and uses a random forest model to predict the interference. Then the scheduler makes the decision with the aim of fully utilizing the cluster resources. Analogously, Liquid \cite{Liquid} also supports fine-grained GPU sharing for further resource utilization improvement using a random forest model. Harmony~\cite{Harmony} applies an RL model to make placement decisions for minimizing the interference and maximizing the throughput for bin packing DL workloads in a GPU datacenter. It contains a reward function for the prediction module and RL placement decision-making module. This reward function aims to maximize the normalized training speed across all the concurrent jobs in a fixed scheduling interval. The training speed estimation of bin-packing jobs can not be directly obtained, and it depends upon a neural network model via supervised learning from historical logs. To stabilize and accelerate RL model training, Harmony adopts several techniques including actor-critic, job-aware action space, and experience replay. Putting them together, Harmony outperforms significantly over heuristic baselines.

\subsubsection{Elastic Training}

In order to maximize the GPU utilization and improve the training efficiency, many novel schedulers support elastic training, which dynamically adjusts the parallelism and resource allocation of workloads to achieve the objectives {\tiny\encircle{\normalsize C1}} {\tiny\encircle{\normalsize T6}}. 

Gandiva \cite{Gandiva} designs a \emph{Grow-Shrink} mechanism which uses the profiling information to estimate each job’s progress rate and then allocates GPUs accordingly. Optimus \cite{Optimus} estimates the loss reduction rate on any placement policies based on a performance model. then it designs a greedy resource allocation scheme to prioritize the maximum marginal loss reduction. This greedy policy successfully maximizes the cluster-wide training throughput. Elan \cite{ELAN} designs several mechanisms to achieve efficient elastic training: \emph{hybrid scaling} can better trade-off the training efficiency and model performance; \emph{concurrent IO-free replication} leverages RDMA to reduce the numbers of IO and CPU-GPU memory copy operations; \emph{asynchronous coordination} avoids the high overhead of start and initialization during re-adjustments. With the integration of the FIFO and backfill scheduling algorithms, Elan successfully improves the cluster resource utilization and reduces the job pending time. AFS \cite{AFS} is proposed based on the insight that the scheduler should proactively prepare for the resource contention in the future by utilizing the current resources. It considers both resource efficiency and job length for resource allocation while amortizing the cost of future jobs into the calculation of the current share. Besides, a DL training system framework, CoDDL, is also implemented to deliver automatic job parallelization and efficient re-adjustments. EDL \cite{EDL} also supports elasticity in DL job scheduling. It implements \emph{stop-free scaling} and \emph{graceful exit} to minimize the scale-out and scale-in overheads respectively. Furthermore, EDL optimizes the data allocation pipeline by on-demand and pre-fetching data. MARBLE \cite{Marble} enables elastic DL training in HPC systems. It determines the optimal number of GPUs through offline profiling and employs a FIFO-based policy for scheduling. Vaibhav \etal \cite{Effective20} designs a job scalability analyzer and a dynamic programming based optimizer to determine the batch sizes and GPU counts for DL jobs. OASiS \cite{OASiS} introduces a primal-dual framework for optimizing the distributed PS-based jobs, which is coupled with efficient dual subroutines to achieve good long-term performance guarantees with polynomial time complexity. During the training, OASiS dynamically scales in or scales out the number of workers and parameter servers for the best resource utilization and training expedition.

Some online scheduling algorithms adopt the elastic training mechanism for datacenter optimization. For instance, GADGET \cite{GADGET} formulates a resource scheduling analytical model for ring-all-reduce DL and uses a greedy approach to maximize the utility of unfinished jobs. It obtains provable guarantee for high performance. AOnline \cite{AOnline,AOnlineTPDS} uses the integer linear program to formulate the maximum weighted schedule problem. It schedules a job if its weight is higher than its estimated serving cost to maximize the total weight of scheduled jobs.

A number of works apply RL to optimize the elastic training policy. Specifically, RIFLING \cite{RIFLING} adopts K-means to divide concurrent jobs into several groups based on the computation-communication ratio similarity. The group operation reduces the state space and accelerates the convergence speed of the RL model. The RL model only determines the number of GPUs and nodes for each job. This can effectively reduce the action space. A reward function is designed to minimize the resource fragmentation and improve the job throughput. RIFLING chooses the Q-Learning algorithm and allows the RL model to perform online update from historical logs to adapt to the workload variation. $DL^2$~\cite{DL2} is another RL scheduler focusing on the PS architecture and dynamically adjusts the resources allocated to the parameter server and workers. It mitigates the optimization instability by a combination of offline supervised learning and online actor-critic reinforcement learning. The RL model also takes the job state and resource state as input and then makes the resource allocation decision for each job.
The reward function targets the cluster-wide normalized epoch progress.
These techniques enable $DL^2$ to present satisfactory job latency reduction even for unseen job types.

Some works focus on the optimization of elasticity implementation in practical schedulers, e.g., Kubernetes. Wang \etal \cite{NonIntrusive} developed an elastic scheduling framework as plugins in Kubernetes. It uses the training job progress information to allocate and reallocate GPUs to minimize JCT.
It efficiently reallocates GPUs based on a \emph{SideCar} process, which introduces an early initialization mechanism for fast reshaping down and achieves non-intrusion to DL training frameworks.
DynamoML \cite{DynamoML} is a Kubernetes platform which combines KubeShare \cite{KubeShare} and Dragon \cite{Dragon} for DL workload scheduling. Dragon~\cite{Dragon} fills the gap that existing Kubernetes schedulers fail to manage the distributed training workloads. It resolves this issue by introducing three enhancements including gang-scheduling, locality-aware placement and autoscaling of training workloads. DynamoML also supports scheduling optimization for inference jobs, which will be discussed in Sec. \ref{sec_inference}.

In addition to the elasticity of GPU resources, DL job configurations can also be dynamically adjusted {\tiny\encircle{\normalsize C3}}. Pollux \cite{Pollux} aims to achieve higher utilization by automatically configuring the DL training jobs and co-adaptively allocating resources to them. Specifically, it defines \emph{goodput}, a metric for comprehensively measuring training performance including system throughput and statistical efficiency. It designs a joint scheduling architecture to maximize the goodput. At the job-level granularity, Pollux dynamically tunes the batch size and learning rate for the best utilization of the allocated resources. At the cluster-level granularity, Pollux dynamically (re-)allocates resources based on the goodput of all the jobs sharing the cluster as well as other cluster-level objectives (e.g., fairness, JCT). Aryl \cite{Aryl} further extends Pollux by dynamically loaning idle inference GPU nodes to training jobs. It brings higher cluster utilization and lower queuing time. Similar to Pollux, ONES \cite{ONES} automatically manages the elasticity of each job based on the training batch size. It designs an online evolutionary search algorithm to continuously optimize the scheduling decisions, which achieves superior performance compared with greedy strategies. More recently, Microsoft presents Singularity \cite{Singularity}, an epochal distributed scheduling system for Azure DL training and inference workloads. It achieves transparent preemption, migration and elasticity across a global fleet of AI accelerators (e.g., GPUs, FPGAs). It implements \emph{device proxy} for the decoupled execution and elastic scheduling across the workers. Although it is developed for public cloud services, the promising techniques are also effective in managing private GPU datacenters.

\subsection{Implications}
The scheduling objective plays an important role in designing schedulers for GPU datacenters. A majority of schedulers consider timing-efficiency and fairness. In contrast, other objectives including deadline guarantee, cost efficiency and accuracy efficiency are not fully explored yet, although they have been thoroughly considered in the conventional cloud and HPC systems. Modern cloud providers are accelerating the pace of building GPU platforms to support a sizable number of training workloads. We anticipate these objectives are also important for training workload management. This inspires researchers and developers to jointly optimize their objectives with the constraints of deadline guarantee and cost.

According to the unique resource consumption features of DL training jobs, datacenter managers can enhance the overall resource utilization and improve users' experience through designing more efficient resource allocation mechanisms, e.g., fine-grained job placement on GPUs, dynamic job parallelism adjustment, adaptive CPU allocation, etc. However, these approaches have their limitations that can hinder their deployment in practice. For instance, adaptive training could change jobs' batch size, learning rate and GPU amount, which can cause model convergence issues. Its generalization for more application scenarios also requires more validations. Job colocation can cause potential performance degradation and fault tolerance issue, which can make users unwilling to adopt this feature. How to address these practical issues is a promising and challenging future direction. We look forward to seeing more progress in this topic.

Although different scheduling algorithms for conventional workloads and systems have been extensively studied for decades, it still requires more efforts to design effective scheduling algorithms for large-scale GPU datacenters to reduce the operational cost and improve the job throughput. The rapid development of AI technology motivates researchers to investigate the possibility of using machine learning to optimize scheduler designs. From our summary, ML-based schedulers have shown their effectiveness in some scenarios. However, the datacenter managers are still concerned about the reliability and scalability of these ML-based schedulers. We expect more research works will be performed to address these concerns and make these schedulers more practical.

\section{Scheduling DL Inference Workloads}
\label{sec_inference}

\begin{table*}[t]
    \centering
    \caption{\textbf{Summary of schedulers for DL inference workloads in GPU Datacenters.}}
    \vspace{-10pt}
    \resizebox{\textwidth}{!}{

        \begin{tabular}{@{}cccccccccc@{}}
            \toprule
            \textbf{Year}         & \textbf{Scheduler}                           & \textbf{Objectives}            & \textbf{Approaches}                              & \textbf{Advantages}                                    & \textbf{Batching} & \textbf{Colocate} & \textbf{Cloud} & \textbf{\makecell[c]{Exp.             \\ Scale}} & \textbf{\makecell[c]{Source \\ Code}} \\ \midrule
            2017                  & Clipper \cite{Daniel2017clipper}             & \ding{168}\ding{170}\ding{171} & Query-level caching; Layered   architecture      & General abstraction for model selection                & \ding{52}         & -                 & -              & M                         & \ding{52} \\ \midrule
            \multirow{3}{*}{2018} & Space-Time \cite{jain2018dynamic}            & \ding{171}\ding{95}            & GPU sharing across space and time                & Performance isolation under sharing                    & \ding{52}         & \ding{52}         & -              & S                         & -         \\
                                  & Ease.ml   \cite{li2018easeml}                & \ding{168}                     & Multi-tenant model selection                     & Homogeneous declarative   inference platform           & -                 & -                 & -              & -                         & \ding{52} \\
                                  & HiveMind \cite{HiveMind}                     & \ding{171}                     & Sharings of pipelines, weights,   and layers     & Multi-model training and inference                     & \ding{52}         & \ding{52}         & -              & S                         & -         \\ \midrule
            \multirow{8}{*}{2019} & MArk \cite{zhang2019mark}                    & \ding{170}\ding{169}           & Predictive autoscaling on serverless instances   & Flexible to burst requests                             & \ding{52}         & -                 & \ding{52}      & S                         & \ding{52} \\
                                  & Tolerance Tiers   \cite{Halpern2019onesize}  & \ding{168}\ding{170}\ding{169} & Service Version Ensembling                       & Explicit accuracy-latency   trade-off in requests      & -                 & -                 & \ding{52}      & S                         & -         \\
                                  & ParM \cite{ParM}                             & \ding{170}                     & Coded-computation via a learning-based approach  & Erasure-coded resilience for inference                 & \ding{52}         & -                 & -              & M                         & \ding{52} \\
                                  & Gilman et al.   \cite{Challenges19}          & \ding{170}\ding{171}           & Preloading model into GPU memory                 & DNN model execution caching                            & -                 & \ding{52}         & -              & -                         & -         \\
                                  & Nanily \cite{tang2019nanily}                 & \ding{170}\ding{171}           & Adaptive batching; autoscaling                   & Batch size adjustment by remaining time                & \ding{52}         & -                 & -              & S                         & -         \\
                                  & RRL   \cite{qin2019swift}                    & \ding{170}                     & Region-based Reinforcement Learning              & Parallelism configuration tuning                       & \ding{52}         & \ding{52}         & -              & M                         & \ding{52} \\
                                  & TrIMS   \cite{dakkak2019TrIMS}               & \ding{170}\ding{171}\ding{95}  & Multi-layered caching across FaaS                & Memory efficiency by sharing                           & \ding{52}         & \ding{52}         & \ding{52}      & S                         & \ding{52} \\
                                  & Ebird \cite{Ebird}                           & \ding{170}\ding{171}\ding{95}  & CUDA stream paralleism; GPU-side   memory pool   & Transfer-computation overlapping                       & \ding{52}         & \ding{52}         & -              & S                         & \ding{52} \\ \midrule
            \multirow{7}{*}{2020} & GSLICE \cite{dhakal2020GSLICE}               & \ding{171}\ding{95}            & Dynamic GPU resource   apportioning              & Efficient fine-grained sharing                         & \ding{52}         & \ding{52}         & -              & S                         & -         \\
                                  & Clockwork   \cite{Arpan2020clockwork}        & \ding{170}\ding{171}           & Consolidating choice                             & Predictable E2E performance                            & \ding{52}         & -                 & -              & M                         & \ding{52} \\
                                  & Irina \cite{wu2020irina}                     & \ding{170}\ding{171}\ding{95}  & Batching, colocation and preemption              & Graceful general preemption                            & \ding{52}         & \ding{52}         & -              & S                         & -         \\
                                  & PERSEUS   \cite{lemay2020Perseus}            & \ding{170}\ding{169}\ding{171} & Performance and cost characterization on Cloud   & Cost savings under GPU instances                       & \ding{52}         & -                 & \ding{52}      & S                         & \ding{52} \\
                                  & AutoDeep \cite{AutoDeep}                     & \ding{170}\ding{169}\ding{171} & BO and DRL                                       & Cloud configuration and device placement               & -                 & \ding{52}         & \ding{52}      & S                         & -         \\
                                  & DyBatch   \cite{zhang2020dybatch}            & \ding{170}\ding{171}           & Task slicing and reordering                      & Fine-grained batching;   Fairness-driven scheduling    & \ding{52}         & \ding{52}         & -              & S                         & -         \\
                                  & Inferline   \cite{crankshaw2020inferline}    & \ding{170}\ding{169}           & Low-frequency planner; high-frequency tuner      & Near-optimal scaling   cost-efficiency                 & \ding{52}         & -                 & \ding{52}      & M                         & \ding{52} \\ \midrule
            \multirow{5}{*}{2021} & INFaaS \cite{INFaaS}                         & \ding{170}\ding{169}\ding{171} & VM- and model-level autoscaling                  & Automatical model variants selection                   & -                 & \ding{52}         & \ding{52}      & -                         & \ding{52} \\
                                  & Mendoza et al.   \cite{Interference-Aware21} & \ding{170}                     & Latency degradation prediction during colocation & Safe colocation                                        & -                 & \ding{52}         & -              & M                         & -         \\
                                  & Morphling \cite{Morphling}                   & \ding{169}\ding{171}           & Model-agnostic meta-learning                     & Performance prediction under   different configuration & \ding{52}         & \ding{52}         & \ding{52}      & -                         & \ding{52} \\
                                  & Abacus   \cite{cui2021abacus}                & \ding{170}\ding{171}           & Runtime operator scheduling                      & Deterministic latency under   colocation               & -                 & \ding{52}         & -              & M                         & \ding{52} \\
                                  & MIG-SERVING   \cite{tan2021MIG-SERVING}      & \ding{170}\ding{169}           & Greedy algorithm, GA, and MCTS                   & MIG enabled inference scheduling                       & \ding{52}         & \ding{52}         & -              & S                         & -         \\ \midrule
            2022                  & Cocktail \cite{Jashwant2021Cocktail}         & \ding{168}\ding{170}\ding{169} & Weighted majority voting policy                  & Ensembling-based model selection                       & -                 & -                 & \ding{52}      & -                         & -         \\ \bottomrule
        \end{tabular}
    }

    \begin{flushleft}
        \begin{tablenotes}[para,flushleft]
            \scriptsize
            \textbf{Objectives:} \ding{95} Utilization \ding{168} Accuracy \ding{169} Cost \ding{170} Latency  \ding{171} Throughput; \hspace{10pt}  \textbf{Experiment Scale:} S (Single Node) M (Multi Nodes) -: no evaluation on a physical cluster or not clearly specified.
        \end{tablenotes}
    \end{flushleft}
    \vspace{-15pt}
    \label{tab_inference}
\end{table*}

As more DL-based applications are released as online services in our daily life, it becomes more critical to manage and schedule large-scale inference workloads in the GPU datacenter. 
Different from the resource-intensive and long-term training workloads, inference jobs have unique characteristics and requirements (Sec. \ref{sec_background-inference}), which demand new scheduling solutions.
Similar as Sec. \ref{sec_training}, we categorize these inference scheduling techniques based on their objectives, and resource consumption features. Then we give some implications from these works at the end of this section. Table \ref{tab_inference} summaries the relevant papers and their features.

\subsection{Scheduling Objectives}

We first review prior works based on the scheduling objectives.

\subsubsection{Efficiency}
\label{sec_accuracy_latency_cost}

As discussed in Sec. \ref{sec_inference_challenge}, the main objective for scheduling an inference workload is to improve its efficiency. This includes the reduction of inference latency and cost, and improvement of the prediction accuracy {\tiny\encircle{\normalsize I2}}. The challenge here is that there exist trade-offs among different efficiency goals. Here we discuss the techniques to improve each goal as well as to jointly balance and improve them.

\textbf{1) Accuracy efficiency.} Improving the prediction accuracy is a perpetual objective in an inference system. To achieve this, one approach is to collect a set of models, and select the best one to predict the result for each input query.
The scheduling decision made includes model selection and resource allocation among different candidates.
Ease.ml~\cite{li2018easeml} leverages the input and output shape information of the query sample to automatically select the model. It estimates the potential accuracy improvement of each candidate model and then picks the highest one for actual inference. It also formulates the cost-aware model selection process under both single-tenant and multi-tenant settings with multi-armed bandit and Bayesian Optimization. Another effective approach is model ensemble, which combines the prediction results from multiple models to improve the prediction accuracy and generalization. Clipper~\cite{Daniel2017clipper} examines the benefits brought from the model ensemble in computer vision tasks and applies a linear ensemble method to compute a weighted average of the base model predictions. The linear weights are decided by bandit- and learning-based approaches. Rafiki~\cite{wang2018rafiki} leverages an RL model to determine the model set for the ensemble. This model is also used to identify the final optimal model combinations, and tune critical parameters, e.g., batch size.

\textbf{2) Latency efficiency.}
An inference system should have a satisfactory response time, even for burst and fluctuating query requests {\tiny\encircle{\normalsize C7}}.
The latency requirement poses challenges for the scheduler to decide which jobs to be prioritized in the job assignment and rearrangement process.
This objective can be achieved via carefully optimizing the resource allocation.

It is common to launch multiple inference execution instances concurrently to meet the corresponding latency requirement as much as possible due to the low GPU utilization for each request {\tiny\encircle{\normalsize C5}}. Therefore, the inference scheduler can make scheduling decisions aiming at scaling up resources according to the request density to maintain a satisfactory latency.
Clipper~\cite{Daniel2017clipper} conducts linear scaling of inference instances and uses separate docker containers to isolate different models. It replicates the model containers according to the number of queries and applies adaptive batching independently for each model due to the varied execution time.
MArk~\cite{zhang2019mark,zhang2020mark} scales the inference instances with the cloud services. It selects and combines different cloud services like AWS EC2 and Lambda in order based on their prices and scaling abilities. Also, it monitors the system loads and request queuing situations proactively and leverages Lambda to scale up instances when there exist requests violating the latency demands. InferLine~\cite{crankshaw2020inferline} targets the pipelined inference workloads with multiple stages.
It monitors the frequency of queries to each model and makes the scaling decisions of each component separately, to maintain the latency SLOs even during sharp bursts.

A number of works aim to provide bounded latency for inference execution at the system level considering its deterministic execution {\tiny\encircle{\normalsize I1}}. Clockwork~\cite{Arpan2020clockwork} discovers that many DL inference models have deterministic performance because of the underlying deterministic computations. Thus, it guarantees deterministic latency by alleviating the uncertainty introduced by other components of the system. To overcome the uncertainty from memory and cache usages, hardware interactions, and other uncontrollable external performance variations, Clockwork consolidates the configurations among all the system layers during the inference execution, by proactively controlling the memory allocation and deallocation, and disabling concurrent execution of multiple inference workloads to eliminate the interaction.
Reducing the parallelism of execution eliminates the interference from other tasks, but inevitably brings lower throughput and resource utilization. To address this issue, Abacus~\cite{cui2021abacus} tries to guarantee SLO for query requests under the GPU co-location scenarios. It controls the execution sequence and the co-location situation proactively, rather than the default random-ordered execution overlap. Given the explicit order and specific co-location operators on GPUs, Abacus could predict the estimated running time under co-location from the early offline profiling stage. Based on the estimation, the query controller schedules all the simultaneous inference workloads to guarantee the QoS by searching the optimal execution sequence of DNN operators. ParM~\cite{Kosaian2019parity} migrates the concept of erasure codes from distributed computing to model inference systems, and uses learning-based coded computation to introduce redundancy and thus supports the recovery of inference executions with tail latency or failures.

Some solutions proactively schedules the inference workloads and rearranges the execution sequence at the job level. Irina~\cite{wu2020irina} is the first online inference scheduling system, modeling the satisfaction of latency demands as a scheduling problem. By leveraging preemption for DL inference workloads, Irina dynamically decides whether to preempt the ongoing query and launch the later arrived one, which brings significant reduction of average completion time for inference workloads. The main challenge is that existing ML frameworks are not designed and suitable for preemption during execution. Irina carefully manages the preemption process by adding an exit node to the existing dataflow graph of the inference workload, thus enabling safe preemption at arbitrary moments. It is necessary to have more runtime information about the inference workloads for effective scheduling.
Kube-Knots~\cite{Thinakaran2019kubeknots} makes predictions about the resource utilization of each inference workload from two aspects. From the spatial aspect, Kube-Knots discovers the correlations across different resource utilization metrics, and then forecasts the future resource utilization. From the temporal aspect, Kube-Knots predicts the peak inference usage and tries to avoid co-locating jobs which could attain peak consumption of the same resources simultaneously.

\textbf{3) Cost-efficiency.}
The monetary cost becomes one of the main concerns when using public cloud resources to deploy DL inference workloads. Considering the varied compute capabilities and prices for different types of resources and services, a couple of schedulers implement many mechanisms to achieve cost-efficient inference.
MArk~\cite{zhang2019mark,zhang2020mark}
analyzes the cost of utilizing different levels of resource abstractions in Amazon Web Services (AWS) and Google Cloud Platform (GCP) for inference. It finds that the Infrastructure-as-a-Service (IaaS) provides better cost efficiency than Content-as-a-Service (CaaS), while Function-as-a-Service (FaaS) could compensate for the relatively long cold start latency of IaaS at the cost of increased costs. Small instances with advanced CPUs and burstable instances are also recommended. For GPU instances, the cost can be greatly reduced by batch processing as well. Given different levels of capability, scalability, and pricing, MArk greedily selects the most cost-effective type of instances and leverages the spot instances for cost-saving.
AutoDeep~\cite{AutoDeep} considers not only the resource type in the cloud but also the device placement for DL inference. It leverages Bayesian Optimization for the nearly optimal cloud configuration and Deep Reinforcement Learning for the nearly optimal device placement. Constrained by the SLO requirements, AutoDeep performs joint optimization to minimize the total cost of the inference execution. Cocktail~\cite{Jashwant2021Cocktail} develops a distributed weighted auto-scaling policy and leverages the spot instances to minimize the cost.

\textbf{4) Trade-offs between accuracy, latency and cost.}
The objectives of accuracy, latency and cost are not independent {\tiny\encircle{\normalsize C6}}. Improving one goal may possibly compromise another goal if the solution is not designed properly. Besides, users may also have their specific expectations about different objectives. This motivates researchers to explore the trade-offs between these objectives, and devise more flexible and comprehensive scheduling systems.

The adoption of multiple models can improve the model inference accuracy, but might also increase the response latency and cost. Several works track the latency and prediction accuracy of different models and implement mechanisms to select the most appropriate ones determined by the schedulers.
Clipper~\cite{Daniel2017clipper} introduces a model selection abstraction, which supports both single model selection and model ensemble selection. It executes the inference for all the models and combines their results. It observes the corresponding accuracy and latency feedback continuously to make the selection with a best-effort search method. Model-Switching~\cite{jeff2020modelswitching} pursues the trade-offs between computational cost and service accuracy by switching different model variants proactively, to improve the accuracy of responses under the latency constraint. By maximizing the ratio of correct predictions returned within the deadline, it makes selections among model variations with different computation demands and accuracy.  Cocktail~\cite{Jashwant2021Cocktail} balances the cost with accuracy and latency on the public cloud via the optimization of the model ensemble. With a dynamic model selection policy that searches models tightly with the accuracy and latency bounds, Cocktail reduces the candidates in the ensemble and accomplishes fast and efficient model selection.

Some schedulers allow users to specify their demands about accuracy, latency and cost, and make scheduling decisions directly according to the demands. Tolerance Tiers~\cite{Halpern2019onesize} discloses the efforts the system can offer to achieve each objective, and makes users programmatically select their demands. Observing that improving the accuracy of some extreme requests can increase the latency greatly, Tolerance Tiers relaxes and sacrifices the accuracy demand to improve the latency and service cost. Each tier defines an error tolerance to indicate the tolerable accuracy loss, and an optimization objective. Then, Tolerance Tiers optimizes the objective under the constraint of the maximum error tolerance. INFaaS~\cite{Neeraja2019INFaaSw,INFaaS} also asks for the performance and accuracy demands from the users.
It generates some variants from the existing models with different specific parameters (e.g., batch size, hardware configurations,  hardware-specific parameters). After one-time profiling for each variant, INFaaS selects the model variant based on the resource consumption and performance information from the profiling to serve users' requests. Since each model variant may have different performance and monetary costs during execution, INFaaS makes the selection via a heuristic-based approach, which selects the variant with the minimum cost while meeting the SLO constraint, or upgrades existing variants with higher performance to fulfill the burst queries.

\subsubsection{System Throughput}
\label{sec_throughput}

Another important objective for scheduling inference workloads is to improve its throughput capability. The techniques to achieve this goal is summarized as follows.

\textbf{1) Batching execution.}
One common approach is to batch multiple inference queries, and execute them concurrently. Handling individual inference queries usually leads to GPU underutilization {\tiny\encircle{\normalsize C5}}. Hence, batching inference can efficiently improve the utilization and reduce the system overhead. Like job queuing in parallel job scheduling, batching multiple queries can delay the execution of the requests that come earlier, and possibly jeopardize the SLO requirement. Setting a proper batch size is critical to balance such delays and system throughput. Most schedulers dynamically adjust this hyperparameter based on the actual SLO requirement and queuing situation.

First, some schedulers adopt heuristic methods to tune the batch size. Clipper~\cite{Daniel2017clipper} and Rafiki~\cite{wang2018rafiki} apply the practical Additive-Increase-Multiplicative-Decrease (AIMD) algorithm to adjust the inference batch size. Specifically, the batch size is additively increased by a fixed amount until the latency of processing a batch exceeds the latency requirement and then multiplicatively decreased by a fixed percent. Clipper evaluates that AIMD is simple yet effective and adaptive to the changing throughput of a model in special cases. It also aggressively delays the execution of queries under moderate loads to the subsequent batch, which can bring a significant throughput increase for some models. GSLICE~\cite{dhakal2020GSLICE} also applies a similar adaptive and self-learning approach to determine the optimal batch size. It carefully tracks the execution time and increases the batch size until the execution of the last batch exceeds the SLO requirement. Ebird~\cite{Ebird,E2bird} proposes a novel elastic batching mechanism, which runs different CUDA streams concurrently for different batches. Motivated by the observation that processing multiple queries in a large batch has similar performance as multiple CUDA streams with smaller batch sizes, Ebird dynamically adjusts the batch size per stream to utilize the spare GPU resources and fulfill the whole GPU. In each scheduling round, it selects and launches a job based on its batch size and remaining GPU resources in a best-effort manner, squeezing the full GPU utilization and minimizing inference queuing delay.

Second, some schedulers propose optimization-based methods to balance the inference delay and throughput.
MArk~\cite{zhang2019mark,zhang2020mark} considers the maximum time of delaying a query, profiles the processing rate without batching, and searches for the optimal batch size under the SLO constraint. Nanily~\cite{tang2019nanily} presents the upper bound of the batch size by retrieving the maximum remaining time for the requests, calculated as the remaining time towards the deadline subtracted by the least queuing time for the available resources. It then derives the corresponding batch size, which makes the inference execution time equal or close to the maximum remaining time. DyBatch~\cite{zhang2020dybatch} considers the fairness of the delay for each independent workload when batching. It implements fine-grained batching schemes along with fairness-driven scheduling, which can compensate for the deviation of slowdown for small inference workloads. DyBatch organizes the workload batches in a time-sharing manner and selects the batch with the lowest resource utilization for running, thus maintaining fairness and minimizing the slowdown of each workload.

\textbf{2) Caching and reusing. } Another widely-used strategy for throughput improvement is caching and reusing the prediction results across different requests {\tiny\encircle{\normalsize C7}}.
The scheduler selects the request that benefits most from caching and allocates proper resources. This can be done at two levels.

The first direction is to perform optimization at the query level. To provide fast responses to different queries, the inference system can cache the inference execution and prediction results for burst queries. Clipper~\cite{Daniel2017clipper} maintains a prediction cache for requests with the same target model and the query input. Then it can produce the results for some queries without evaluating the model, thus increasing the inference throughput. Clipper also applies an LRU cache eviction policy to optimize the caching efficiency. However, this approach may be less efficient when the queries do not have high similarities in practical scenarios, which leads to high cache miss rates and evictions.

The second direction is to perform optimization at the device level.  Gillman \etal~\cite{Challenges19} proposed to cache the DL models instead of the inference results. It schedules models to be loaded into the limited GPU memory to maximize the probability of servicing an incoming request without swapping the models in and out of the memory, thus accelerating the inference by eliminating the cold start latency with cache hits. The caching and eviction policy considers many runtime aspects of DL inference workloads, including model size, frequency, model accuracy, and speed. This work also discusses some future directions for more dynamic caching mechanisms and policies, like framework-level GPU memory-friendly optimization, proactively loading and evicting, and cluster-level GPU memory allocation. To address the limitation of GPU memory, GSLICE~\cite{dhakal2020GSLICE} and HiveMind~\cite{narayanan2018accelerating} explore the common GPU memory component in different inference models, and propose to save GPU resources via memory sharing. Particularly, GSLICE enables efficient GPU memory sharing by allowing the reuse of model parameters via modifications to the DL framework, exposing the CUDA address to different instances. Therefore, it supports loading the inference model-related parameters only once to the GPUs, resulting in faster module loading. HiveMind extends the shared content and brings more possibilities of shared model weights and layers across different inference workloads, saving the overhead from both model loading and inference evaluation. TrIMS~\cite{dakkak2019TrIMS} organizes the memory sharing of different models in a more systematic design. The model resource manager in TrIMS offers a multi-tiered cache for DL models to be shared across users' FaaS functions and enables DL model sharing across all levels of the memory hierarchy in the GPU, CPU, local storage, and remote storage in the cloud. TrIMS reconciles the lifecycle of model memory consumption and carefully handles the cache misses and evictions. It also considers multi-node, isolation, and fairness problems during sharing. Extensive evaluations on different models show its general abilities to improve the inference performance by mitigating the model loading overhead.

\textbf{3) System configuration tuning. }
Besides the optimization techniques detailed above, there exist some schedulers leveraging end-to-end configuration tuning to improve the system throughput. Morphling~\cite{wang2021Morphling} formulates the optimal configuration search as a few-shot learning problem. Then it adopts model-agnostic meta-learning (MAML)~\cite{finn2017model} to combine offline meta-model training for inference serving performance modeling under varied hardware and runtime configurations, and performs online few-shot learning to predict the service performance. Based on the prediction, Morphling auto-tunes the resource provisioning configurations and makes better scheduling decisions.
RRL~\cite{qin2019swift} concentrates on optimizing the parallelism configurations from different levels, including request level parallelism and intra-request level (inter-op and intra-op) parallelism, which have strong impacts on the latency of the entire system. RRL utilizes a region-based RL method to tune the parallelism configurations and reduce the inference processing latency, based on the system performance similarity between different configurations within a similar parallelism setting.

\subsection{Resource Consumption Feature}
\label{sec_resource}
Similar to training workloads, the inference schedulers can also be categorized based on the resource consumption feature. Below we detail the scheduling solutions designed to target these features.

\subsubsection{Colocation and resource sharing}
\label{sec:inference_colocation}

From Challenge {\tiny\encircle{\normalsize C5}} in Sec. \ref{sec_inference_challenge}, executing one inference request can lead to GPU resource underutilization. The recent development of GPU architecture designs motivates the GPU sharing from both the hardware perspective~\cite{mps,nvidia2020a100whitepaper} and software perspective~\cite{shi2012vCUDA,Salus}. GPU sharing across different inference requests can significantly improve the resource utilization by better leveraging GPUs' parallel compute capability. However, it can also incur the risks of violating the latency requirements due to the uncertain running time.

One line of research works adopt the static colocation strategy to guarantee the inference latency. Space-Time~\cite{jain2018dynamic} calls for both space-sharing and time-sharing
for DL inference with GPUs. It preserves the predictability and isolation during virtualization by monitoring the inference latency per kernel. Then it improves the utilization by merging concurrent small kernels into larger super-kernels that can fill up the GPU utilization under the time-sharing mechanism. Irina~\cite{wu2020irina} only considers a safe GPU colocation situation based on the peak GPU requirement of the workloads, when their total GPU requirement does not exceed the capacity. It assumes there is no interference and slowdown on the job completion time and heuristically places the newly arrived job on the GPU with the smallest JCT. Nexus~\cite{shen2019nexus} also applies a heuristic approach to select the requests to be co-located on the same GPU. First, it identifies the most appropriate batch size for throughput and SLO needs of the existing inference workload. Afterward, it establishes all possible combinations within a GPU’s duty cycle on a single GPU in a best-fit manner, and maximizes the utilization without violating the latency requirement.

Some other works introduce dynamic colocation mechanisms for managing the inference workloads.
GSLICE~\cite{dhakal2020GSLICE} is an inference system to systematically achieve safe and efficient GPU sharing. It leverages spatial GPU multiplexing for different inference requests on top of the state-of-the-art GPU spatial multiplexing framework MPS. It intensively evaluates the colocation performance with and without MPS, and with the resource provisioning limits. It discovers that the colocation interference could be amortized under the careful configuration of the resource limits. The performance improvement reaches a point of diminishing returns (i.e., kneepoint) after certain configurations, which has a non-linear relationship with the throughput and latency of the inference. Based on these observations, GSLICE tracks the kneepoint of different inference workloads and partition the whole GPU according to their kneepoints. It also designs a hot-standby mechanism to dynamically adjust the resource limit configuration of the specific inference job, along with other implementation optimizations including minimizing the data transfer overhead. MIG-SERVING~\cite{tan2021MIG-SERVING} leverages the hardware support for GPU virtualization (i.e., MIG \cite{mig} on NVIDIA A100\cite{nvidia2020a100whitepaper}) for efficient GPU colocation.
With MIG, an A100 card could be dynamically partitioned into several instances with smaller hardware capacities under some hard constraints. MIG-SERVING discovers that the throughput of most models does not grow linearly with the increase of resources on different instances.
It establishes a reconfigurable scheduling problem and applies a generic algorithm to find a sub-optimal and feasible solution, and improve it via a search-based method.

Since the colocation of multiple inference workloads multiplexes the GPU, it is hard to measure and predict their running time due to the colocation interference.
Therefore, several inference scheduling systems focus on estimating and handling the colocation interference. INFaaS~\cite{Neeraja2019INFaaSw,INFaaS} targets the inference services in the cloud. It identifies the colocation interference caused by the shared hardware resources.
Then it allocates the available resources to the interfered instances by migration or VM-level scaling. The evaluation shows that INFaaS can save the monetary cost by GPU sharing and satisfy the latency requirements by VM-level scaling. Mendoza \etal~\cite{Interference-Aware21} proposed an interference-aware scheduler for inference workloads, which proactively considers the impact of interference on the latency from co-location. By predicting the performance degradation, it minimizes the latency degradation and reduces the SLO violations. It improves the prediction accuracy by exploiting the similarity of co-location configurations, including inference model attributes and machine types. PERSEUS~\cite{lemay2020Perseus} compares the cost and throughput under exclusive execution and colocation for inference workloads. It concludes that the mixed ratio of different models affects the cost efficiency of colocation. The interference on the model and data loading time during cold start also differs across different models because of the cache and data transfer requirements under colocation.

\subsubsection{Heterogeneous Resources.}
Several works exploit both CPU and GPU machines of different sizes for DL inference, especially for the cloud environment. MArk \cite{zhang2019mark,zhang2020mark} characterizes and compares the cost and performance of inference workloads on different types of instances available in GCP and AWS. It concludes that smaller instances with advanced CPU models achieve a higher performance-cost ratio. It also suggests that GPU instances can achieve lower per-request cost and smaller inference latency than CPU instances with appropriate batch sizes. PERSEUS \cite{lemay2020Perseus} performs a similar cost-efficiency characterization, considering instances with multiple onboard GPUs. It examines that some DL models with high GPU utilization could introduce intensive interference with other inference workloads in multi-GPU instances. AutoDeep \cite{AutoDeep} and Cocktail \cite{Jashwant2021Cocktail} focus more on the price of different cloud configurations and search for the best configuration according to the pricing information.

\subsection{Implications}

In Sec. \ref{sec_accuracy_latency_cost} we mainly discuss the monetary cost of deploying DL inference workloads. In a GPU datacenter, how to improve the efficiency of energy cost is also critical. While this objective has been extensively explored for training workloads (Sec. \ref{sec_training_efficiency}), it is relatively less studied for the inference workloads. Some works \cite{facebook,FaceBookTrace} provide some energy characterizations of production DL inference clusters. Kube-Knots~\cite{Thinakaran2019kubeknots} presents simple energy efficiency comparisons of inference workloads between GPUs and CPUs. It is necessary to comprehensively explore the energy optimization of different DL inference models with different types of compute resources, and design more sophisticated energy-saving mechanisms with the consideration of latency and resource utilization. This will be a promising research direction in the future.

Most of the above works treat the inference jobs as a black box for management and optimization. In reality, an inference pipeline may consist of several separate stages to fulfill the query. It is interesting to consider the characteristics of internal stages to optimize the execution in a white-box manner. PRETZEL~\cite{PRETZEL} leverages this idea, which stores and re-uses the model parameters and computation among similar white-box representations of pipeline stages to reduce the resource utilization. We expect more future works will focus on the optimization of inference workloads at this sub-request level.

As the scale of DL models grows faster than the GPU compute capacity, it is more difficult to accommodate single models on a single GPU or machine. This problem becomes more serious when cloud users adopt the resource-constraint serverless platform for inference services. A natural solution to this problem is model partitioning,
Gillis~\cite{yu2021Gillis} splits the network layers and minimizes the inference latency via dynamic programming. It also adopts some searching methods like Bayesian Optimization and RL to minimize the cost. AMPS-Inf~\cite{Jananie2021AMPS-Inf} jointly considers the model partitioning and resource allocation by calculating the optimal execution and resource provisioning plans under the constraint of the response time in the serverless platforms. It is worth more effort to explore the methodologies of serving large models with limited resources for different objectives.

Some inference systems for CPU clusters predict the future trends of query requests to satisfy the latency requirement. For instance, BARISTA~\cite{Bhattacharjee2019BARISTA} predicts the future workload patterns based on the historical data and estimates the required resources for the application to maintain the SLOs. Then it makes resource provisions based on the difference between the desired throughput and latency requirement with the current ones. Swayam~\cite{gujarati2017Swayam} ensures an appropriate replica of service instances by predicting the load and auto-scales the service in a fully distributed way along with the self-reclamation of resources. It adopts linear regression to predict the request arrival rate over short periods to tolerate the rapid changes in time. Since these solutions are hardware-agnostic, they can be applied to the GPU clusters as well. We also expect this strategy can inspire more solutions dedicated to the GPU clusters.

\section{Scheduling Other Types of DL Workloads}
\label{sec_other_workload}

The previous two sections categorize the scheduling of general training and inference workloads, respectively. In this section, we discuss the works for optimizing other types of DL workloads.
Table~\ref{tab_other} presents the past works for hyperparameter optimization workloads as well as hybrid training and inference workloads, with the detailed description as below. 

\subsection{Hyperparameter Optimization Workloads}

A Hyperparameter Optimization (HPO) job aims to identify the best hyperparameters for a DL task. Technically speaking, HPO belongs to the category of DL training workloads. Here we discuss it separately because it has unique features compared to the general DL training jobs. Specifically, a HPO job typically needs to search a set of hyperparameter configurations. Each configuration is associated with a \emph{trial} \cite{RubberBand}, which is a common DL training workload containing training and evaluation procedures. However, in the HPO context, these trails are extremely similar and orchestrated by a HPO scheduling policy. To accelerate the HPO process, the policy can kill poor-performing trails through early stopping and allocate more resources to promising trails. These optimizations on HPO workloads can deliver remarkable acceleration.

\textbf{Accuracy efficiency.}
Hermes~\cite{Hermes} is a scheduler to expedite HPO workloads in GPU datacenters. It provides a container preemption mechanism to enable migration between DL jobs with minimal overhead. Besides, it also considers the algorithmic property of HPO workloads and devises a convergence-aware scheduling algorithm to favor promising hyperparameter configurations.
HyperSched~\cite{HyperSched} aims to boost the accuracy performance of HPO workloads within the given time and resource budgets. In an HPO workload, the promising hyperparameter trial generally has higher accuracy at the early training stage. Then, HyperSched allocates more resources to the promising trials and terminates unpromising ones. With this technique, HyperSched can parallel orders of magnitude more hyperparameter trials and therefore maximize the final accuracy substantially. As an extension of HyperSched~\cite{HyperSched}, SEER~\cite{SEER} replaces the resource budget with the monetary cost budget in the cloud. Without the resource constraint, SEER can launch numerous training trials at the early training stage to explore promising hyperparameter combinations. Then it will terminate poor training trials and allocate more cost budget to promising trials during the training progress. The accommodation of more trials at the hyperparameter exploration stage can improve the final model accuracy. HyperDrive~\cite{HyperDrive} is another scheduler framework for hyperparameter exploration as well.
It develops an accuracy curve fitting model to extrapolate the accuracy in the subsequent training iterations. The accuracy prediction engine allows HyperDrive to terminate low-quality jobs early and adjust the resource allocation dynamically. Moreover, Fluid \cite{Fluid} further leverages the job packing mechanism to improve resource utilization and accelerate the HPO process.

\textbf{Cost efficiency.}
RubberBand~\cite{RubberBand} aims to reduce the monetary cost of an HPO job with the constraint of timing budget in the cloud. Since the optimal amount of resources for an HPO workload differs in early and later stages of the hyperparameter search and model optimization processes, RubberBand needs to jointly accommodate the job throughput, resource amount and cloud pricing. By building a profiling model, RubberBand can predict the job throughput and corresponding cost for a given resource allocation policy. Then it uses this model to generate an optimal cost-efficient solution for satisfactory cost reduction.

\begin{table*}[t]
    \centering
    \caption{\textbf{Summary of schedulers for other workloads in GPU Datacenters.}}
    \vspace{-10pt}
    \resizebox{\textwidth}{!}{
        \begin{tabular}{@{}cccccccc@{}}
            \toprule
            \textbf{Type}           & \textbf{Year}               & \textbf{Scheduler}                        & \textbf{Objectives}                                                                                                                                                             & \textbf{Approaches}                           & \textbf{Advantages}                         & \textbf{\makecell[c]{Exp.             \\ Scale}} & \textbf{\makecell[c]{Source \\ Code}} \\ \midrule
            \multirow{6}{*}{HPO}    & 2017                        & HyperDrive \cite{HyperDrive}              & \ding{168}                                                                                                                                                                      & Dynamic Probabilistic Accuracy Prediction     & JCT Reduction                               & S                         & -         \\
                                    & 2019                        & HyperSched \cite{HyperSched}              & \ding{171}                                                                                                                                                                      & ASHA; Dynamic Resource Allocation             & Efficient HPO under DDL                     & S                         & \ding{52} \\
                                    & 2021                        & Hermes \cite{Hermes}                      & \ding{168}\ding{169}                                                                                                                                                            & Time-sharing Execution with Low Overhead      & JCT Reduction                               & S                         & -         \\
                                    & 2021                        & RubberBand \cite{RubberBand}              & \ding{168}\ding{169}                                                                                                                                                            & Model JCT and Cost Prior to Runtime           & Cost-Effectiveness                          & S                         & -         \\
                                    & 2021                        & SEER \cite{SEER}                          & \ding{168}\ding{169}                                                                                                                                                            & Dynamic Resource Allocation                   & Cost Constraint HPO                         & S                         & -         \\
                                    & 2021                        & Fluid \cite{Fluid}                        & \ding{95}                                                                                                                                                  \ding{168}\ding{169} & Job Packing; Elastic
            Training                & Better Resource Utilization & S                                         & \ding{52}                                                                                                                                                                                                                                                                                                             \\ \midrule
            \multirow{4}{*}{Hybrid} & 2018                        & Rafiki \cite{wang2018rafiki}              & \ding{170}\ding{169}                                                                                                                                                            & RL to optimize accuracy and latency           & Unified platform for training and inference & S                         & \ding{52} \\
                                    & 2019                        & Kube-Knots \cite{Thinakaran2019kubeknots} & \ding{169}\ding{95}                                                                                                                                                             & Dynamic container resizing                    & GPU utilization-aware colocation            & S                         & -         \\
                                    & 2020                        & CMS \cite{li2020cms}                      & \ding{170}\ding{95}                                                                                                                                                             & Common architecture for trainers and modelets & Continuous learning                         & S                         & -         \\
                                    & 2022                        & Aryl \cite{Aryl}                          & \ding{95}\ding{170}                                                                                                                                                             & Capacity Loaning                              & Cluster Size Extension                      & L                         & -         \\ \bottomrule
        \end{tabular}
    }

    \begin{flushleft}
        \begin{tablenotes}[para,flushleft]
            \scriptsize
            \textbf{Objectives:} \ding{95} Utilization \ding{168} Makespan \ding{169} Cost \ding{170} Accuracy \ding{171} DDL; \hspace{10pt}  \textbf{Experiment GPU Scales:} the scale of physical testbed. S $(0, 30]$ M $(30, 60]$ L $(60, 120]$ XL $(120, \infty]$ -: no evaluation on a physical cluster or not clearly specified.
        \end{tablenotes}
    \end{flushleft}

    \vspace{-10pt}
    \label{tab_other}
\end{table*}

\subsection{Hybrid of Training and Inference Workloads}
In Sec. \ref{sec_training} and Sec. \ref{sec_inference} we consider the scheduling techniques for training and inference solutions separately. Actually, there are some DL systems that host these two workloads in a unified manner. Due to the distinct features of the two types of workloads, new solutions are needed for efficient scheduling in GPU datacenters. Below we review and summarize these works.

\textbf{End-to-end DL development.}
Some works consider the entire pipeline of DL development and deployment, including model training, inference, as well as periodically retraining and updating.
CMS~\cite{li2020cms} designs a continuous machine learning and serving platform, which orchestrates the model training executions, model deployments, and model update services. It unifies different training jobs into a simple trainer contract and provides common essential pipeline abstracts of training. Then the scheduler monitors the resource consumption of these resource-intensive training tasks to avoid contention. Other techniques including model validation are also applied to guarantee the quality of newly-trained models.
Rafiki~\cite{wang2018rafiki} proposes to reuse the datasets and parameters across different training and inference jobs. It implements a unified distributed dataset storage and a parameter server. The parameter server is kept in memory and shares model parameters across different trials in the training and dumped for later reuse in the inference. Other common underlying components are also shared across training and inference workloads to reduce the operational cost, e.g., storage, communication protocols, and compute resources.

\textbf{Mix of training and inference development.}
Some works optimize the datacenter with mixed workloads of DL training and inference.
Kube-Knots~\cite{Thinakaran2019kubeknots} minimizes the resource waste by allocating offline batch jobs to better utilize the spare resources. It discovers that the GPU energy efficiency increases with its utilization. Therefore, achieving higher GPU utilization indicates higher energy efficiency and less resource wastes. To this end, Kube-Knots colocates the predictable GPU batch jobs with the online inference workloads, as the inference jobs generally underutilize the GPUs. During the scheduling, it also predicts the peak resource utilization of the jobs from their online resource usages.
Aryl~\cite{Aryl} also observes the resource can be wasted due to the resource over-provisioning for burst inference requests. It proposes the concept of capacity loaning, which allows the inference cluster to loan the idle GPU servers during low-traffic periods to run training jobs. Since the preemption cost is not negligible, it minimizes the total number of preemptions during the scheduling. Combined with the elasticity on resource demands for part of training jobs, Aryl successfully guarantees the timely resource allocation to inference jobs via a heuristic method.

\subsection{Implications}

In addition to optimizing general DL workloads, there are some opportunities for further resource efficiency improvement. One direction is to apply specific optimizations for hyperparameter search workloads. Compared with the workload-agnostic manner, it can deliver over one magnitude resource and time conservation. However, how to integrate this mechanism well into the scheduler and coordinate with general DL workloads is a challenging topic, requiring more future research investigation. Besides, considering the whole DL model development pipeline instead of focusing on a certain stage can bring extra system performance enhancement. For instance, breaking the shackles between the training and inference cluster resource not only considerably diminishes the queuing delay of training workloads but also improves the model serving quality. These directions deserve more attention in future research on GPU datacenter scheduling systems.

\section{Conclusions and Outlooks}
\label{sec_conclusion}

The scheduler design is an ever-lasting topic in the system research community. The prosperity of DL workloads considerably pushes forward the progress of this research area in all stages of scheduling.
The unique features exhibited by DL workloads advocate novel DL scheduler design to manage the GPU resources and jobs in a more intelligent and efficient manner. Our comprehensive summary draws three conclusions. First, new works prefer to adopt advanced algorithms (e.g., RL, MAML), which can significantly improve the scheduling performance. Second, it is necessary to take advantages of emerging hardware resources (e.g., heterogeneous GPU, GPU colocation and sharing, elastic training) when designing efficient schedulers. Third, new scheduling systems are motivated by the emerging DL workloads and applications, as well as users' new requirements. 

DL workload scheduling in GPU datacenters remains premature. There are multiple interesting future research directions, as summarized below.

\textbf{DL workloads}. The diverse DL workloads pose different challenges to the scheduler. Domain-specific schedulers are required for extreme efficiency for specific applications. Therefore, a set of novel DL workloads with special resource requirements (e.g., HPO, hybrid workloads) call for more research efforts. Searching-based DL workloads like HPO eagerly rely on being served as early as possible to get search results earlier and thus optimize the search direction. Training extremely large models like Transformers needs extensive and high-performance resources. Surging needs from DL jobs for debugging purposes should also be balanced with those production jobs under hybrid situations. Otherwise, the under-efficiency problem will arise. Another important direction is the co-design of both the scheduling system and DL framework. Better scheduling decisions could be made by negotiating the fine-grained resource demand of workloads and delegating framework-level control to the scheduling system.

\textbf{Scheduling decision making}. Many existing schedulers may encounter problems with GPU datacenters at scale. First, a lot of scheduling systems require additional information about the workload from users or online profiling, posing great challenges when facing numerous workloads and resources. Second, some schedulers form the decision-making process as an optimation problem, which cannot be solved within an acceptable time online. Other systems such as online monitoring of DL workloads, resource management, and coordination also add difficulty to the operations of large-scale GPU datacenters.

\textbf{Underlying hardware resources}. GPU datacenters are also growing at an alarming speed. It is common for modern GPU datacenters to contain heterogeneous and complex generations of GPUs and other accelerators. Schedulers need to make scheduling decisions based on different affinities between workloads and GPUs. GPUs may also reveal different capabilities for serving the workloads, e.g., hardware-level support for multiplexing, advanced support for low-precision ALU. Other resources like emerging networking topology also draw the attention of schedulers for performance efficiency.

\bibliographystyle{ACM-Reference-Format}
\bibliography{csur}


\begin{thebibliography}{184}


\ifx \showCODEN    \undefined \def \showCODEN     #1{\unskip}     \fi
\ifx \showDOI      \undefined \def \showDOI       #1{#1}\fi
\ifx \showISBNx    \undefined \def \showISBNx     #1{\unskip}     \fi
\ifx \showISBNxiii \undefined \def \showISBNxiii  #1{\unskip}     \fi
\ifx \showISSN     \undefined \def \showISSN      #1{\unskip}     \fi
\ifx \showLCCN     \undefined \def \showLCCN      #1{\unskip}     \fi
\ifx \shownote     \undefined \def \shownote      #1{#1}          \fi
\ifx \showarticletitle \undefined \def \showarticletitle #1{#1}   \fi
\ifx \showURL      \undefined \def \showURL       {\relax}        \fi
\providecommand\bibfield[2]{#2}
\providecommand\bibinfo[2]{#2}
\providecommand\natexlab[1]{#1}
\providecommand\showeprint[2][]{arXiv:#2}

\bibitem[\protect\citeauthoryear{??}{MXN}{2022}]%
        {MXNET-mms}
 \bibinfo{year}{2022}\natexlab{}.
\newblock \bibinfo{title}{Multi Model Server: a tool for serving neural net
  models for inference}.
\newblock
  \bibinfo{howpublished}{\url{https://github.com/awslabs/multi-model-server}}.
\newblock


\bibitem[\protect\citeauthoryear{??}{nvi}{2022}]%
        {nvidia2020a100whitepaper}
 \bibinfo{year}{2022}\natexlab{}.
\newblock \bibinfo{title}{NVIDIA A100}.
\newblock
  \bibinfo{howpublished}{\url{https://www.nvidia.com/en-sg/data-center/a100/}}.
\newblock


\bibitem[\protect\citeauthoryear{??}{mig}{2022}]%
        {mig}
 \bibinfo{year}{2022}\natexlab{}.
\newblock \bibinfo{title}{NVIDIA Multi-Instance GPU}.
\newblock
  \bibinfo{howpublished}{\url{https://www.nvidia.com/en-us/technologies/multi-instance-gpu/}}.
\newblock


\bibitem[\protect\citeauthoryear{??}{mps}{2022}]%
        {mps}
 \bibinfo{year}{2022}\natexlab{}.
\newblock \bibinfo{title}{NVIDIA Multi-Process Service}.
\newblock
  \bibinfo{howpublished}{\url{https://docs.nvidia.com/deploy/mps/index.html}}.
\newblock


\bibitem[\protect\citeauthoryear{??}{Ope}{2022}]%
        {OpenPBS}
 \bibinfo{year}{2022}\natexlab{}.
\newblock \bibinfo{title}{OpenPBS}.
\newblock \bibinfo{howpublished}{\url{https://www.openpbs.org/}}.
\newblock


\bibitem[\protect\citeauthoryear{Abadi, Barham, Chen, Chen, Davis, Dean, Devin,
  Ghemawat, Irving, Isard, Kudlur, Levenberg, Monga, Moore, Murray, Steiner,
  Tucker, Vasudevan, Warden, Wicke, Yu, and Zheng}{Abadi et~al\mbox{.}}{2016}]%
        {TensorFlow}
\bibfield{author}{\bibinfo{person}{Mart{\'\i}n Abadi}, \bibinfo{person}{Paul
  Barham}, \bibinfo{person}{Jianmin Chen}, \bibinfo{person}{Zhifeng Chen},
  \bibinfo{person}{Andy Davis}, \bibinfo{person}{Jeffrey Dean},
  \bibinfo{person}{Matthieu Devin}, \bibinfo{person}{Sanjay Ghemawat},
  \bibinfo{person}{Geoffrey Irving}, \bibinfo{person}{Michael Isard},
  \bibinfo{person}{Manjunath Kudlur}, \bibinfo{person}{Josh Levenberg},
  \bibinfo{person}{Rajat Monga}, \bibinfo{person}{Sherry Moore},
  \bibinfo{person}{Derek~G. Murray}, \bibinfo{person}{Benoit Steiner},
  \bibinfo{person}{Paul Tucker}, \bibinfo{person}{Vijay Vasudevan},
  \bibinfo{person}{Pete Warden}, \bibinfo{person}{Martin Wicke},
  \bibinfo{person}{Yuan Yu}, {and} \bibinfo{person}{Xiaoqiang Zheng}.}
  \bibinfo{year}{2016}\natexlab{}.
\newblock \showarticletitle{TensorFlow: A System for Large-Scale Machine
  Learning}. In \bibinfo{booktitle}{\emph{12th {USENIX} Symposium on Operating
  Systems Design and Implementation}} \emph{(\bibinfo{series}{OSDI '16})}.
\newblock


\bibitem[\protect\citeauthoryear{Amaral, Polo, Carrera, Seelam, and
  Steinder}{Amaral et~al\mbox{.}}{2017}]%
        {Topology-Aware}
\bibfield{author}{\bibinfo{person}{Marcelo Amaral}, \bibinfo{person}{Jord\`{a}
  Polo}, \bibinfo{person}{David Carrera}, \bibinfo{person}{Seetharami Seelam},
  {and} \bibinfo{person}{Malgorzata Steinder}.}
  \bibinfo{year}{2017}\natexlab{}.
\newblock \showarticletitle{Topology-Aware GPU Scheduling for Learning
  Workloads in Cloud Environments}. In \bibinfo{booktitle}{\emph{Proceedings of
  the International Conference for High Performance Computing, Networking,
  Storage and Analysis}} \emph{(\bibinfo{series}{SC '17})}.
\newblock


\bibitem[\protect\citeauthoryear{Arpaci-Dusseau and
  Arpaci-Dusseau}{Arpaci-Dusseau and Arpaci-Dusseau}{2018}]%
        {arpaci2018operating}
\bibfield{author}{\bibinfo{person}{Remzi~H Arpaci-Dusseau} {and}
  \bibinfo{person}{Andrea~C Arpaci-Dusseau}.} \bibinfo{year}{2018}\natexlab{}.
\newblock \bibinfo{booktitle}{\emph{Operating systems: Three easy pieces}}.
\newblock \bibinfo{publisher}{Arpaci-Dusseau Books LLC}.
\newblock


\bibitem[\protect\citeauthoryear{Bao, Peng, and Wu}{Bao et~al\mbox{.}}{2019}]%
        {Harmony}
\bibfield{author}{\bibinfo{person}{Yixin Bao}, \bibinfo{person}{Yanghua Peng},
  {and} \bibinfo{person}{Chuan Wu}.} \bibinfo{year}{2019}\natexlab{}.
\newblock \showarticletitle{Deep Learning-based Job Placement in Distributed
  Machine Learning Clusters}. In \bibinfo{booktitle}{\emph{IEEE Conference on
  Computer Communications}} \emph{(\bibinfo{series}{INFOCOM '19})}.
\newblock


\bibitem[\protect\citeauthoryear{Bao, Peng, Wu, and Li}{Bao
  et~al\mbox{.}}{2018}]%
        {OASiS}
\bibfield{author}{\bibinfo{person}{Yixin Bao}, \bibinfo{person}{Yanghua Peng},
  \bibinfo{person}{Chuan Wu}, {and} \bibinfo{person}{Zongpeng Li}.}
  \bibinfo{year}{2018}\natexlab{}.
\newblock \showarticletitle{Online job scheduling in distributed machine
  learning clusters}. In \bibinfo{booktitle}{\emph{IEEE INFOCOM 2018-IEEE
  Conference on Computer Communications}} \emph{(\bibinfo{series}{INFOCOM
  '18})}.
\newblock


\bibitem[\protect\citeauthoryear{Bhattacharjee, Chhokra, Kang, Sun, Gokhale,
  and Karsai}{Bhattacharjee et~al\mbox{.}}{2019}]%
        {Bhattacharjee2019BARISTA}
\bibfield{author}{\bibinfo{person}{Anirban Bhattacharjee},
  \bibinfo{person}{Ajay~Dev Chhokra}, \bibinfo{person}{Zhuangwei Kang},
  \bibinfo{person}{Hongyang Sun}, \bibinfo{person}{Aniruddha Gokhale}, {and}
  \bibinfo{person}{Gabor Karsai}.} \bibinfo{year}{2019}\natexlab{}.
\newblock \showarticletitle{BARISTA: Efficient and Scalable Serverless Serving
  System for Deep Learning Prediction Services}. In
  \bibinfo{booktitle}{\emph{2019 IEEE International Conference on Cloud
  Engineering (IC2E)}}.
\newblock


\bibitem[\protect\citeauthoryear{Bian, Li, Wang, and You}{Bian
  et~al\mbox{.}}{2021}]%
        {ONES}
\bibfield{author}{\bibinfo{person}{Zhengda Bian}, \bibinfo{person}{Shenggui
  Li}, \bibinfo{person}{Wei Wang}, {and} \bibinfo{person}{Yang You}.}
  \bibinfo{year}{2021}\natexlab{}.
\newblock \showarticletitle{Online evolutionary batch size orchestration for
  scheduling deep learning workloads in GPU clusters}. In
  \bibinfo{booktitle}{\emph{Proceedings of the International Conference for
  High Performance Computing, Networking, Storage and Analysis}}
  \emph{(\bibinfo{series}{SC '21})}.
\newblock


\bibitem[\protect\citeauthoryear{Bl\"{o}cher, Wang, Eugster, and
  Schmidt}{Bl\"{o}cher et~al\mbox{.}}{2021}]%
        {HIRE}
\bibfield{author}{\bibinfo{person}{Marcel Bl\"{o}cher}, \bibinfo{person}{Lin
  Wang}, \bibinfo{person}{Patrick Eugster}, {and} \bibinfo{person}{Max
  Schmidt}.} \bibinfo{year}{2021}\natexlab{}.
\newblock \showarticletitle{Switches for HIRE: Resource Scheduling for Data
  Center in-Network Computing}. In \bibinfo{booktitle}{\emph{Proceedings of the
  26th ACM International Conference on Architectural Support for Programming
  Languages and Operating Systems}} \emph{(\bibinfo{series}{ASPLOS '21})}.
\newblock


\bibitem[\protect\citeauthoryear{Burns, Grant, Oppenheimer, Brewer, and
  Wilkes}{Burns et~al\mbox{.}}{2016}]%
        {BorgK8S}
\bibfield{author}{\bibinfo{person}{Brendan Burns}, \bibinfo{person}{Brian
  Grant}, \bibinfo{person}{David Oppenheimer}, \bibinfo{person}{Eric Brewer},
  {and} \bibinfo{person}{John Wilkes}.} \bibinfo{year}{2016}\natexlab{}.
\newblock \showarticletitle{Borg, Omega, and Kubernetes: Lessons Learned from
  Three Container-Management Systems over a Decade}.
\newblock \bibinfo{journal}{\emph{Queue}} (\bibinfo{year}{2016}).
\newblock


\bibitem[\protect\citeauthoryear{Chahal, Mishra, Palepu, and Singhal}{Chahal
  et~al\mbox{.}}{2021}]%
        {chahal2021performance}
\bibfield{author}{\bibinfo{person}{Dheeraj Chahal}, \bibinfo{person}{Mayank
  Mishra}, \bibinfo{person}{Surya Palepu}, {and} \bibinfo{person}{Rekha
  Singhal}.} \bibinfo{year}{2021}\natexlab{}.
\newblock \showarticletitle{Performance and Cost Comparison of Cloud Services
  for Deep Learning Workload}. In \bibinfo{booktitle}{\emph{Companion of the
  ACM/SPEC International Conference on Performance Engineering}}.
\newblock


\bibitem[\protect\citeauthoryear{Chaudhary, Ramjee, Sivathanu, Kwatra, and
  Viswanatha}{Chaudhary et~al\mbox{.}}{2020}]%
        {Gandivafair}
\bibfield{author}{\bibinfo{person}{Shubham Chaudhary},
  \bibinfo{person}{Ramachandran Ramjee}, \bibinfo{person}{Muthian Sivathanu},
  \bibinfo{person}{Nipun Kwatra}, {and} \bibinfo{person}{Srinidhi Viswanatha}.}
  \bibinfo{year}{2020}\natexlab{}.
\newblock \showarticletitle{Balancing Efficiency and Fairness in Heterogeneous
  GPU Clusters for Deep Learning}. In \bibinfo{booktitle}{\emph{Proceedings of
  the Fifteenth European Conference on Computer Systems}}
  \emph{(\bibinfo{series}{EuroSys '20})}.
\newblock


\bibitem[\protect\citeauthoryear{Chen, Weng, Wang, Li, and Li}{Chen
  et~al\mbox{.}}{2020c}]%
        {LB-BSP}
\bibfield{author}{\bibinfo{person}{Chen Chen}, \bibinfo{person}{Qizhen Weng},
  \bibinfo{person}{Wei Wang}, \bibinfo{person}{Baochun Li}, {and}
  \bibinfo{person}{Bo Li}.} \bibinfo{year}{2020}\natexlab{c}.
\newblock \showarticletitle{Semi-dynamic load balancing: efficient distributed
  learning in non-dedicated environments}. In
  \bibinfo{booktitle}{\emph{Proceedings of the 11th ACM Symposium on Cloud
  Computing}} \emph{(\bibinfo{series}{SoCC '20})}.
\newblock


\bibitem[\protect\citeauthoryear{Chen}{Chen}{[n.d.]}]%
        {RIFLING}
\bibfield{author}{\bibinfo{person}{Zhaoyun Chen}.}
  \bibinfo{year}{[n.d.]}\natexlab{}.
\newblock \showarticletitle{RIFLING: A reinforcement learning-based GPU
  scheduler for deep learning research and development platforms}.
\newblock \bibinfo{journal}{\emph{Software: Practice and Experience}}
  (\bibinfo{year}{[n.\,d.]}).
\newblock


\bibitem[\protect\citeauthoryear{Chen, Quan, Wen, Fang, Yu, Zhang, and
  Luo}{Chen et~al\mbox{.}}{2020a}]%
        {DLQoSSched}
\bibfield{author}{\bibinfo{person}{Zhaoyun Chen}, \bibinfo{person}{Wei Quan},
  \bibinfo{person}{Mei Wen}, \bibinfo{person}{Jianbin Fang},
  \bibinfo{person}{Jie Yu}, \bibinfo{person}{Chunyuan Zhang}, {and}
  \bibinfo{person}{Lei Luo}.} \bibinfo{year}{2020}\natexlab{a}.
\newblock \showarticletitle{Deep Learning Research and Development Platform:
  Characterizing and Scheduling with QoS Guarantees on GPU Clusters}.
\newblock \bibinfo{journal}{\emph{IEEE Transactions on Parallel and Distributed
  Systems}} (\bibinfo{year}{2020}).
\newblock


\bibitem[\protect\citeauthoryear{Chen, Quan, Wen, Fang, Yu, Zhang, and
  Luo}{Chen et~al\mbox{.}}{2020b}]%
        {GENIE}
\bibfield{author}{\bibinfo{person}{Zhaoyun Chen}, \bibinfo{person}{Wei Quan},
  \bibinfo{person}{Mei Wen}, \bibinfo{person}{Jianbin Fang},
  \bibinfo{person}{Jie Yu}, \bibinfo{person}{Chunyuan Zhang}, {and}
  \bibinfo{person}{Lei Luo}.} \bibinfo{year}{2020}\natexlab{b}.
\newblock \showarticletitle{Deep Learning Research and Development Platform:
  Characterizing and Scheduling with QoS Guarantees on GPU Clusters}.
\newblock \bibinfo{journal}{\emph{IEEE Transactions on Parallel and Distributed
  Systems}} (\bibinfo{year}{2020}).
\newblock


\bibitem[\protect\citeauthoryear{Chiang and Chou}{Chiang and Chou}{2021}]%
        {DynamoML}
\bibfield{author}{\bibinfo{person}{Min-Chi Chiang} {and} \bibinfo{person}{Jerry
  Chou}.} \bibinfo{year}{2021}\natexlab{}.
\newblock \showarticletitle{DynamoML: Dynamic Resource Management Operators for
  Machine Learning Workloads.}. In \bibinfo{booktitle}{\emph{CLOSER}}
  \emph{(\bibinfo{series}{CLOSER '21})}.
\newblock


\bibitem[\protect\citeauthoryear{Chowdhury and Stoica}{Chowdhury and
  Stoica}{2015}]%
        {chowdhury2015efficient}
\bibfield{author}{\bibinfo{person}{Mosharaf Chowdhury} {and}
  \bibinfo{person}{Ion Stoica}.} \bibinfo{year}{2015}\natexlab{}.
\newblock \showarticletitle{Efficient coflow scheduling without prior
  knowledge}.
\newblock \bibinfo{journal}{\emph{ACM SIGCOMM Computer Communication Review}}
  (\bibinfo{year}{2015}).
\newblock


\bibitem[\protect\citeauthoryear{Corbat{\'o}, Merwin-Daggett, and
  Daley}{Corbat{\'o} et~al\mbox{.}}{1962}]%
        {corbato1962experimental}
\bibfield{author}{\bibinfo{person}{Fernando~J Corbat{\'o}},
  \bibinfo{person}{Marjorie Merwin-Daggett}, {and} \bibinfo{person}{Robert~C
  Daley}.} \bibinfo{year}{1962}\natexlab{}.
\newblock \showarticletitle{An experimental time-sharing system}. In
  \bibinfo{booktitle}{\emph{spring joint computer conference}}.
\newblock


\bibitem[\protect\citeauthoryear{Crankshaw, Sela, Mo, Zumar, Stoica, Gonzalez,
  and Tumanov}{Crankshaw et~al\mbox{.}}{2020}]%
        {crankshaw2020inferline}
\bibfield{author}{\bibinfo{person}{Daniel Crankshaw}, \bibinfo{person}{Gur-Eyal
  Sela}, \bibinfo{person}{Xiangxi Mo}, \bibinfo{person}{Corey Zumar},
  \bibinfo{person}{Ion Stoica}, \bibinfo{person}{Joseph Gonzalez}, {and}
  \bibinfo{person}{Alexey Tumanov}.} \bibinfo{year}{2020}\natexlab{}.
\newblock \showarticletitle{InferLine: Latency-Aware Provisioning and Scaling
  for Prediction Serving Pipelines}. In \bibinfo{booktitle}{\emph{Proceedings
  of the 11th ACM Symposium on Cloud Computing}} \emph{(\bibinfo{series}{SoCC
  '20})}.
\newblock


\bibitem[\protect\citeauthoryear{Crankshaw, Wang, Zhou, Franklin, Gonzalez, and
  Stoica}{Crankshaw et~al\mbox{.}}{2017}]%
        {Daniel2017clipper}
\bibfield{author}{\bibinfo{person}{Daniel Crankshaw}, \bibinfo{person}{Xin
  Wang}, \bibinfo{person}{Guilio Zhou}, \bibinfo{person}{Michael~J. Franklin},
  \bibinfo{person}{Joseph~E. Gonzalez}, {and} \bibinfo{person}{Ion Stoica}.}
  \bibinfo{year}{2017}\natexlab{}.
\newblock \showarticletitle{Clipper: A {Low-Latency} Online Prediction Serving
  System}. In \bibinfo{booktitle}{\emph{14th USENIX Symposium on Networked
  Systems Design and Implementation (NSDI 17)}}.
\newblock


\bibitem[\protect\citeauthoryear{Cui, Chen, Zhao, Wei, Tang, and Guo}{Cui
  et~al\mbox{.}}{2021a}]%
        {E2bird}
\bibfield{author}{\bibinfo{person}{Weihao Cui}, \bibinfo{person}{Quan Chen},
  \bibinfo{person}{Han Zhao}, \bibinfo{person}{Mengze Wei},
  \bibinfo{person}{Xiaoxin Tang}, {and} \bibinfo{person}{Minyi Guo}.}
  \bibinfo{year}{2021}\natexlab{a}.
\newblock \showarticletitle{E2bird: Enhanced Elastic Batch for Improving
  Responsiveness and Throughput of Deep Learning Services}.
\newblock \bibinfo{journal}{\emph{IEEE Transactions on Parallel and Distributed
  Systems}} (\bibinfo{year}{2021}).
\newblock


\bibitem[\protect\citeauthoryear{Cui, Wei, Chen, Tang, Leng, Li, and Guo}{Cui
  et~al\mbox{.}}{2019}]%
        {Ebird}
\bibfield{author}{\bibinfo{person}{Weihao Cui}, \bibinfo{person}{Mengze Wei},
  \bibinfo{person}{Quan Chen}, \bibinfo{person}{Xiaoxin Tang},
  \bibinfo{person}{Jingwen Leng}, \bibinfo{person}{Li Li}, {and}
  \bibinfo{person}{Mingyi Guo}.} \bibinfo{year}{2019}\natexlab{}.
\newblock \showarticletitle{Ebird: Elastic Batch for Improving Responsiveness
  and Throughput of Deep Learning Services}. In \bibinfo{booktitle}{\emph{2019
  IEEE 37th International Conference on Computer Design (ICCD)}}.
\newblock


\bibitem[\protect\citeauthoryear{Cui, Zhao, Chen, Zheng, Leng, Zhao, Song, Ma,
  Yang, Li, and Guo}{Cui et~al\mbox{.}}{2021b}]%
        {cuiEnable2021}
\bibfield{author}{\bibinfo{person}{Weihao Cui}, \bibinfo{person}{Han Zhao},
  \bibinfo{person}{Quan Chen}, \bibinfo{person}{Ningxin Zheng},
  \bibinfo{person}{Jingwen Leng}, \bibinfo{person}{Jieru Zhao},
  \bibinfo{person}{Zhuo Song}, \bibinfo{person}{Tao Ma}, \bibinfo{person}{Yong
  Yang}, \bibinfo{person}{Chao Li}, {and} \bibinfo{person}{Minyi Guo}.}
  \bibinfo{year}{2021}\natexlab{b}.
\newblock \showarticletitle{Enable simultaneous DNN services based on
  deterministic operator overlap and precise latency prediction}. In
  \bibinfo{booktitle}{\emph{Proceedings of the International Conference for
  High Performance Computing, Networking, Storage and Analysis}}
  \emph{(\bibinfo{series}{SC '21})}.
\newblock


\bibitem[\protect\citeauthoryear{Cui, Zhao, Chen, Zheng, Leng, Zhao, Song, Ma,
  Yang, Li, and Guo}{Cui et~al\mbox{.}}{2021c}]%
        {cui2021abacus}
\bibfield{author}{\bibinfo{person}{Weihao Cui}, \bibinfo{person}{Han Zhao},
  \bibinfo{person}{Quan Chen}, \bibinfo{person}{Ningxin Zheng},
  \bibinfo{person}{Jingwen Leng}, \bibinfo{person}{Jieru Zhao},
  \bibinfo{person}{Zhuo Song}, \bibinfo{person}{Tao Ma}, \bibinfo{person}{Yong
  Yang}, \bibinfo{person}{Chao Li}, {and} \bibinfo{person}{Minyi Guo}.}
  \bibinfo{year}{2021}\natexlab{c}.
\newblock \showarticletitle{Enable Simultaneous DNN Services Based on
  Deterministic Operator Overlap and Precise Latency Prediction}. In
  \bibinfo{booktitle}{\emph{Proceedings of the International Conference for
  High Performance Computing, Networking, Storage and Analysis}}
  \emph{(\bibinfo{series}{SC '21})}.
\newblock


\bibitem[\protect\citeauthoryear{Dakkak, Li, Garcia~de Gonzalo, Xiong, and
  Hwu}{Dakkak et~al\mbox{.}}{2019}]%
        {dakkak2019TrIMS}
\bibfield{author}{\bibinfo{person}{Abdul Dakkak}, \bibinfo{person}{Cheng Li},
  \bibinfo{person}{Simon Garcia~de Gonzalo}, \bibinfo{person}{Jinjun Xiong},
  {and} \bibinfo{person}{Wen-mei Hwu}.} \bibinfo{year}{2019}\natexlab{}.
\newblock \showarticletitle{TrIMS: Transparent and Isolated Model Sharing for
  Low Latency Deep Learning Inference in Function-as-a-Service}. In
  \bibinfo{booktitle}{\emph{2019 IEEE 12th International Conference on Cloud
  Computing (CLOUD)}}.
\newblock


\bibitem[\protect\citeauthoryear{Dhakal, Kulkarni, and Ramakrishnan}{Dhakal
  et~al\mbox{.}}{2020}]%
        {dhakal2020GSLICE}
\bibfield{author}{\bibinfo{person}{Aditya Dhakal}, \bibinfo{person}{Sameer~G
  Kulkarni}, {and} \bibinfo{person}{K.~K. Ramakrishnan}.}
  \bibinfo{year}{2020}\natexlab{}.
\newblock \showarticletitle{GSLICE: controlled spatial sharing of GPUs for a
  scalable inference platform}. In \bibinfo{booktitle}{\emph{Proceedings of the
  11th ACM Symposium on Cloud Computing}} \emph{(\bibinfo{series}{SoCC '20})}.
\newblock


\bibitem[\protect\citeauthoryear{Ding, Zhu, Jia, Pekhimenko, and Han}{Ding
  et~al\mbox{.}}{2021}]%
        {IOS}
\bibfield{author}{\bibinfo{person}{Yaoyao Ding}, \bibinfo{person}{Ligeng Zhu},
  \bibinfo{person}{Zhihao Jia}, \bibinfo{person}{Gennady Pekhimenko}, {and}
  \bibinfo{person}{Song Han}.} \bibinfo{year}{2021}\natexlab{}.
\newblock \showarticletitle{IOS: Inter-Operator Scheduler for CNN
  Acceleration}. In \bibinfo{booktitle}{\emph{Proceedings of Machine Learning
  and Systems}} \emph{(\bibinfo{series}{MLSys '21})}.
\newblock


\bibitem[\protect\citeauthoryear{Dunlap, Kandasamy, Misra, Liaw, Jordan,
  Stoica, and Gonzalez}{Dunlap et~al\mbox{.}}{2021}]%
        {SEER}
\bibfield{author}{\bibinfo{person}{Lisa Dunlap}, \bibinfo{person}{Kirthevasan
  Kandasamy}, \bibinfo{person}{Ujval Misra}, \bibinfo{person}{Richard Liaw},
  \bibinfo{person}{Michael Jordan}, \bibinfo{person}{Ion Stoica}, {and}
  \bibinfo{person}{Joseph~E. Gonzalez}.} \bibinfo{year}{2021}\natexlab{}.
\newblock \showarticletitle{Elastic Hyperparameter Tuning on the Cloud}. In
  \bibinfo{booktitle}{\emph{Proceedings of the ACM Symposium on Cloud
  Computing}} \emph{(\bibinfo{series}{SoCC '21})}.
\newblock


\bibitem[\protect\citeauthoryear{Epema}{Epema}{1995}]%
        {epema1995analysis}
\bibfield{author}{\bibinfo{person}{Dick~HJ Epema}.}
  \bibinfo{year}{1995}\natexlab{}.
\newblock \showarticletitle{An analysis of decay-usage scheduling in
  multiprocessors}.
\newblock \bibinfo{journal}{\emph{ACM SIGMETRICS Performance Evaluation
  Review}} (\bibinfo{year}{1995}).
\newblock


\bibitem[\protect\citeauthoryear{Feitelson}{Feitelson}{1996}]%
        {gang}
\bibfield{author}{\bibinfo{person}{Dror~G. Feitelson}.}
  \bibinfo{year}{1996}\natexlab{}.
\newblock \showarticletitle{Packing schemes for gang scheduling}. In
  \bibinfo{booktitle}{\emph{Job Scheduling Strategies for Parallel
  Processing}}.
\newblock


\bibitem[\protect\citeauthoryear{Feitelson}{Feitelson}{1997}]%
        {feitelson1997job}
\bibfield{author}{\bibinfo{person}{Dror~G Feitelson}.}
  \bibinfo{year}{1997}\natexlab{}.
\newblock \showarticletitle{Job scheduling in multiprogrammed parallel
  systems}.
\newblock \bibinfo{journal}{\emph{IBM Research Report}} (\bibinfo{year}{1997}).
\newblock


\bibitem[\protect\citeauthoryear{Feitelson and Rudolph}{Feitelson and
  Rudolph}{1995}]%
        {feitelson1995parallel}
\bibfield{author}{\bibinfo{person}{Dror~G Feitelson} {and}
  \bibinfo{person}{Larry Rudolph}.} \bibinfo{year}{1995}\natexlab{}.
\newblock \showarticletitle{Parallel job scheduling: Issues and approaches}. In
  \bibinfo{booktitle}{\emph{Workshop on Job Scheduling Strategies for Parallel
  Processing}}.
\newblock


\bibitem[\protect\citeauthoryear{Feitelson, Rudolph, Schwiegelshohn, Sevcik,
  and Wong}{Feitelson et~al\mbox{.}}{1997}]%
        {feitelson1997theory}
\bibfield{author}{\bibinfo{person}{Dror~G Feitelson}, \bibinfo{person}{Larry
  Rudolph}, \bibinfo{person}{Uwe Schwiegelshohn}, \bibinfo{person}{Kenneth~C
  Sevcik}, {and} \bibinfo{person}{Parkson Wong}.}
  \bibinfo{year}{1997}\natexlab{}.
\newblock \showarticletitle{Theory and practice in parallel job scheduling}. In
  \bibinfo{booktitle}{\emph{Workshop on Job Scheduling Strategies for Parallel
  Processing}}.
\newblock


\bibitem[\protect\citeauthoryear{Filippini, Ardagna, Lattuada, Amaldi, Riedl,
  Materka, Skrzypek, Ciavotta, Magugliani, and Cicala}{Filippini
  et~al\mbox{.}}{2021}]%
        {ANDREAS}
\bibfield{author}{\bibinfo{person}{Federica Filippini}, \bibinfo{person}{Danilo
  Ardagna}, \bibinfo{person}{Marco Lattuada}, \bibinfo{person}{Edoardo Amaldi},
  \bibinfo{person}{Maciek Riedl}, \bibinfo{person}{Katarzyna Materka},
  \bibinfo{person}{Paweł Skrzypek}, \bibinfo{person}{Michele Ciavotta},
  \bibinfo{person}{Fabrizio Magugliani}, {and} \bibinfo{person}{Marco Cicala}.}
  \bibinfo{year}{2021}\natexlab{}.
\newblock \showarticletitle{ANDREAS: Artificial intelligence traiNing scheDuler
  foR accElerAted resource clusterS}. In \bibinfo{booktitle}{\emph{2021 8th
  International Conference on Future Internet of Things and Cloud}}
  \emph{(\bibinfo{series}{FiCloud '21})}.
\newblock


\bibitem[\protect\citeauthoryear{Finn, Abbeel, and Levine}{Finn
  et~al\mbox{.}}{2017}]%
        {finn2017model}
\bibfield{author}{\bibinfo{person}{Chelsea Finn}, \bibinfo{person}{Pieter
  Abbeel}, {and} \bibinfo{person}{Sergey Levine}.}
  \bibinfo{year}{2017}\natexlab{}.
\newblock \showarticletitle{Model-agnostic meta-learning for fast adaptation of
  deep networks}. In \bibinfo{booktitle}{\emph{International conference on
  machine learning}}.
\newblock


\bibitem[\protect\citeauthoryear{Gao, Yu, Wu, and Li}{Gao
  et~al\mbox{.}}{2018}]%
        {gao2018low}
\bibfield{author}{\bibinfo{person}{Pin Gao}, \bibinfo{person}{Lingfan Yu},
  \bibinfo{person}{Yongwei Wu}, {and} \bibinfo{person}{Jinyang Li}.}
  \bibinfo{year}{2018}\natexlab{}.
\newblock \showarticletitle{Low Latency RNN Inference with Cellular Batching}.
  In \bibinfo{booktitle}{\emph{Proceedings of the Thirteenth EuroSys
  Conference}} \emph{(\bibinfo{series}{EuroSys '18})}.
\newblock


\bibitem[\protect\citeauthoryear{Gao, Ye, Sun, Wen, and Zhang}{Gao
  et~al\mbox{.}}{2021}]%
        {Chronus}
\bibfield{author}{\bibinfo{person}{Wei Gao}, \bibinfo{person}{Zhisheng Ye},
  \bibinfo{person}{Peng Sun}, \bibinfo{person}{Yonggang Wen}, {and}
  \bibinfo{person}{Tianwei Zhang}.} \bibinfo{year}{2021}\natexlab{}.
\newblock \showarticletitle{Chronus: A Novel Deadline-aware Scheduler for Deep
  Learning Training Jobs}. In \bibinfo{booktitle}{\emph{Proceedings of the ACM
  Symposium on Cloud Computing}} \emph{(\bibinfo{series}{SoCC '21})}.
\newblock


\bibitem[\protect\citeauthoryear{Ghodsi, Zaharia, Hindman, Konwinski, Shenker,
  and Stoica}{Ghodsi et~al\mbox{.}}{2011}]%
        {DRF}
\bibfield{author}{\bibinfo{person}{Ali Ghodsi}, \bibinfo{person}{Matei
  Zaharia}, \bibinfo{person}{Benjamin Hindman}, \bibinfo{person}{Andy
  Konwinski}, \bibinfo{person}{Scott Shenker}, {and} \bibinfo{person}{Ion
  Stoica}.} \bibinfo{year}{2011}\natexlab{}.
\newblock \showarticletitle{{Dominant Resource Fairness: Fair Allocation of
  Multiple Resource Types}}. In \bibinfo{booktitle}{\emph{Proceedings of the
  8th USENIX Conference on Networked Systems Design and Implementation}}
  \emph{(\bibinfo{series}{NSDI '11})}.
\newblock


\bibitem[\protect\citeauthoryear{Gholami, Kim, Dong, Yao, Mahoney, and
  Keutzer}{Gholami et~al\mbox{.}}{2021}]%
        {SurveyInfer21}
\bibfield{author}{\bibinfo{person}{Amir Gholami}, \bibinfo{person}{Sehoon Kim},
  \bibinfo{person}{Zhen Dong}, \bibinfo{person}{Zhewei Yao},
  \bibinfo{person}{Michael~W. Mahoney}, {and} \bibinfo{person}{Kurt Keutzer}.}
  \bibinfo{year}{2021}\natexlab{}.
\newblock \showarticletitle{A Survey of Quantization Methods for Efficient
  Neural Network Inference}.
\newblock \bibinfo{journal}{\emph{CoRR}} (\bibinfo{year}{2021}).
\newblock


\bibitem[\protect\citeauthoryear{Gilman, Ogden, Walls, and Guo}{Gilman
  et~al\mbox{.}}{2019}]%
        {Challenges19}
\bibfield{author}{\bibinfo{person}{Guin~R. Gilman}, \bibinfo{person}{Samuel~S.
  Ogden}, \bibinfo{person}{Robert~J. Walls}, {and} \bibinfo{person}{Tian Guo}.}
  \bibinfo{year}{2019}\natexlab{}.
\newblock \showarticletitle{Challenges and Opportunities of DNN Model Execution
  Caching}. In \bibinfo{booktitle}{\emph{Proceedings of the Workshop on
  Distributed Infrastructures for Deep Learning}} \emph{(\bibinfo{series}{DIDL
  '19})}.
\newblock


\bibitem[\protect\citeauthoryear{Gu, Chowdhury, Shin, Zhu, Jeon, Qian, Liu, and
  Guo}{Gu et~al\mbox{.}}{2019}]%
        {Tiresias}
\bibfield{author}{\bibinfo{person}{Juncheng Gu}, \bibinfo{person}{Mosharaf
  Chowdhury}, \bibinfo{person}{Kang~G. Shin}, \bibinfo{person}{Yibo Zhu},
  \bibinfo{person}{Myeongjae Jeon}, \bibinfo{person}{Junjie Qian},
  \bibinfo{person}{Hongqiang Liu}, {and} \bibinfo{person}{Chuanxiong Guo}.}
  \bibinfo{year}{2019}\natexlab{}.
\newblock \showarticletitle{Tiresias: A {GPU} Cluster Manager for Distributed
  Deep Learning}. In \bibinfo{booktitle}{\emph{16th {USENIX} Symposium on
  Networked Systems Design and Implementation}} \emph{(\bibinfo{series}{NSDI
  '19})}.
\newblock


\bibitem[\protect\citeauthoryear{Gu, Chen, Liu, Dai, Chen, Zhang, Che, and
  Huang}{Gu et~al\mbox{.}}{2021}]%
        {Liquid}
\bibfield{author}{\bibinfo{person}{Rong Gu}, \bibinfo{person}{Yuquan Chen},
  \bibinfo{person}{Shuai Liu}, \bibinfo{person}{Haipeng Dai},
  \bibinfo{person}{Guihai Chen}, \bibinfo{person}{Kai Zhang},
  \bibinfo{person}{Yang Che}, {and} \bibinfo{person}{Yihua Huang}.}
  \bibinfo{year}{2021}\natexlab{}.
\newblock \showarticletitle{Liquid: Intelligent Resource Estimation and
  Network-Efficient Scheduling for Deep Learning Jobs on Distributed GPU
  Clusters}.
\newblock \bibinfo{journal}{\emph{IEEE Transactions on Parallel and Distributed
  Systems}} (\bibinfo{year}{2021}).
\newblock


\bibitem[\protect\citeauthoryear{Gu, Holly, Lillicrap, and Levine}{Gu
  et~al\mbox{.}}{2017}]%
        {gu2017deep}
\bibfield{author}{\bibinfo{person}{Shixiang Gu}, \bibinfo{person}{Ethan Holly},
  \bibinfo{person}{Timothy Lillicrap}, {and} \bibinfo{person}{Sergey Levine}.}
  \bibinfo{year}{2017}\natexlab{}.
\newblock \showarticletitle{Deep reinforcement learning for robotic
  manipulation with asynchronous off-policy updates}. In
  \bibinfo{booktitle}{\emph{2017 IEEE international conference on robotics and
  automation (ICRA)}}. IEEE, \bibinfo{pages}{3389--3396}.
\newblock


\bibitem[\protect\citeauthoryear{Guerreiro, Ilic, Roma, and Tomas}{Guerreiro
  et~al\mbox{.}}{2018}]%
        {DVFS-AWARE}
\bibfield{author}{\bibinfo{person}{Joao Guerreiro}, \bibinfo{person}{Aleksandar
  Ilic}, \bibinfo{person}{Nuno Roma}, {and} \bibinfo{person}{Pedro Tomas}.}
  \bibinfo{year}{2018}\natexlab{}.
\newblock \showarticletitle{GPGPU Power Modeling for Multi-domain
  Voltage-Frequency Scaling}. In \bibinfo{booktitle}{\emph{2018 IEEE
  International Symposium on High Performance Computer Architecture}}
  \emph{(\bibinfo{series}{HPCA '18})}.
\newblock


\bibitem[\protect\citeauthoryear{Gujarati, Elnikety, He, McKinley, and
  Brandenburg}{Gujarati et~al\mbox{.}}{2017}]%
        {gujarati2017Swayam}
\bibfield{author}{\bibinfo{person}{Arpan Gujarati}, \bibinfo{person}{Sameh
  Elnikety}, \bibinfo{person}{Yuxiong He}, \bibinfo{person}{Kathryn~S.
  McKinley}, {and} \bibinfo{person}{Bj{\"o}rn~B. Brandenburg}.}
  \bibinfo{year}{2017}\natexlab{}.
\newblock \showarticletitle{Swayam: distributed autoscaling to meet SLAs of
  machine learning inference services with resource efficiency}. In
  \bibinfo{booktitle}{\emph{Proceedings of the 18th ACM/IFIP/USENIX Middleware
  Conference}} \emph{(\bibinfo{series}{Middleware '17})}.
\newblock


\bibitem[\protect\citeauthoryear{Gujarati, Karimi, Alzayat, Hao, Kaufmann,
  Vigfusson, and Mace}{Gujarati et~al\mbox{.}}{2020}]%
        {Arpan2020clockwork}
\bibfield{author}{\bibinfo{person}{Arpan Gujarati}, \bibinfo{person}{Reza
  Karimi}, \bibinfo{person}{Safya Alzayat}, \bibinfo{person}{Wei Hao},
  \bibinfo{person}{Antoine Kaufmann}, \bibinfo{person}{Ymir Vigfusson}, {and}
  \bibinfo{person}{Jonathan Mace}.} \bibinfo{year}{2020}\natexlab{}.
\newblock \showarticletitle{Serving {DNNs} like Clockwork: Performance
  Predictability from the Bottom Up}. In \bibinfo{booktitle}{\emph{14th USENIX
  Symposium on Operating Systems Design and Implementation (OSDI 20)}}.
\newblock


\bibitem[\protect\citeauthoryear{Gunasekaran, Mishra, Thinakaran, Kandemir, and
  Das}{Gunasekaran et~al\mbox{.}}{2021}]%
        {Jashwant2021Cocktail}
\bibfield{author}{\bibinfo{person}{Jashwant~Raj Gunasekaran},
  \bibinfo{person}{Cyan~Subhra Mishra}, \bibinfo{person}{Prashanth Thinakaran},
  \bibinfo{person}{Mahmut~Taylan Kandemir}, {and} \bibinfo{person}{Chita~R.
  Das}.} \bibinfo{year}{2021}\natexlab{}.
\newblock \showarticletitle{Cocktail: Leveraging Ensemble Learning for
  Optimized Model Serving in Public Cloud}.
\newblock \bibinfo{journal}{\emph{CoRR}} (\bibinfo{year}{2021}).
\newblock


\bibitem[\protect\citeauthoryear{Gunasekaran, Thinakaran, Nachiappan, Kandemir,
  and Das}{Gunasekaran et~al\mbox{.}}{2020}]%
        {Gunasekaran2020Fifer}
\bibfield{author}{\bibinfo{person}{Jashwant~Raj Gunasekaran},
  \bibinfo{person}{Prashanth Thinakaran}, \bibinfo{person}{Nachiappan~C.
  Nachiappan}, \bibinfo{person}{Mahmut~Taylan Kandemir}, {and}
  \bibinfo{person}{Chita~R. Das}.} \bibinfo{year}{2020}\natexlab{}.
\newblock \showarticletitle{Fifer: Tackling Resource Underutilization in the
  Serverless Era}. In \bibinfo{booktitle}{\emph{Proceedings of the 21st
  International Middleware Conference}} \emph{(\bibinfo{series}{Middleware
  '20})}.
\newblock


\bibitem[\protect\citeauthoryear{Gupta, Hsia, Saraph, Wang, Reagen, Wei, Lee,
  Brooks, and Wu}{Gupta et~al\mbox{.}}{2020}]%
        {gupta2020deeprecsys}
\bibfield{author}{\bibinfo{person}{Udit Gupta}, \bibinfo{person}{Samuel Hsia},
  \bibinfo{person}{Vikram Saraph}, \bibinfo{person}{Xiaodong Wang},
  \bibinfo{person}{Brandon Reagen}, \bibinfo{person}{Gu-Yeon Wei},
  \bibinfo{person}{Hsien-Hsin~S. Lee}, \bibinfo{person}{David Brooks}, {and}
  \bibinfo{person}{Carole-Jean Wu}.} \bibinfo{year}{2020}\natexlab{}.
\newblock \showarticletitle{DeepRecSys: A System for Optimizing End-To-End
  At-Scale Neural Recommendation Inference}. In \bibinfo{booktitle}{\emph{2020
  ACM/IEEE 47th Annual International Symposium on Computer Architecture
  (ISCA)}}.
\newblock


\bibitem[\protect\citeauthoryear{Halpern, Boroujerdian, Mummert, Duesterwald,
  and Janapa~Reddi}{Halpern et~al\mbox{.}}{2019}]%
        {Halpern2019onesize}
\bibfield{author}{\bibinfo{person}{Matthew Halpern}, \bibinfo{person}{Behzad
  Boroujerdian}, \bibinfo{person}{Todd Mummert}, \bibinfo{person}{Evelyn
  Duesterwald}, {and} \bibinfo{person}{Vijay Janapa~Reddi}.}
  \bibinfo{year}{2019}\natexlab{}.
\newblock \showarticletitle{One Size Does Not Fit All: Quantifying and Exposing
  the Accuracy-Latency Trade-Off in Machine Learning Cloud Service APIs via
  Tolerance Tiers}. In \bibinfo{booktitle}{\emph{2019 IEEE International
  Symposium on Performance Analysis of Systems and Software (ISPASS)}}.
\newblock


\bibitem[\protect\citeauthoryear{Han, Rafique, Xu, Butt, Lim, and
  Vazhkudai}{Han et~al\mbox{.}}{2020a}]%
        {Marble}
\bibfield{author}{\bibinfo{person}{Jingoo Han}, \bibinfo{person}{M~Mustafa
  Rafique}, \bibinfo{person}{Luna Xu}, \bibinfo{person}{Ali~R Butt},
  \bibinfo{person}{Seung-Hwan Lim}, {and} \bibinfo{person}{Sudharshan~S
  Vazhkudai}.} \bibinfo{year}{2020}\natexlab{a}.
\newblock \showarticletitle{Marble: A multi-gpu aware job scheduler for deep
  learning on hpc systems}. In \bibinfo{booktitle}{\emph{2020 20th IEEE/ACM
  International Symposium on Cluster, Cloud and Internet Computing}}
  \emph{(\bibinfo{series}{CCGRID '20})}.
\newblock


\bibitem[\protect\citeauthoryear{Han, Tan, Jiang, Fu, Cao, and Lau}{Han
  et~al\mbox{.}}{2020b}]%
        {SPIN}
\bibfield{author}{\bibinfo{person}{Zhenhua Han}, \bibinfo{person}{Haisheng
  Tan}, \bibinfo{person}{Shaofeng H-C Jiang}, \bibinfo{person}{Xiaoming Fu},
  \bibinfo{person}{Wanli Cao}, {and} \bibinfo{person}{Francis~CM Lau}.}
  \bibinfo{year}{2020}\natexlab{b}.
\newblock \showarticletitle{Scheduling Placement-Sensitive BSP Jobs with
  Inaccurate Execution Time Estimation}. In \bibinfo{booktitle}{\emph{IEEE
  INFOCOM 2020-IEEE Conference on Computer Communications}}
  \emph{(\bibinfo{series}{INFOCOM '20})}.
\newblock


\bibitem[\protect\citeauthoryear{Hazelwood, Bird, Brooks, Chintala, Diril,
  Dzhulgakov, Fawzy, Jia, Jia, Kalro, et~al\mbox{.}}{Hazelwood
  et~al\mbox{.}}{2018}]%
        {FaceBookTrace}
\bibfield{author}{\bibinfo{person}{Kim Hazelwood}, \bibinfo{person}{Sarah
  Bird}, \bibinfo{person}{David Brooks}, \bibinfo{person}{Soumith Chintala},
  \bibinfo{person}{Utku Diril}, \bibinfo{person}{Dmytro Dzhulgakov},
  \bibinfo{person}{Mohamed Fawzy}, \bibinfo{person}{Bill Jia},
  \bibinfo{person}{Yangqing Jia}, \bibinfo{person}{Aditya Kalro},
  {et~al\mbox{.}}} \bibinfo{year}{2018}\natexlab{}.
\newblock \showarticletitle{Applied machine learning at facebook: A datacenter
  infrastructure perspective}. In \bibinfo{booktitle}{\emph{2018 IEEE
  International Symposium on High Performance Computer Architecture}}
  \emph{(\bibinfo{series}{HPCA '18})}.
\newblock


\bibitem[\protect\citeauthoryear{Hindman, Konwinski, Zaharia, Ghodsi, Joseph,
  Katz, Shenker, and Stoica}{Hindman et~al\mbox{.}}{2011}]%
        {Mesos}
\bibfield{author}{\bibinfo{person}{Benjamin Hindman}, \bibinfo{person}{Andy
  Konwinski}, \bibinfo{person}{Matei Zaharia}, \bibinfo{person}{Ali Ghodsi},
  \bibinfo{person}{Anthony~D. Joseph}, \bibinfo{person}{Randy Katz},
  \bibinfo{person}{Scott Shenker}, {and} \bibinfo{person}{Ion Stoica}.}
  \bibinfo{year}{2011}\natexlab{}.
\newblock \showarticletitle{Mesos: A Platform for Fine-Grained Resource Sharing
  in the Data Center}. In \bibinfo{booktitle}{\emph{8th {USENIX} Symposium on
  Networked Systems Design and Implementation}} \emph{(\bibinfo{series}{NSDI
  '11})}.
\newblock


\bibitem[\protect\citeauthoryear{Holmes, Mawhirter, He, Yan, and Wu}{Holmes
  et~al\mbox{.}}{2019}]%
        {holmes2019grnn}
\bibfield{author}{\bibinfo{person}{Connor Holmes}, \bibinfo{person}{Daniel
  Mawhirter}, \bibinfo{person}{Yuxiong He}, \bibinfo{person}{Feng Yan}, {and}
  \bibinfo{person}{Bo Wu}.} \bibinfo{year}{2019}\natexlab{}.
\newblock \showarticletitle{GRNN: Low-Latency and Scalable RNN Inference on
  GPUs}. In \bibinfo{booktitle}{\emph{Proceedings of the Fourteenth EuroSys
  Conference 2019}} \emph{(\bibinfo{series}{EuroSys '19})}.
\newblock


\bibitem[\protect\citeauthoryear{Hong, Spence, and Nikolopoulos}{Hong
  et~al\mbox{.}}{2017}]%
        {SurvGPUVirtual}
\bibfield{author}{\bibinfo{person}{Cheol-Ho Hong}, \bibinfo{person}{Ivor
  Spence}, {and} \bibinfo{person}{Dimitrios~S. Nikolopoulos}.}
  \bibinfo{year}{2017}\natexlab{}.
\newblock \showarticletitle{GPU Virtualization and Scheduling Methods: A
  Comprehensive Survey}.
\newblock \bibinfo{journal}{\emph{Comput. Surveys}} (\bibinfo{year}{2017}).
\newblock


\bibitem[\protect\citeauthoryear{Hu, Sun, Yan, Wen, and Zhang}{Hu
  et~al\mbox{.}}{2021}]%
        {Helios}
\bibfield{author}{\bibinfo{person}{Qinghao Hu}, \bibinfo{person}{Peng Sun},
  \bibinfo{person}{Shengen Yan}, \bibinfo{person}{Yonggang Wen}, {and}
  \bibinfo{person}{Tianwei Zhang}.} \bibinfo{year}{2021}\natexlab{}.
\newblock \showarticletitle{Characterization and Prediction of Deep Learning
  Workloads in Large-Scale GPU Datacenters}. In
  \bibinfo{booktitle}{\emph{Proceedings of the International Conference for
  High Performance Computing, Networking, Storage and Analysis}}
  \emph{(\bibinfo{series}{SC '21})}.
\newblock


\bibitem[\protect\citeauthoryear{Hutt, Viswanathan, and Nadolski}{Hutt
  et~al\mbox{.}}{2019}]%
        {aws2019}
\bibfield{author}{\bibinfo{person}{Gadi Hutt}, \bibinfo{person}{Vibhav
  Viswanathan}, {and} \bibinfo{person}{Adam Nadolski}.}
  \bibinfo{year}{2019}\natexlab{}.
\newblock \bibinfo{title}{Deliver high performance ML inference with AWS
  Inferentia}.
\newblock
\newblock


\bibitem[\protect\citeauthoryear{Hwang, Kim, Kim, Shin, and Park}{Hwang
  et~al\mbox{.}}{2021}]%
        {AFS}
\bibfield{author}{\bibinfo{person}{Changho Hwang}, \bibinfo{person}{Taehyun
  Kim}, \bibinfo{person}{Sunghyun Kim}, \bibinfo{person}{Jinwoo Shin}, {and}
  \bibinfo{person}{KyoungSoo Park}.} \bibinfo{year}{2021}\natexlab{}.
\newblock \showarticletitle{Elastic Resource Sharing for Distributed Deep
  Learning}. In \bibinfo{booktitle}{\emph{18th {USENIX} Symposium on Networked
  Systems Design and Implementation}} \emph{(\bibinfo{series}{NSDI '21})}.
\newblock


\bibitem[\protect\citeauthoryear{Hwang, Kim, Kwon, and Rhu}{Hwang
  et~al\mbox{.}}{2020}]%
        {hwang2020Centaur}
\bibfield{author}{\bibinfo{person}{Ranggi Hwang}, \bibinfo{person}{Taehun Kim},
  \bibinfo{person}{Youngeun Kwon}, {and} \bibinfo{person}{Minsoo Rhu}.}
  \bibinfo{year}{2020}\natexlab{}.
\newblock \showarticletitle{Centaur: a chiplet-based, hybrid sparse-dense
  accelerator for personalized recommendations}. In
  \bibinfo{booktitle}{\emph{Proceedings of the ACM/IEEE 47th Annual
  International Symposium on Computer Architecture}}
  \emph{(\bibinfo{series}{ISCA '20})}.
\newblock


\bibitem[\protect\citeauthoryear{Ishakian, Muthusamy, and Slominski}{Ishakian
  et~al\mbox{.}}{2018}]%
        {Ishakian2018Serving}
\bibfield{author}{\bibinfo{person}{Vatche Ishakian}, \bibinfo{person}{Vinod
  Muthusamy}, {and} \bibinfo{person}{Aleksander Slominski}.}
  \bibinfo{year}{2018}\natexlab{}.
\newblock \showarticletitle{Serving Deep Learning Models in a Serverless
  Platform}. In \bibinfo{booktitle}{\emph{2018 IEEE International Conference on
  Cloud Engineering (IC2E)}}.
\newblock


\bibitem[\protect\citeauthoryear{Jahani, Lattuada, Ciavotta, Ardagna, Amaldi,
  and Zhang}{Jahani et~al\mbox{.}}{2019}]%
        {Jahani}
\bibfield{author}{\bibinfo{person}{Arezoo Jahani}, \bibinfo{person}{Marco
  Lattuada}, \bibinfo{person}{Michele Ciavotta}, \bibinfo{person}{Danilo
  Ardagna}, \bibinfo{person}{Edoardo Amaldi}, {and} \bibinfo{person}{Li
  Zhang}.} \bibinfo{year}{2019}\natexlab{}.
\newblock \showarticletitle{Optimizing on-demand gpus in the cloud for deep
  learning applications training}. In \bibinfo{booktitle}{\emph{2019 4th
  International Conference on Computing, Communications and Security}}
  \emph{(\bibinfo{series}{ICCCS '18})}.
\newblock


\bibitem[\protect\citeauthoryear{Jain, Mo, Jain, Subbaraj, Durrani, Tumanov,
  Gonzalez, and Stoica}{Jain et~al\mbox{.}}{2018}]%
        {jain2018dynamic}
\bibfield{author}{\bibinfo{person}{Paras Jain}, \bibinfo{person}{Xiangxi Mo},
  \bibinfo{person}{Ajay Jain}, \bibinfo{person}{Harikaran Subbaraj},
  \bibinfo{person}{Rehan~Sohail Durrani}, \bibinfo{person}{Alexey Tumanov},
  \bibinfo{person}{Joseph Gonzalez}, {and} \bibinfo{person}{Ion Stoica}.}
  \bibinfo{year}{2018}\natexlab{}.
\newblock \showarticletitle{Dynamic space-time scheduling for gpu inference}.
\newblock \bibinfo{journal}{\emph{arXiv preprint arXiv:1901.00041}}
  (\bibinfo{year}{2018}).
\newblock


\bibitem[\protect\citeauthoryear{Jarachanthan, Chen, Xu, and Li}{Jarachanthan
  et~al\mbox{.}}{2021}]%
        {Jananie2021AMPS-Inf}
\bibfield{author}{\bibinfo{person}{Jananie Jarachanthan}, \bibinfo{person}{Li
  Chen}, \bibinfo{person}{Fei Xu}, {and} \bibinfo{person}{Bo Li}.}
  \bibinfo{year}{2021}\natexlab{}.
\newblock \showarticletitle{AMPS-Inf: Automatic Model Partitioning for
  Serverless Inference with Cost Efficiency}. In \bibinfo{booktitle}{\emph{50th
  International Conference on Parallel Processing}}
  \emph{(\bibinfo{series}{ICPP 2021})}.
\newblock


\bibitem[\protect\citeauthoryear{Jayaram, Muthusamy, Dube, Ishakian, Wang,
  Herta, Boag, Arroyo, Tantawi, Verma, Pollok, and Khalaf}{Jayaram
  et~al\mbox{.}}{2019}]%
        {FfDL}
\bibfield{author}{\bibinfo{person}{K.~R. Jayaram}, \bibinfo{person}{Vinod
  Muthusamy}, \bibinfo{person}{Parijat Dube}, \bibinfo{person}{Vatche
  Ishakian}, \bibinfo{person}{Chen Wang}, \bibinfo{person}{Benjamin Herta},
  \bibinfo{person}{Scott Boag}, \bibinfo{person}{Diana Arroyo},
  \bibinfo{person}{Asser Tantawi}, \bibinfo{person}{Archit Verma},
  \bibinfo{person}{Falk Pollok}, {and} \bibinfo{person}{Rania Khalaf}.}
  \bibinfo{year}{2019}\natexlab{}.
\newblock \showarticletitle{FfDL: A Flexible Multi-tenant Deep Learning
  Platform}. In \bibinfo{booktitle}{\emph{Proceedings of the 20th International
  Middleware Conference}} \emph{(\bibinfo{series}{Middleware '19})}.
\newblock


\bibitem[\protect\citeauthoryear{Jeon, Venkataraman, Phanishayee, Qian, Xiao,
  and Yang}{Jeon et~al\mbox{.}}{2019}]%
        {Philly}
\bibfield{author}{\bibinfo{person}{Myeongjae Jeon}, \bibinfo{person}{Shivaram
  Venkataraman}, \bibinfo{person}{Amar Phanishayee}, \bibinfo{person}{Junjie
  Qian}, \bibinfo{person}{Wencong Xiao}, {and} \bibinfo{person}{Fan Yang}.}
  \bibinfo{year}{2019}\natexlab{}.
\newblock \showarticletitle{Analysis of Large-Scale Multi-Tenant {GPU} Clusters
  for {DNN} Training Workloads}. In \bibinfo{booktitle}{\emph{2019 {USENIX}
  Annual Technical Conference}} \emph{(\bibinfo{series}{{USENIX} {ATC} '19})}.
\newblock


\bibitem[\protect\citeauthoryear{Jiang, He, Zhang, Preu\ss~er, Zeng, Feng,
  Zhang, Liu, Li, Zhou, Zhang, and Alonso}{Jiang et~al\mbox{.}}{2021a}]%
        {jiang2021MicroRec}
\bibfield{author}{\bibinfo{person}{Wenqi Jiang}, \bibinfo{person}{Zhenhao He},
  \bibinfo{person}{Shuai Zhang}, \bibinfo{person}{Thomas~B. Preu\ss~er},
  \bibinfo{person}{Kai Zeng}, \bibinfo{person}{Liang Feng},
  \bibinfo{person}{Jiansong Zhang}, \bibinfo{person}{Tongxuan Liu},
  \bibinfo{person}{Yong Li}, \bibinfo{person}{Jingren Zhou},
  \bibinfo{person}{Ce Zhang}, {and} \bibinfo{person}{Gustavo Alonso}.}
  \bibinfo{year}{2021}\natexlab{a}.
\newblock \showarticletitle{MicroRec: Efficient Recommendation Inference by
  Hardware and Data Structure Solutions}. In
  \bibinfo{booktitle}{\emph{Proceedings of Machine Learning and Systems}}.
\newblock


\bibitem[\protect\citeauthoryear{Jiang, He, Zhang, Zeng, Feng, Zhang, Liu, Li,
  Zhou, Zhang, and Alonso}{Jiang et~al\mbox{.}}{2021b}]%
        {jiang2021FleetRec}
\bibfield{author}{\bibinfo{person}{Wenqi Jiang}, \bibinfo{person}{Zhenhao He},
  \bibinfo{person}{Shuai Zhang}, \bibinfo{person}{Kai Zeng},
  \bibinfo{person}{Liang Feng}, \bibinfo{person}{Jiansong Zhang},
  \bibinfo{person}{Tongxuan Liu}, \bibinfo{person}{Yong Li},
  \bibinfo{person}{Jingren Zhou}, \bibinfo{person}{Ce Zhang}, {and}
  \bibinfo{person}{Gustavo Alonso}.} \bibinfo{year}{2021}\natexlab{b}.
\newblock \showarticletitle{FleetRec: Large-Scale Recommendation Inference on
  Hybrid GPU-FPGA Clusters}. In \bibinfo{booktitle}{\emph{Proceedings of the
  27th ACM SIGKDD Conference on Knowledge Discovery \& Data Mining}}
  \emph{(\bibinfo{series}{KDD '21})}.
\newblock


\bibitem[\protect\citeauthoryear{Jiang, Zhu, Lan, Yi, Cui, and Guo}{Jiang
  et~al\mbox{.}}{2020}]%
        {BytePS}
\bibfield{author}{\bibinfo{person}{Yimin Jiang}, \bibinfo{person}{Yibo Zhu},
  \bibinfo{person}{Chang Lan}, \bibinfo{person}{Bairen Yi},
  \bibinfo{person}{Yong Cui}, {and} \bibinfo{person}{Chuanxiong Guo}.}
  \bibinfo{year}{2020}\natexlab{}.
\newblock \showarticletitle{A Unified Architecture for Accelerating Distributed
  {DNN} Training in Heterogeneous GPU/CPU Clusters}. In
  \bibinfo{booktitle}{\emph{14th {USENIX} Symposium on Operating Systems Design
  and Implementation}} \emph{(\bibinfo{series}{OSDI '20})}.
\newblock


\bibitem[\protect\citeauthoryear{Kang, Mathur, Veeramacheneni, Bailis, and
  Zaharia}{Kang et~al\mbox{.}}{2020}]%
        {kang2020jointly}
\bibfield{author}{\bibinfo{person}{Daniel Kang}, \bibinfo{person}{Ankit
  Mathur}, \bibinfo{person}{Teja Veeramacheneni}, \bibinfo{person}{Peter
  Bailis}, {and} \bibinfo{person}{Matei Zaharia}.}
  \bibinfo{year}{2020}\natexlab{}.
\newblock \showarticletitle{Jointly Optimizing Preprocessing and Inference for
  DNN-Based Visual Analytics}.
\newblock \bibinfo{journal}{\emph{Proc. VLDB Endow.}} (\bibinfo{year}{2020}).
\newblock


\bibitem[\protect\citeauthoryear{Kim and Kim}{Kim and Kim}{2020}]%
        {Co-scheML}
\bibfield{author}{\bibinfo{person}{Sejin Kim} {and} \bibinfo{person}{Yoonhee
  Kim}.} \bibinfo{year}{2020}\natexlab{}.
\newblock \showarticletitle{Co-scheML: Interference-aware Container
  Co-scheduling Scheme Using Machine Learning Application Profiles for GPU
  Clusters}. In \bibinfo{booktitle}{\emph{2020 IEEE International Conference on
  Cluster Computing}} \emph{(\bibinfo{series}{CLUSTER '20})}.
\newblock


\bibitem[\protect\citeauthoryear{Kosaian, Rashmi, and Venkataraman}{Kosaian
  et~al\mbox{.}}{2019a}]%
        {ParM}
\bibfield{author}{\bibinfo{person}{Jack Kosaian}, \bibinfo{person}{K.~V.
  Rashmi}, {and} \bibinfo{person}{Shivaram Venkataraman}.}
  \bibinfo{year}{2019}\natexlab{a}.
\newblock \showarticletitle{Parity Models: Erasure-Coded Resilience for
  Prediction Serving Systems}. In \bibinfo{booktitle}{\emph{Proceedings of the
  27th ACM Symposium on Operating Systems Principles}}
  \emph{(\bibinfo{series}{SOSP '19})}.
\newblock


\bibitem[\protect\citeauthoryear{Kosaian, Rashmi, and Venkataraman}{Kosaian
  et~al\mbox{.}}{2019b}]%
        {Kosaian2019parity}
\bibfield{author}{\bibinfo{person}{Jack Kosaian}, \bibinfo{person}{K.~V.
  Rashmi}, {and} \bibinfo{person}{Shivaram Venkataraman}.}
  \bibinfo{year}{2019}\natexlab{b}.
\newblock \showarticletitle{Parity Models: Erasure-Coded Resilience for
  Prediction Serving Systems} \emph{(\bibinfo{series}{SOSP '19})}.
\newblock


\bibitem[\protect\citeauthoryear{Kumar, Subramanian, Venkataraman, and
  Akella}{Kumar et~al\mbox{.}}{2021}]%
        {JIGSAW}
\bibfield{author}{\bibinfo{person}{Adarsh Kumar}, \bibinfo{person}{Kausik
  Subramanian}, \bibinfo{person}{Shivaram Venkataraman}, {and}
  \bibinfo{person}{Aditya Akella}.} \bibinfo{year}{2021}\natexlab{}.
\newblock \showarticletitle{Doing more by doing less: how structured partial
  backpropagation improves deep learning clusters}. In
  \bibinfo{booktitle}{\emph{Proceedings of the 2nd ACM International Workshop
  on Distributed Machine Learning}}.
\newblock


\bibitem[\protect\citeauthoryear{Le, Sun, Chowdhury, and Liu}{Le
  et~al\mbox{.}}{2020}]%
        {Allox}
\bibfield{author}{\bibinfo{person}{Tan~N. Le}, \bibinfo{person}{Xiao Sun},
  \bibinfo{person}{Mosharaf Chowdhury}, {and} \bibinfo{person}{Zhenhua Liu}.}
  \bibinfo{year}{2020}\natexlab{}.
\newblock \showarticletitle{AlloX: Compute Allocation in Hybrid Clusters}. In
  \bibinfo{booktitle}{\emph{Proceedings of the Fifteenth European Conference on
  Computer Systems}} \emph{(\bibinfo{series}{EuroSys '20})}.
\newblock


\bibitem[\protect\citeauthoryear{L\'{e}cuyer, Spahn, Vodrahalli, Geambasu, and
  Hsu}{L\'{e}cuyer et~al\mbox{.}}{2019}]%
        {Sage}
\bibfield{author}{\bibinfo{person}{Mathias L\'{e}cuyer}, \bibinfo{person}{Riley
  Spahn}, \bibinfo{person}{Kiran Vodrahalli}, \bibinfo{person}{Roxana
  Geambasu}, {and} \bibinfo{person}{Daniel Hsu}.}
  \bibinfo{year}{2019}\natexlab{}.
\newblock \showarticletitle{Privacy Accounting and Quality Control in the Sage
  Differentially Private ML Platform}. In \bibinfo{booktitle}{\emph{Proceedings
  of the 27th ACM Symposium on Operating Systems Principles}}
  \emph{(\bibinfo{series}{SOSP '19})}.
\newblock


\bibitem[\protect\citeauthoryear{Lee, Scolari, Chun, Santambrogio, Weimer, and
  Interlandi}{Lee et~al\mbox{.}}{2018}]%
        {PRETZEL}
\bibfield{author}{\bibinfo{person}{Yunseong Lee}, \bibinfo{person}{Alberto
  Scolari}, \bibinfo{person}{Byung-Gon Chun}, \bibinfo{person}{Marco~Domenico
  Santambrogio}, \bibinfo{person}{Markus Weimer}, {and} \bibinfo{person}{Matteo
  Interlandi}.} \bibinfo{year}{2018}\natexlab{}.
\newblock \showarticletitle{{PRETZEL}: Opening the Black Box of Machine
  Learning Prediction Serving Systems}. In \bibinfo{booktitle}{\emph{13th
  {USENIX} Symposium on Operating Systems Design and Implementation}}
  \emph{(\bibinfo{series}{OSDI '18})}.
\newblock


\bibitem[\protect\citeauthoryear{LeMay, Li, and Guo}{LeMay
  et~al\mbox{.}}{2020}]%
        {lemay2020Perseus}
\bibfield{author}{\bibinfo{person}{Matthew LeMay}, \bibinfo{person}{Shijian
  Li}, {and} \bibinfo{person}{Tian Guo}.} \bibinfo{year}{2020}\natexlab{}.
\newblock \showarticletitle{PERSEUS: Characterizing Performance and Cost of
  Multi-Tenant Serving for CNN Models}. In \bibinfo{booktitle}{\emph{2020 IEEE
  International Conference on Cloud Engineering (IC2E)}}.
\newblock


\bibitem[\protect\citeauthoryear{Li, Sun, Li, and Xu}{Li
  et~al\mbox{.}}{2020c}]%
        {OJPP}
\bibfield{author}{\bibinfo{person}{Hongliang Li}, \bibinfo{person}{Ting Sun},
  \bibinfo{person}{Xiang Li}, {and} \bibinfo{person}{Haixiao Xu}.}
  \bibinfo{year}{2020}\natexlab{c}.
\newblock \showarticletitle{Job Placement Strategy with Opportunistic Resource
  Sharing for Distributed Deep Learning Clusters}. In
  \bibinfo{booktitle}{\emph{2020 IEEE 22nd International Conference on High
  Performance Computing and Communications}} \emph{(\bibinfo{series}{HPCC
  '20})}.
\newblock


\bibitem[\protect\citeauthoryear{Li, Xu, Zhu, Liu, Guo, and Wang}{Li
  et~al\mbox{.}}{2022}]%
        {Aryl}
\bibfield{author}{\bibinfo{person}{Jiamin Li}, \bibinfo{person}{Hong Xu},
  \bibinfo{person}{Yibo Zhu}, \bibinfo{person}{Zherui Liu},
  \bibinfo{person}{Chuanxiong Guo}, {and} \bibinfo{person}{Cong Wang}.}
  \bibinfo{year}{2022}\natexlab{}.
\newblock \showarticletitle{Aryl: An Elastic Cluster Scheduler for Deep
  Learning}.
\newblock \bibinfo{journal}{\emph{CoRR}} (\bibinfo{year}{2022}).
\newblock


\bibitem[\protect\citeauthoryear{Li and Gui}{Li and Gui}{2020}]%
        {li2020cms}
\bibfield{author}{\bibinfo{person}{KeDi Li} {and} \bibinfo{person}{Ning Gui}.}
  \bibinfo{year}{2020}\natexlab{}.
\newblock \showarticletitle{CMS: A continuous machine-learning and serving
  platform for industrial big data}.
\newblock \bibinfo{journal}{\emph{Future Internet}} (\bibinfo{year}{2020}).
\newblock


\bibitem[\protect\citeauthoryear{Li, Zhong, Liu, Wu, and Zhang}{Li
  et~al\mbox{.}}{2018}]%
        {li2018easeml}
\bibfield{author}{\bibinfo{person}{Tian Li}, \bibinfo{person}{Jie Zhong},
  \bibinfo{person}{Ji Liu}, \bibinfo{person}{Wentao Wu}, {and}
  \bibinfo{person}{Ce Zhang}.} \bibinfo{year}{2018}\natexlab{}.
\newblock \showarticletitle{Ease.Ml: Towards Multi-Tenant Resource Sharing for
  Machine Learning Workloads}.
\newblock \bibinfo{journal}{\emph{Proc. VLDB Endow.}} (\bibinfo{year}{2018}).
\newblock


\bibitem[\protect\citeauthoryear{Li, Chen, Li, Qi, Xu, and Zhang}{Li
  et~al\mbox{.}}{2020a}]%
        {Parrot}
\bibfield{author}{\bibinfo{person}{Wenxin Li}, \bibinfo{person}{Sheng Chen},
  \bibinfo{person}{Keqiu Li}, \bibinfo{person}{Heng Qi},
  \bibinfo{person}{Renhai Xu}, {and} \bibinfo{person}{Song Zhang}.}
  \bibinfo{year}{2020}\natexlab{a}.
\newblock \showarticletitle{Efficient Online Scheduling for Coflow-aware
  Machine Learning Clusters}.
\newblock \bibinfo{journal}{\emph{IEEE Transactions on Cloud Computing}}
  (\bibinfo{year}{2020}).
\newblock


\bibitem[\protect\citeauthoryear{Li, Han, Zhang, Li, and Tan}{Li
  et~al\mbox{.}}{2020b}]%
        {AutoDeep}
\bibfield{author}{\bibinfo{person}{Yang Li}, \bibinfo{person}{Zhenhua Han},
  \bibinfo{person}{Quanlu Zhang}, \bibinfo{person}{Zhenhua Li}, {and}
  \bibinfo{person}{Haisheng Tan}.} \bibinfo{year}{2020}\natexlab{b}.
\newblock \showarticletitle{Automating Cloud Deployment for Deep Learning
  Inference of Real-time Online Services}. In \bibinfo{booktitle}{\emph{IEEE
  Conference on Computer Communications}} \emph{(\bibinfo{series}{INFOCOM
  '20})}.
\newblock


\bibitem[\protect\citeauthoryear{Liang, Glossner, Wang, Shi, and Zhang}{Liang
  et~al\mbox{.}}{2021}]%
        {SurveyPruneQuant}
\bibfield{author}{\bibinfo{person}{Tailin Liang}, \bibinfo{person}{John
  Glossner}, \bibinfo{person}{Lei Wang}, \bibinfo{person}{Shaobo Shi}, {and}
  \bibinfo{person}{Xiaotong Zhang}.} \bibinfo{year}{2021}\natexlab{}.
\newblock \showarticletitle{Pruning and quantization for deep neural network
  acceleration: A survey}.
\newblock \bibinfo{journal}{\emph{Neurocomputing}} (\bibinfo{year}{2021}).
\newblock


\bibitem[\protect\citeauthoryear{Liaw, Bhardwaj, Dunlap, Zou, Gonzalez, Stoica,
  and Tumanov}{Liaw et~al\mbox{.}}{2019}]%
        {HyperSched}
\bibfield{author}{\bibinfo{person}{Richard Liaw}, \bibinfo{person}{Romil
  Bhardwaj}, \bibinfo{person}{Lisa Dunlap}, \bibinfo{person}{Yitian Zou},
  \bibinfo{person}{Joseph~E. Gonzalez}, \bibinfo{person}{Ion Stoica}, {and}
  \bibinfo{person}{Alexey Tumanov}.} \bibinfo{year}{2019}\natexlab{}.
\newblock \showarticletitle{HyperSched: Dynamic Resource Reallocation for Model
  Development on a Deadline}. In \bibinfo{booktitle}{\emph{Proceedings of the
  ACM Symposium on Cloud Computing}} \emph{(\bibinfo{series}{SoCC '19})}.
\newblock


\bibitem[\protect\citeauthoryear{Lin, Yeh, and Chou}{Lin et~al\mbox{.}}{2019}]%
        {Dragon}
\bibfield{author}{\bibinfo{person}{Chan-Yi Lin}, \bibinfo{person}{Ting-An Yeh},
  {and} \bibinfo{person}{Jerry Chou}.} \bibinfo{year}{2019}\natexlab{}.
\newblock \showarticletitle{DRAGON: A Dynamic Scheduling and Scaling Controller
  for Managing Distributed Deep Learning Jobs in Kubernetes Cluster}. In
  \bibinfo{booktitle}{\emph{CLOSER}} \emph{(\bibinfo{series}{CLOSER '19})}.
\newblock


\bibitem[\protect\citeauthoryear{Liu, Gao, Li, Liao, Xiong, Chen, Wang, Yang,
  Zha, Dong, Dou, and Xiong}{Liu et~al\mbox{.}}{2021a}]%
        {liu2021jizhi}
\bibfield{author}{\bibinfo{person}{Hao Liu}, \bibinfo{person}{Qian Gao},
  \bibinfo{person}{Jiang Li}, \bibinfo{person}{Xiaochao Liao},
  \bibinfo{person}{Hao Xiong}, \bibinfo{person}{Guangxing Chen},
  \bibinfo{person}{Wenlin Wang}, \bibinfo{person}{Guobao Yang},
  \bibinfo{person}{Zhiwei Zha}, \bibinfo{person}{Daxiang Dong},
  \bibinfo{person}{Dejing Dou}, {and} \bibinfo{person}{Haoyi Xiong}.}
  \bibinfo{year}{2021}\natexlab{a}.
\newblock \showarticletitle{JIZHI: A Fast and Cost-Effective Model-As-A-Service
  System for Web-Scale Online Inference at Baidu}. In
  \bibinfo{booktitle}{\emph{Proceedings of the 27th ACM SIGKDD Conference on
  Knowledge Discovery \& Data Mining}} \emph{(\bibinfo{series}{KDD '21})}.
\newblock


\bibitem[\protect\citeauthoryear{Liu, Krishnan, Elmore, and Franklin}{Liu
  et~al\mbox{.}}{2021b}]%
        {Pack}
\bibfield{author}{\bibinfo{person}{Rui Liu}, \bibinfo{person}{Sanjay Krishnan},
  \bibinfo{person}{Aaron~J. Elmore}, {and} \bibinfo{person}{Michael~J.
  Franklin}.} \bibinfo{year}{2021}\natexlab{b}.
\newblock \showarticletitle{Understanding and optimizing packed neural network
  training for hyperparameter tuning}. In \bibinfo{booktitle}{\emph{Proceedings
  of the Fifth Workshop on Data Management for End-To-End Machine Learning}}
  \emph{(\bibinfo{series}{DEEM '21})}.
\newblock


\bibitem[\protect\citeauthoryear{Luan, Chen, Zhao, Yang, and Dai}{Luan
  et~al\mbox{.}}{2019}]%
        {Sched2}
\bibfield{author}{\bibinfo{person}{Yunteng Luan}, \bibinfo{person}{Xukun Chen},
  \bibinfo{person}{Hanyu Zhao}, \bibinfo{person}{Zhi Yang}, {and}
  \bibinfo{person}{Yafei Dai}.} \bibinfo{year}{2019}\natexlab{}.
\newblock \showarticletitle{SCHED$^2$: Scheduling Deep Learning Training via
  Deep Reinforcement Learning}. In \bibinfo{booktitle}{\emph{2019 IEEE Global
  Communications Conference}} \emph{(\bibinfo{series}{GLOBECOM '19})}.
\newblock


\bibitem[\protect\citeauthoryear{Luo, Pan, Tholoniat, Cidon, Geambasu, and
  L{\'e}cuyer}{Luo et~al\mbox{.}}{2021}]%
        {DPF}
\bibfield{author}{\bibinfo{person}{Tao Luo}, \bibinfo{person}{Mingen Pan},
  \bibinfo{person}{Pierre Tholoniat}, \bibinfo{person}{Asaf Cidon},
  \bibinfo{person}{Roxana Geambasu}, {and} \bibinfo{person}{Mathias
  L{\'e}cuyer}.} \bibinfo{year}{2021}\natexlab{}.
\newblock \showarticletitle{Privacy Budget Scheduling}. In
  \bibinfo{booktitle}{\emph{15th {USENIX} Symposium on Operating Systems Design
  and Implementation}} \emph{(\bibinfo{series}{OSDI '21})}.
\newblock


\bibitem[\protect\citeauthoryear{Mahajan, Balasubramanian, Singhvi,
  Venkataraman, Akella, Phanishayee, and Chawla}{Mahajan et~al\mbox{.}}{2020}]%
        {Themis}
\bibfield{author}{\bibinfo{person}{Kshiteej Mahajan}, \bibinfo{person}{Arjun
  Balasubramanian}, \bibinfo{person}{Arjun Singhvi}, \bibinfo{person}{Shivaram
  Venkataraman}, \bibinfo{person}{Aditya Akella}, \bibinfo{person}{Amar
  Phanishayee}, {and} \bibinfo{person}{Shuchi Chawla}.}
  \bibinfo{year}{2020}\natexlab{}.
\newblock \showarticletitle{Themis: Fair and Efficient {GPU} Cluster
  Scheduling}. In \bibinfo{booktitle}{\emph{17th {USENIX} Symposium on
  Networked Systems Design and Implementation}} \emph{(\bibinfo{series}{NSDI
  '20})}.
\newblock


\bibitem[\protect\citeauthoryear{Mayer and Jacobsen}{Mayer and
  Jacobsen}{2021}]%
        {SurvScalableDL}
\bibfield{author}{\bibinfo{person}{Ruben Mayer} {and}
  \bibinfo{person}{Hans-Arno Jacobsen}.} \bibinfo{year}{2021}\natexlab{}.
\newblock \showarticletitle{Scalable Deep Learning on Distributed
  Infrastructures: Challenges, Techniques, and Tools}.
\newblock \bibinfo{journal}{\emph{Comput. Surveys}} (\bibinfo{year}{2021}).
\newblock


\bibitem[\protect\citeauthoryear{Mei, Wang, Chu, Liu, Leung, and Li}{Mei
  et~al\mbox{.}}{2021}]%
        {mei2021energy}
\bibfield{author}{\bibinfo{person}{Xinxin Mei}, \bibinfo{person}{Qiang Wang},
  \bibinfo{person}{Xiaowen Chu}, \bibinfo{person}{Hai Liu},
  \bibinfo{person}{Yiu-Wing Leung}, {and} \bibinfo{person}{Zongpeng Li}.}
  \bibinfo{year}{2021}\natexlab{}.
\newblock \showarticletitle{Energy-aware Task Scheduling with Deadline
  Constraint in DVFS-enabled Heterogeneous Clusters}.
\newblock \bibinfo{journal}{\emph{CoRR}} (\bibinfo{year}{2021}).
\newblock


\bibitem[\protect\citeauthoryear{Mendoza, Romero, Li, Yadwadkar, and
  Kozyrakis}{Mendoza et~al\mbox{.}}{2021}]%
        {Interference-Aware21}
\bibfield{author}{\bibinfo{person}{Daniel Mendoza}, \bibinfo{person}{Francisco
  Romero}, \bibinfo{person}{Qian Li}, \bibinfo{person}{Neeraja~J. Yadwadkar},
  {and} \bibinfo{person}{Christos Kozyrakis}.} \bibinfo{year}{2021}\natexlab{}.
\newblock \showarticletitle{Interference-Aware Scheduling for Inference
  Serving}. In \bibinfo{booktitle}{\emph{Proceedings of the 1st Workshop on
  Machine Learning and Systems}} \emph{(\bibinfo{series}{EuroMLSys '21})}.
\newblock


\bibitem[\protect\citeauthoryear{Misra, Liaw, Dunlap, Bhardwaj, Kandasamy,
  Gonzalez, Stoica, and Tumanov}{Misra et~al\mbox{.}}{2021}]%
        {RubberBand}
\bibfield{author}{\bibinfo{person}{Ujval Misra}, \bibinfo{person}{Richard
  Liaw}, \bibinfo{person}{Lisa Dunlap}, \bibinfo{person}{Romil Bhardwaj},
  \bibinfo{person}{Kirthevasan Kandasamy}, \bibinfo{person}{Joseph~E.
  Gonzalez}, \bibinfo{person}{Ion Stoica}, {and} \bibinfo{person}{Alexey
  Tumanov}.} \bibinfo{year}{2021}\natexlab{}.
\newblock \showarticletitle{RubberBand: Cloud-Based Hyperparameter Tuning}. In
  \bibinfo{booktitle}{\emph{Proceedings of the Sixteenth European Conference on
  Computer Systems}} \emph{(\bibinfo{series}{EuroSys '21})}.
\newblock


\bibitem[\protect\citeauthoryear{Mohan, Phanishayee, Kulkarni, and
  Chidambaram}{Mohan et~al\mbox{.}}{2022}]%
        {Synergy}
\bibfield{author}{\bibinfo{person}{Jayashree Mohan}, \bibinfo{person}{Amar
  Phanishayee}, \bibinfo{person}{Janardhan Kulkarni}, {and}
  \bibinfo{person}{Vijay Chidambaram}.} \bibinfo{year}{2022}\natexlab{}.
\newblock \showarticletitle{Synergy: Looking Beyond GPUs for DNN Scheduling on
  Multi-Tenant Clusters}. In \bibinfo{booktitle}{\emph{16th {USENIX} Symposium
  on Operating Systems Design and Implementation}} \emph{(\bibinfo{series}{OSDI
  '22})}.
\newblock


\bibitem[\protect\citeauthoryear{Mu'alem and Feitelson}{Mu'alem and
  Feitelson}{2001}]%
        {mu2001backfilling}
\bibfield{author}{\bibinfo{person}{A.W. Mu'alem} {and} \bibinfo{person}{D.G.
  Feitelson}.} \bibinfo{year}{2001}\natexlab{}.
\newblock \showarticletitle{Utilization, predictability, workloads, and user
  runtime estimates in scheduling the IBM SP2 with backfilling}.
\newblock \bibinfo{journal}{\emph{IEEE Transactions on Parallel and Distributed
  Systems}} (\bibinfo{year}{2001}).
\newblock
\urldef\tempurl%
\url{https://doi.org/10.1109/71.932708}
\showDOI{\tempurl}


\bibitem[\protect\citeauthoryear{Narayanan, Kazhamiaka, Abuzaid, Kraft,
  Agrawal, Kandula, Boyd, and Zaharia}{Narayanan et~al\mbox{.}}{2021}]%
        {POP}
\bibfield{author}{\bibinfo{person}{Deepak Narayanan}, \bibinfo{person}{Fiodar
  Kazhamiaka}, \bibinfo{person}{Firas Abuzaid}, \bibinfo{person}{Peter Kraft},
  \bibinfo{person}{Akshay Agrawal}, \bibinfo{person}{Srikanth Kandula},
  \bibinfo{person}{Stephen Boyd}, {and} \bibinfo{person}{Matei Zaharia}.}
  \bibinfo{year}{2021}\natexlab{}.
\newblock \showarticletitle{Solving Large-Scale Granular Resource Allocation
  Problems Efficiently with POP}. In \bibinfo{booktitle}{\emph{Proceedings of
  the ACM SIGOPS 28th Symposium on Operating Systems Principles}}
  \emph{(\bibinfo{series}{SOSP '21})}.
\newblock


\bibitem[\protect\citeauthoryear{Narayanan, Santhanam, Kazhamiaka, Phanishayee,
  and Zaharia}{Narayanan et~al\mbox{.}}{2020a}]%
        {MLCloudPrice}
\bibfield{author}{\bibinfo{person}{Deepak Narayanan}, \bibinfo{person}{Keshav
  Santhanam}, \bibinfo{person}{Fiodar Kazhamiaka}, \bibinfo{person}{Amar
  Phanishayee}, {and} \bibinfo{person}{Matei Zaharia}.}
  \bibinfo{year}{2020}\natexlab{a}.
\newblock \showarticletitle{Analysis and exploitation of dynamic pricing in the
  public cloud for ml training}. In \bibinfo{booktitle}{\emph{VLDB DISPA
  Workshop 2020}}.
\newblock


\bibitem[\protect\citeauthoryear{Narayanan, Santhanam, Kazhamiaka, Phanishayee,
  and Zaharia}{Narayanan et~al\mbox{.}}{2020b}]%
        {Gavel}
\bibfield{author}{\bibinfo{person}{Deepak Narayanan}, \bibinfo{person}{Keshav
  Santhanam}, \bibinfo{person}{Fiodar Kazhamiaka}, \bibinfo{person}{Amar
  Phanishayee}, {and} \bibinfo{person}{Matei Zaharia}.}
  \bibinfo{year}{2020}\natexlab{b}.
\newblock \showarticletitle{{Heterogeneity-Aware Cluster Scheduling Policies
  for Deep Learning Workloads}}. In \bibinfo{booktitle}{\emph{14th {USENIX}
  Symposium on Operating Systems Design and Implementation}}
  \emph{(\bibinfo{series}{OSDI '20})}.
\newblock


\bibitem[\protect\citeauthoryear{Narayanan, Santhanam, Phanishayee, and
  Zaharia}{Narayanan et~al\mbox{.}}{2018a}]%
        {HiveMind}
\bibfield{author}{\bibinfo{person}{Deepak Narayanan}, \bibinfo{person}{Keshav
  Santhanam}, \bibinfo{person}{Amar Phanishayee}, {and} \bibinfo{person}{Matei
  Zaharia}.} \bibinfo{year}{2018}\natexlab{a}.
\newblock \showarticletitle{Accelerating deep learning workloads through
  efficient multi-model execution}. In \bibinfo{booktitle}{\emph{NeurIPS
  Workshop on Systems for Machine Learning}}.
\newblock


\bibitem[\protect\citeauthoryear{Narayanan, Santhanam, Phanishayee, and
  Zaharia}{Narayanan et~al\mbox{.}}{2018b}]%
        {narayanan2018accelerating}
\bibfield{author}{\bibinfo{person}{Deepak Narayanan}, \bibinfo{person}{Keshav
  Santhanam}, \bibinfo{person}{Amar Phanishayee}, {and} \bibinfo{person}{Matei
  Zaharia}.} \bibinfo{year}{2018}\natexlab{b}.
\newblock \showarticletitle{Accelerating deep learning workloads through
  efficient multi-model execution}. In \bibinfo{booktitle}{\emph{NeurIPS
  Workshop on Systems for Machine Learning}}.
\newblock


\bibitem[\protect\citeauthoryear{Netto, Calheiros, Rodrigues, Cunha, and
  Buyya}{Netto et~al\mbox{.}}{2018}]%
        {SurvHPC}
\bibfield{author}{\bibinfo{person}{Marco A.~S. Netto},
  \bibinfo{person}{Rodrigo~N. Calheiros}, \bibinfo{person}{Eduardo~R.
  Rodrigues}, \bibinfo{person}{Renato L.~F. Cunha}, {and}
  \bibinfo{person}{Rajkumar Buyya}.} \bibinfo{year}{2018}\natexlab{}.
\newblock \showarticletitle{HPC Cloud for Scientific and Business Applications:
  Taxonomy, Vision, and Research Challenges}.
\newblock \bibinfo{journal}{\emph{Comput. Surveys}} (\bibinfo{year}{2018}).
\newblock


\bibitem[\protect\citeauthoryear{Ogden, Kong, and Guo}{Ogden
  et~al\mbox{.}}{2021}]%
        {Ogden2021PieSlicer}
\bibfield{author}{\bibinfo{person}{Samuel~S. Ogden}, \bibinfo{person}{Xiangnan
  Kong}, {and} \bibinfo{person}{Tian Guo}.} \bibinfo{year}{2021}\natexlab{}.
\newblock \showarticletitle{PieSlicer: Dynamically Improving Response Time for
  Cloud-based CNN Inference}. In \bibinfo{booktitle}{\emph{Proceedings of the
  ACM/SPEC International Conference on Performance Engineering}}
  \emph{(\bibinfo{series}{ICPE '21})}.
\newblock


\bibitem[\protect\citeauthoryear{Olston, Fiedel, Gorovoy, Harmsen, Lao, Li,
  Rajashekhar, Ramesh, and Soyke}{Olston et~al\mbox{.}}{2017}]%
        {TensorFlow-Serving}
\bibfield{author}{\bibinfo{person}{Christopher Olston}, \bibinfo{person}{Noah
  Fiedel}, \bibinfo{person}{Kiril Gorovoy}, \bibinfo{person}{Jeremiah Harmsen},
  \bibinfo{person}{Li Lao}, \bibinfo{person}{Fangwei Li}, \bibinfo{person}{Vinu
  Rajashekhar}, \bibinfo{person}{Sukriti Ramesh}, {and} \bibinfo{person}{Jordan
  Soyke}.} \bibinfo{year}{2017}\natexlab{}.
\newblock \showarticletitle{TensorFlow-Serving: Flexible, High-Performance ML
  Serving}.
\newblock \bibinfo{journal}{\emph{CoRR}}  \bibinfo{volume}{abs/1712.06139}
  (\bibinfo{year}{2017}).
\newblock


\bibitem[\protect\citeauthoryear{Ouyang, Dong, Xu, and Xiao}{Ouyang
  et~al\mbox{.}}{2021}]%
        {SurveyDistDL}
\bibfield{author}{\bibinfo{person}{Shuo Ouyang}, \bibinfo{person}{Dezun Dong},
  \bibinfo{person}{Yemao Xu}, {and} \bibinfo{person}{Liquan Xiao}.}
  \bibinfo{year}{2021}\natexlab{}.
\newblock \showarticletitle{Communication optimization strategies for
  distributed deep neural network training: A survey}.
\newblock \bibinfo{journal}{\emph{J. Parallel and Distrib. Comput.}}
  (\bibinfo{year}{2021}).
\newblock


\bibitem[\protect\citeauthoryear{Park, Naumov, Basu, Deng, Kalaiah, Khudia,
  Law, Malani, Malevich, Nadathur, Pino, Schatz, Sidorov, Sivakumar, Tulloch,
  Wang, Wu, Yuen, Diril, Dzhulgakov, Hazelwood, Jia, Jia, Qiao, Rao, Rotem,
  Yoo, and Smelyanskiy}{Park et~al\mbox{.}}{2018}]%
        {facebook}
\bibfield{author}{\bibinfo{person}{Jongsoo Park}, \bibinfo{person}{Maxim
  Naumov}, \bibinfo{person}{Protonu Basu}, \bibinfo{person}{Summer Deng},
  \bibinfo{person}{Aravind Kalaiah}, \bibinfo{person}{Daya Khudia},
  \bibinfo{person}{James Law}, \bibinfo{person}{Parth Malani},
  \bibinfo{person}{Andrey Malevich}, \bibinfo{person}{Satish Nadathur},
  \bibinfo{person}{Juan Pino}, \bibinfo{person}{Martin Schatz},
  \bibinfo{person}{Alexander Sidorov}, \bibinfo{person}{Viswanath Sivakumar},
  \bibinfo{person}{Andrew Tulloch}, \bibinfo{person}{Xiaodong Wang},
  \bibinfo{person}{Yiming Wu}, \bibinfo{person}{Hector Yuen},
  \bibinfo{person}{Utku Diril}, \bibinfo{person}{Dmytro Dzhulgakov},
  \bibinfo{person}{Kim Hazelwood}, \bibinfo{person}{Bill Jia},
  \bibinfo{person}{Yangqing Jia}, \bibinfo{person}{Lin Qiao},
  \bibinfo{person}{Vijay Rao}, \bibinfo{person}{Nadav Rotem},
  \bibinfo{person}{Sungjoo Yoo}, {and} \bibinfo{person}{Mikhail Smelyanskiy}.}
  \bibinfo{year}{2018}\natexlab{}.
\newblock \showarticletitle{Deep Learning Inference in Facebook Data Centers:
  Characterization, Performance Optimizations and Hardware Implications}.
\newblock \bibinfo{journal}{\emph{CoRR}} (\bibinfo{year}{2018}).
\newblock


\bibitem[\protect\citeauthoryear{Paszke, Gross, Massa, Lerer, Bradbury, Chanan,
  Killeen, Lin, Gimelshein, Antiga, Desmaison, Kopf, Yang, DeVito, Raison,
  Tejani, Chilamkurthy, Steiner, Fang, Bai, and Chintala}{Paszke
  et~al\mbox{.}}{2019}]%
        {PyTorch}
\bibfield{author}{\bibinfo{person}{Adam Paszke}, \bibinfo{person}{Sam Gross},
  \bibinfo{person}{Francisco Massa}, \bibinfo{person}{Adam Lerer},
  \bibinfo{person}{James Bradbury}, \bibinfo{person}{Gregory Chanan},
  \bibinfo{person}{Trevor Killeen}, \bibinfo{person}{Zeming Lin},
  \bibinfo{person}{Natalia Gimelshein}, \bibinfo{person}{Luca Antiga},
  \bibinfo{person}{Alban Desmaison}, \bibinfo{person}{Andreas Kopf},
  \bibinfo{person}{Edward Yang}, \bibinfo{person}{Zachary DeVito},
  \bibinfo{person}{Martin Raison}, \bibinfo{person}{Alykhan Tejani},
  \bibinfo{person}{Sasank Chilamkurthy}, \bibinfo{person}{Benoit Steiner},
  \bibinfo{person}{Lu Fang}, \bibinfo{person}{Junjie Bai}, {and}
  \bibinfo{person}{Soumith Chintala}.} \bibinfo{year}{2019}\natexlab{}.
\newblock \showarticletitle{PyTorch: An Imperative Style, High-Performance Deep
  Learning Library}. In \bibinfo{booktitle}{\emph{Advances in Neural
  Information Processing Systems}}.
\newblock


\bibitem[\protect\citeauthoryear{Peng, Bao, Chen, Wu, and Guo}{Peng
  et~al\mbox{.}}{2018}]%
        {Optimus}
\bibfield{author}{\bibinfo{person}{Yanghua Peng}, \bibinfo{person}{Yixin Bao},
  \bibinfo{person}{Yangrui Chen}, \bibinfo{person}{Chuan Wu}, {and}
  \bibinfo{person}{Chuanxiong Guo}.} \bibinfo{year}{2018}\natexlab{}.
\newblock \showarticletitle{Optimus: An Efficient Dynamic Resource Scheduler
  for Deep Learning Clusters}. In \bibinfo{booktitle}{\emph{Proceedings of the
  Thirteenth EuroSys Conference}} \emph{(\bibinfo{series}{EuroSys '18})}.
\newblock


\bibitem[\protect\citeauthoryear{Peng, Bao, Chen, Wu, Meng, and Lin}{Peng
  et~al\mbox{.}}{2021}]%
        {DL2}
\bibfield{author}{\bibinfo{person}{Yanghua Peng}, \bibinfo{person}{Yixin Bao},
  \bibinfo{person}{Yangrui Chen}, \bibinfo{person}{Chuan Wu},
  \bibinfo{person}{Chen Meng}, {and} \bibinfo{person}{Wei Lin}.}
  \bibinfo{year}{2021}\natexlab{}.
\newblock \showarticletitle{DL2: A Deep Learning-Driven Scheduler for Deep
  Learning Clusters}.
\newblock \bibinfo{journal}{\emph{IEEE Transactions on Parallel and Distributed
  Systems}} (\bibinfo{year}{2021}).
\newblock


\bibitem[\protect\citeauthoryear{Qiao, Choe, Subramanya, Neiswanger, Ho, Zhang,
  Ganger, and Xing}{Qiao et~al\mbox{.}}{2021}]%
        {Pollux}
\bibfield{author}{\bibinfo{person}{Aurick Qiao}, \bibinfo{person}{Sang~Keun
  Choe}, \bibinfo{person}{Suhas~Jayaram Subramanya}, \bibinfo{person}{Willie
  Neiswanger}, \bibinfo{person}{Qirong Ho}, \bibinfo{person}{Hao Zhang},
  \bibinfo{person}{Gregory~R. Ganger}, {and} \bibinfo{person}{Eric~P. Xing}.}
  \bibinfo{year}{2021}\natexlab{}.
\newblock \showarticletitle{Pollux: Co-adaptive Cluster Scheduling for
  Goodput-Optimized Deep Learning}. In \bibinfo{booktitle}{\emph{15th {USENIX}
  Symposium on Operating Systems Design and Implementation}}
  \emph{(\bibinfo{series}{OSDI '21})}.
\newblock


\bibitem[\protect\citeauthoryear{Qin, Zawad, Zhou, Yang, Zhao, and Yan}{Qin
  et~al\mbox{.}}{2019}]%
        {qin2019swift}
\bibfield{author}{\bibinfo{person}{Heyang Qin}, \bibinfo{person}{Syed Zawad},
  \bibinfo{person}{Yanqi Zhou}, \bibinfo{person}{Lei Yang},
  \bibinfo{person}{Dongfang Zhao}, {and} \bibinfo{person}{Feng Yan}.}
  \bibinfo{year}{2019}\natexlab{}.
\newblock \showarticletitle{Swift machine learning model serving scheduling: a
  region based reinforcement learning approach}. In
  \bibinfo{booktitle}{\emph{Proceedings of the International Conference for
  High Performance Computing, Networking, Storage and Analysis}}
  \emph{(\bibinfo{series}{SC '19})}.
\newblock


\bibitem[\protect\citeauthoryear{Rasley, He, Yan, Ruwase, and Fonseca}{Rasley
  et~al\mbox{.}}{2017}]%
        {HyperDrive}
\bibfield{author}{\bibinfo{person}{Jeff Rasley}, \bibinfo{person}{Yuxiong He},
  \bibinfo{person}{Feng Yan}, \bibinfo{person}{Olatunji Ruwase}, {and}
  \bibinfo{person}{Rodrigo Fonseca}.} \bibinfo{year}{2017}\natexlab{}.
\newblock \showarticletitle{HyperDrive: Exploring Hyperparameters with POP
  Scheduling}. In \bibinfo{booktitle}{\emph{Proceedings of the 18th
  International Middleware Conference}} \emph{(\bibinfo{series}{Middleware
  '17})}.
\newblock


\bibitem[\protect\citeauthoryear{Reuther, Byun, Arcand, Bestor, Bergeron,
  Hubbell, Jones, Michaleas, Prout, Rosa, and Kepner}{Reuther
  et~al\mbox{.}}{2018}]%
        {SurveyHPCBigData}
\bibfield{author}{\bibinfo{person}{Albert Reuther}, \bibinfo{person}{Chansup
  Byun}, \bibinfo{person}{William Arcand}, \bibinfo{person}{David Bestor},
  \bibinfo{person}{Bill Bergeron}, \bibinfo{person}{Matthew Hubbell},
  \bibinfo{person}{Michael Jones}, \bibinfo{person}{Peter Michaleas},
  \bibinfo{person}{Andrew Prout}, \bibinfo{person}{Antonio Rosa}, {and}
  \bibinfo{person}{Jeremy Kepner}.} \bibinfo{year}{2018}\natexlab{}.
\newblock \showarticletitle{Scalable system scheduling for HPC and big data}.
\newblock \bibinfo{journal}{\emph{J. Parallel and Distrib. Comput.}}
  (\bibinfo{year}{2018}).
\newblock


\bibitem[\protect\citeauthoryear{Romero, Li, Yadwadkar, and Kozyrakis}{Romero
  et~al\mbox{.}}{2021}]%
        {INFaaS}
\bibfield{author}{\bibinfo{person}{Francisco Romero}, \bibinfo{person}{Qian
  Li}, \bibinfo{person}{Neeraja~J. Yadwadkar}, {and} \bibinfo{person}{Christos
  Kozyrakis}.} \bibinfo{year}{2021}\natexlab{}.
\newblock \showarticletitle{INFaaS: Automated Model-less Inference Serving}. In
  \bibinfo{booktitle}{\emph{2021 {USENIX} Annual Technical Conference}}
  \emph{(\bibinfo{series}{{USENIX} {ATC} '21})}.
\newblock


\bibitem[\protect\citeauthoryear{Saxena, Jayaram, Basu, Sabharwal, and
  Verma}{Saxena et~al\mbox{.}}{2020}]%
        {Effective20}
\bibfield{author}{\bibinfo{person}{Vaibhav Saxena}, \bibinfo{person}{K.~R.
  Jayaram}, \bibinfo{person}{Saurav Basu}, \bibinfo{person}{Yogish Sabharwal},
  {and} \bibinfo{person}{Ashish Verma}.} \bibinfo{year}{2020}\natexlab{}.
\newblock \showarticletitle{Effective Elastic Scaling of Deep Learning
  Workloads}. In \bibinfo{booktitle}{\emph{28th International Symposium on
  Modeling, Analysis, and Simulation of Computer and Telecommunication
  Systems}} \emph{(\bibinfo{series}{MASCOTS '20})}.
\newblock


\bibitem[\protect\citeauthoryear{Seo, Cha, Kim, Huh, and Park}{Seo
  et~al\mbox{.}}{2021}]%
        {seo2021SLO_Aware}
\bibfield{author}{\bibinfo{person}{Wonik Seo}, \bibinfo{person}{Sanghoon Cha},
  \bibinfo{person}{Yeonjae Kim}, \bibinfo{person}{Jaehyuk Huh}, {and}
  \bibinfo{person}{Jongse Park}.} \bibinfo{year}{2021}\natexlab{}.
\newblock \showarticletitle{SLO-Aware Inference Scheduler for Heterogeneous
  Processors in Edge Platforms}.
\newblock \bibinfo{journal}{\emph{ACM Trans. Archit. Code Optim.}}
  (\bibinfo{year}{2021}).
\newblock


\bibitem[\protect\citeauthoryear{Sergeev and Balso}{Sergeev and Balso}{2018}]%
        {Horovod}
\bibfield{author}{\bibinfo{person}{Alexander Sergeev} {and}
  \bibinfo{person}{Mike~Del Balso}.} \bibinfo{year}{2018}\natexlab{}.
\newblock \showarticletitle{Horovod: fast and easy distributed deep learning in
  TensorFlow}.
\newblock \bibinfo{journal}{\emph{CoRR}} (\bibinfo{year}{2018}).
\newblock


\bibitem[\protect\citeauthoryear{Shen, Wu, and Suk}{Shen et~al\mbox{.}}{2017}]%
        {shen2017deep}
\bibfield{author}{\bibinfo{person}{Dinggang Shen}, \bibinfo{person}{Guorong
  Wu}, {and} \bibinfo{person}{Heung-Il Suk}.} \bibinfo{year}{2017}\natexlab{}.
\newblock \showarticletitle{Deep learning in medical image analysis}.
\newblock \bibinfo{journal}{\emph{Annual review of biomedical engineering}}
  \bibinfo{volume}{19} (\bibinfo{year}{2017}), \bibinfo{pages}{221--248}.
\newblock


\bibitem[\protect\citeauthoryear{Shen, Chen, Jin, Zhao, Kong, Philipose,
  Krishnamurthy, and Sundaram}{Shen et~al\mbox{.}}{2019}]%
        {shen2019nexus}
\bibfield{author}{\bibinfo{person}{Haichen Shen}, \bibinfo{person}{Lequn Chen},
  \bibinfo{person}{Yuchen Jin}, \bibinfo{person}{Liangyu Zhao},
  \bibinfo{person}{Bingyu Kong}, \bibinfo{person}{Matthai Philipose},
  \bibinfo{person}{Arvind Krishnamurthy}, {and} \bibinfo{person}{Ravi
  Sundaram}.} \bibinfo{year}{2019}\natexlab{}.
\newblock \showarticletitle{Nexus: A GPU Cluster Engine for Accelerating
  DNN-Based Video Analysis}. In \bibinfo{booktitle}{\emph{Proceedings of the
  27th ACM Symposium on Operating Systems Principles}}
  \emph{(\bibinfo{series}{SOSP '19})}.
\newblock


\bibitem[\protect\citeauthoryear{Shi, Chen, Sun, and Li}{Shi
  et~al\mbox{.}}{2012}]%
        {shi2012vCUDA}
\bibfield{author}{\bibinfo{person}{Lin Shi}, \bibinfo{person}{Hao Chen},
  \bibinfo{person}{Jianhua Sun}, {and} \bibinfo{person}{Kenli Li}.}
  \bibinfo{year}{2012}\natexlab{}.
\newblock \showarticletitle{vCUDA: GPU-Accelerated High-Performance Computing
  in Virtual Machines}.
\newblock \bibinfo{journal}{\emph{IEEE Trans. Comput.}} (\bibinfo{year}{2012}).
\newblock


\bibitem[\protect\citeauthoryear{Shishira, Kandasamy, and
  Chandrasekaran}{Shishira et~al\mbox{.}}{2017}]%
        {SurveyCloudSched17}
\bibfield{author}{\bibinfo{person}{S~R Shishira}, \bibinfo{person}{A.
  Kandasamy}, {and} \bibinfo{person}{K. Chandrasekaran}.}
  \bibinfo{year}{2017}\natexlab{}.
\newblock \showarticletitle{Workload scheduling in cloud: A comprehensive
  survey and future research directions}. In
  \bibinfo{booktitle}{\emph{International Conference on Cloud Computing, Data
  Science Engineering - Confluence}}.
\newblock


\bibitem[\protect\citeauthoryear{Shukla, Sivathanu, Viswanatha, Gulavani,
  Nehme, Agrawal, Chen, Kwatra, Ramjee, Sharma, et~al\mbox{.}}{Shukla
  et~al\mbox{.}}{2022}]%
        {Singularity}
\bibfield{author}{\bibinfo{person}{Dharma Shukla}, \bibinfo{person}{Muthian
  Sivathanu}, \bibinfo{person}{Srinidhi Viswanatha}, \bibinfo{person}{Bhargav
  Gulavani}, \bibinfo{person}{Rimma Nehme}, \bibinfo{person}{Amey Agrawal},
  \bibinfo{person}{Chen Chen}, \bibinfo{person}{Nipun Kwatra},
  \bibinfo{person}{Ramachandran Ramjee}, \bibinfo{person}{Pankaj Sharma},
  {et~al\mbox{.}}} \bibinfo{year}{2022}\natexlab{}.
\newblock \showarticletitle{Singularity: Planet-Scale, Preemptible, Elastic
  Scheduling of AI Workloads}.
\newblock \bibinfo{journal}{\emph{CoRR}} (\bibinfo{year}{2022}).
\newblock


\bibitem[\protect\citeauthoryear{Silver, Schrittwieser, Simonyan, Antonoglou,
  Huang, Guez, Hubert, Baker, Lai, Bolton, et~al\mbox{.}}{Silver
  et~al\mbox{.}}{2017}]%
        {silver2017mastering}
\bibfield{author}{\bibinfo{person}{David Silver}, \bibinfo{person}{Julian
  Schrittwieser}, \bibinfo{person}{Karen Simonyan}, \bibinfo{person}{Ioannis
  Antonoglou}, \bibinfo{person}{Aja Huang}, \bibinfo{person}{Arthur Guez},
  \bibinfo{person}{Thomas Hubert}, \bibinfo{person}{Lucas Baker},
  \bibinfo{person}{Matthew Lai}, \bibinfo{person}{Adrian Bolton},
  {et~al\mbox{.}}} \bibinfo{year}{2017}\natexlab{}.
\newblock \showarticletitle{Mastering the game of go without human knowledge}.
\newblock \bibinfo{journal}{\emph{nature}} \bibinfo{volume}{550},
  \bibinfo{number}{7676} (\bibinfo{year}{2017}), \bibinfo{pages}{354--359}.
\newblock


\bibitem[\protect\citeauthoryear{Son, Yoo, Kim, Kim, Lee, and Park}{Son
  et~al\mbox{.}}{2021}]%
        {Hermes}
\bibfield{author}{\bibinfo{person}{Jaewon Son}, \bibinfo{person}{Yonghyuk Yoo},
  \bibinfo{person}{Khu-rai Kim}, \bibinfo{person}{Youngjae Kim},
  \bibinfo{person}{Kwonyong Lee}, {and} \bibinfo{person}{Sungyong Park}.}
  \bibinfo{year}{2021}\natexlab{}.
\newblock \showarticletitle{A GPU Scheduling Framework to Accelerate
  Hyper-Parameter Optimization in Deep Learning Clusters}.
\newblock \bibinfo{journal}{\emph{Electronics}} (\bibinfo{year}{2021}).
\newblock


\bibitem[\protect\citeauthoryear{Sultana, Chen, Xu, and Yuan}{Sultana
  et~al\mbox{.}}{2020}]%
        {E-LAS}
\bibfield{author}{\bibinfo{person}{Abeda Sultana}, \bibinfo{person}{Li Chen},
  \bibinfo{person}{Fei Xu}, {and} \bibinfo{person}{Xu Yuan}.}
  \bibinfo{year}{2020}\natexlab{}.
\newblock \showarticletitle{E-LAS: Design and Analysis of Completion-Time
  Agnostic Scheduling for Distributed Deep Learning Cluster}. In
  \bibinfo{booktitle}{\emph{49th International Conference on Parallel
  Processing}} \emph{(\bibinfo{series}{ICPP '20})}.
\newblock


\bibitem[\protect\citeauthoryear{Sun, Wen, Ta, and Yan}{Sun
  et~al\mbox{.}}{2017}]%
        {Dorm}
\bibfield{author}{\bibinfo{person}{Peng Sun}, \bibinfo{person}{Yonggang Wen},
  \bibinfo{person}{Nguyen Binh~Duong Ta}, {and} \bibinfo{person}{Shengen Yan}.}
  \bibinfo{year}{2017}\natexlab{}.
\newblock \showarticletitle{Towards distributed machine learning in shared
  clusters: A dynamically-partitioned approach}. In
  \bibinfo{booktitle}{\emph{2017 IEEE International Conference on Smart
  Computing}} \emph{(\bibinfo{series}{SMARTCOMP '17})}.
\newblock


\bibitem[\protect\citeauthoryear{Ta}{Ta}{2019}]%
        {FC2}
\bibfield{author}{\bibinfo{person}{Nguyen Binh~Duong Ta}.}
  \bibinfo{year}{2019}\natexlab{}.
\newblock \showarticletitle{FC2: cloud-based cluster provisioning for
  distributed machine learning}.
\newblock \bibinfo{journal}{\emph{Cluster Computing}} (\bibinfo{year}{2019}).
\newblock


\bibitem[\protect\citeauthoryear{Tan, Li, Zhang, Cao, Qi, Liu, Zhu, and
  Guo}{Tan et~al\mbox{.}}{2021}]%
        {tan2021MIG-SERVING}
\bibfield{author}{\bibinfo{person}{Cheng Tan}, \bibinfo{person}{Zhichao Li},
  \bibinfo{person}{Jian Zhang}, \bibinfo{person}{Yu Cao},
  \bibinfo{person}{Sikai Qi}, \bibinfo{person}{Zherui Liu},
  \bibinfo{person}{Yibo Zhu}, {and} \bibinfo{person}{Chuanxiong Guo}.}
  \bibinfo{year}{2021}\natexlab{}.
\newblock \showarticletitle{Serving DNN Models with Multi-Instance GPUs: A Case
  of the Reconfigurable Machine Scheduling Problem}.
\newblock \bibinfo{journal}{\emph{CoRR}} (\bibinfo{year}{2021}).
\newblock


\bibitem[\protect\citeauthoryear{Tang, Wang, Liu, Wang, and Han}{Tang
  et~al\mbox{.}}{2019}]%
        {tang2019nanily}
\bibfield{author}{\bibinfo{person}{Xuehai Tang}, \bibinfo{person}{Peng Wang},
  \bibinfo{person}{Qiuyang Liu}, \bibinfo{person}{Wang Wang}, {and}
  \bibinfo{person}{Jizhong Han}.} \bibinfo{year}{2019}\natexlab{}.
\newblock \showarticletitle{Nanily: A QoS-Aware Scheduling for DNN Inference
  Workload in Clouds}. In \bibinfo{booktitle}{\emph{2019 IEEE 21st
  International Conference on High Performance Computing and Communications;
  IEEE 17th International Conference on Smart City; IEEE 5th International
  Conference on Data Science and Systems (HPCC/SmartCity/DSS)}}.
\newblock


\bibitem[\protect\citeauthoryear{Thinakaran, Gunasekaran, Sharma, Kandemir, and
  Das}{Thinakaran et~al\mbox{.}}{2019}]%
        {Thinakaran2019kubeknots}
\bibfield{author}{\bibinfo{person}{Prashanth Thinakaran},
  \bibinfo{person}{Jashwant~Raj Gunasekaran}, \bibinfo{person}{Bikash Sharma},
  \bibinfo{person}{Mahmut~Taylan Kandemir}, {and} \bibinfo{person}{Chita~R.
  Das}.} \bibinfo{year}{2019}\natexlab{}.
\newblock \showarticletitle{Kube-Knots: Resource Harvesting through Dynamic
  Container Orchestration in GPU-based Datacenters}. In
  \bibinfo{booktitle}{\emph{2019 IEEE International Conference on Cluster
  Computing (CLUSTER)}}.
\newblock


\bibitem[\protect\citeauthoryear{Vavilapalli, Murthy, Douglas, Agarwal, Konar,
  Evans, Graves, Lowe, Shah, Seth, Saha, Curino, O'Malley, Radia, Reed, and
  Baldeschwieler}{Vavilapalli et~al\mbox{.}}{2013}]%
        {YARN}
\bibfield{author}{\bibinfo{person}{Vinod~Kumar Vavilapalli},
  \bibinfo{person}{Arun~C. Murthy}, \bibinfo{person}{Chris Douglas},
  \bibinfo{person}{Sharad Agarwal}, \bibinfo{person}{Mahadev Konar},
  \bibinfo{person}{Robert Evans}, \bibinfo{person}{Thomas Graves},
  \bibinfo{person}{Jason Lowe}, \bibinfo{person}{Hitesh Shah},
  \bibinfo{person}{Siddharth Seth}, \bibinfo{person}{Bikas Saha},
  \bibinfo{person}{Carlo Curino}, \bibinfo{person}{Owen O'Malley},
  \bibinfo{person}{Sanjay Radia}, \bibinfo{person}{Benjamin Reed}, {and}
  \bibinfo{person}{Eric Baldeschwieler}.} \bibinfo{year}{2013}\natexlab{}.
\newblock \showarticletitle{Apache Hadoop YARN: Yet Another Resource
  Negotiator}. In \bibinfo{booktitle}{\emph{Proceedings of the 4th Annual
  Symposium on Cloud Computing}} \emph{(\bibinfo{series}{SoCC '13})}.
\newblock


\bibitem[\protect\citeauthoryear{Verbraeken, Wolting, Katzy, Kloppenburg,
  Verbelen, and Rellermeyer}{Verbraeken et~al\mbox{.}}{2020}]%
        {SurveyDML}
\bibfield{author}{\bibinfo{person}{Joost Verbraeken}, \bibinfo{person}{Matthijs
  Wolting}, \bibinfo{person}{Jonathan Katzy}, \bibinfo{person}{Jeroen
  Kloppenburg}, \bibinfo{person}{Tim Verbelen}, {and} \bibinfo{person}{Jan~S.
  Rellermeyer}.} \bibinfo{year}{2020}\natexlab{}.
\newblock \showarticletitle{A Survey on Distributed Machine Learning}.
\newblock \bibinfo{journal}{\emph{Comput. Surveys}} (\bibinfo{year}{2020}).
\newblock


\bibitem[\protect\citeauthoryear{Wang, Zhang, Lai, Hao, and Wang}{Wang
  et~al\mbox{.}}{2021c}]%
        {GPOEO}
\bibfield{author}{\bibinfo{person}{Farui Wang}, \bibinfo{person}{Weizhe Zhang},
  \bibinfo{person}{Shichao Lai}, \bibinfo{person}{Meng Hao}, {and}
  \bibinfo{person}{Zheng Wang}.} \bibinfo{year}{2021}\natexlab{c}.
\newblock \showarticletitle{Dynamic GPU Energy Optimization for Machine
  Learning Training Workloads}.
\newblock \bibinfo{journal}{\emph{IEEE Transactions on Parallel and Distributed
  Systems}} (\bibinfo{year}{2021}).
\newblock


\bibitem[\protect\citeauthoryear{Wang, Liu, and Shen}{Wang
  et~al\mbox{.}}{2020b}]%
        {MLFS}
\bibfield{author}{\bibinfo{person}{Haoyu Wang}, \bibinfo{person}{Zetian Liu},
  {and} \bibinfo{person}{Haiying Shen}.} \bibinfo{year}{2020}\natexlab{b}.
\newblock \showarticletitle{Job scheduling for large-scale machine learning
  clusters}. In \bibinfo{booktitle}{\emph{Proceedings of the 16th International
  Conference on emerging Networking EXperiments and Technologies}}
  \emph{(\bibinfo{series}{CoNEXT '20})}.
\newblock


\bibitem[\protect\citeauthoryear{Wang, Yang, Yu, Wang, Li, Sun, He, and
  Zhang}{Wang et~al\mbox{.}}{2021a}]%
        {Morphling}
\bibfield{author}{\bibinfo{person}{Luping Wang}, \bibinfo{person}{Lingyun
  Yang}, \bibinfo{person}{Yinghao Yu}, \bibinfo{person}{Wei Wang},
  \bibinfo{person}{Bo Li}, \bibinfo{person}{Xianchao Sun},
  \bibinfo{person}{Jian He}, {and} \bibinfo{person}{Liping Zhang}.}
  \bibinfo{year}{2021}\natexlab{a}.
\newblock \showarticletitle{Morphling: Fast, Near-Optimal Auto-Configuration
  for Cloud-Native Model Serving}. In \bibinfo{booktitle}{\emph{Proceedings of
  the ACM Symposium on Cloud Computing}} \emph{(\bibinfo{series}{SoCC '21})}.
\newblock


\bibitem[\protect\citeauthoryear{Wang, Yang, Yu, Wang, Li, Sun, He, and
  Zhang}{Wang et~al\mbox{.}}{2021b}]%
        {wang2021Morphling}
\bibfield{author}{\bibinfo{person}{Luping Wang}, \bibinfo{person}{Lingyun
  Yang}, \bibinfo{person}{Yinghao Yu}, \bibinfo{person}{Wei Wang},
  \bibinfo{person}{Bo Li}, \bibinfo{person}{Xianchao Sun},
  \bibinfo{person}{Jian He}, {and} \bibinfo{person}{Liping Zhang}.}
  \bibinfo{year}{2021}\natexlab{b}.
\newblock \showarticletitle{Morphling: Fast, Near-Optimal Auto-Configuration
  for Cloud-Native Model Serving}. In \bibinfo{booktitle}{\emph{Proceedings of
  the ACM Symposium on Cloud Computing}} \emph{(\bibinfo{series}{SoCC '21})}.
\newblock


\bibitem[\protect\citeauthoryear{Wang, Meng, Long, Wu, Yang, Lin, and Jia}{Wang
  et~al\mbox{.}}{2019}]%
        {AliPAI}
\bibfield{author}{\bibinfo{person}{Mengdi Wang}, \bibinfo{person}{Chen Meng},
  \bibinfo{person}{Guoping Long}, \bibinfo{person}{Chuan Wu},
  \bibinfo{person}{Jun Yang}, \bibinfo{person}{Wei Lin}, {and}
  \bibinfo{person}{Yangqing Jia}.} \bibinfo{year}{2019}\natexlab{}.
\newblock \showarticletitle{Characterizing Deep Learning Training Workloads on
  Alibaba-PAI}. In \bibinfo{booktitle}{\emph{Proceedings of the 2019 IEEE
  International Symposium on Workload Characterization}}
  \emph{(\bibinfo{series}{IISWC '19})}.
\newblock


\bibitem[\protect\citeauthoryear{Wang and Chu}{Wang and Chu}{2020}]%
        {DVFS2020}
\bibfield{author}{\bibinfo{person}{Qiang Wang} {and} \bibinfo{person}{Xiaowen
  Chu}.} \bibinfo{year}{2020}\natexlab{}.
\newblock \showarticletitle{GPGPU Performance Estimation With Core and Memory
  Frequency Scaling}.
\newblock \bibinfo{journal}{\emph{IEEE Transactions on Parallel and Distributed
  Systems}} (\bibinfo{year}{2020}).
\newblock


\bibitem[\protect\citeauthoryear{Wang, Shi, Wang, and Chu}{Wang
  et~al\mbox{.}}{2020c}]%
        {Ada-SRSF}
\bibfield{author}{\bibinfo{person}{Qiang Wang}, \bibinfo{person}{Shaohuai Shi},
  \bibinfo{person}{Canhui Wang}, {and} \bibinfo{person}{Xiaowen Chu}.}
  \bibinfo{year}{2020}\natexlab{c}.
\newblock \showarticletitle{Communication Contention Aware Scheduling of
  Multiple Deep Learning Training Jobs}.
\newblock \bibinfo{journal}{\emph{CoRR}} (\bibinfo{year}{2020}).
\newblock


\bibitem[\protect\citeauthoryear{Wang, Gonzalez, Zhou, Williams, Friedman,
  Havemann, and Woo}{Wang et~al\mbox{.}}{2020a}]%
        {NonIntrusive}
\bibfield{author}{\bibinfo{person}{Shaoqi Wang}, \bibinfo{person}{Oscar~J
  Gonzalez}, \bibinfo{person}{Xiaobo Zhou}, \bibinfo{person}{Thomas Williams},
  \bibinfo{person}{Brian~D Friedman}, \bibinfo{person}{Martin Havemann}, {and}
  \bibinfo{person}{Thomas Woo}.} \bibinfo{year}{2020}\natexlab{a}.
\newblock \showarticletitle{An Efficient and Non-Intrusive GPU Scheduling
  Framework for Deep Learning Training Systems}. In
  \bibinfo{booktitle}{\emph{International Conference for High Performance
  Computing, Networking, Storage and Analysis}} \emph{(\bibinfo{series}{SC
  '20})}.
\newblock


\bibitem[\protect\citeauthoryear{Wang, Gao, Zhang, Wang, Chen, Ng, Ooi, Shao,
  and Reyad}{Wang et~al\mbox{.}}{2018}]%
        {wang2018rafiki}
\bibfield{author}{\bibinfo{person}{Wei Wang}, \bibinfo{person}{Jinyang Gao},
  \bibinfo{person}{Meihui Zhang}, \bibinfo{person}{Sheng Wang},
  \bibinfo{person}{Gang Chen}, \bibinfo{person}{Teck~Khim Ng},
  \bibinfo{person}{Beng~Chin Ooi}, \bibinfo{person}{Jie Shao}, {and}
  \bibinfo{person}{Moaz Reyad}.} \bibinfo{year}{2018}\natexlab{}.
\newblock \showarticletitle{Rafiki: Machine Learning as an Analytics Service
  System}.
\newblock \bibinfo{journal}{\emph{Proc. VLDB Endow.}} (\bibinfo{year}{2018}).
\newblock


\bibitem[\protect\citeauthoryear{Weng, Xiao, Yu, Wang, Wang, He, Li, Zhang,
  Lin, and Ding}{Weng et~al\mbox{.}}{2022}]%
        {MLaaS}
\bibfield{author}{\bibinfo{person}{Qizhen Weng}, \bibinfo{person}{Wencong
  Xiao}, \bibinfo{person}{Yinghao Yu}, \bibinfo{person}{Wei Wang},
  \bibinfo{person}{Cheng Wang}, \bibinfo{person}{Jian He},
  \bibinfo{person}{Yong Li}, \bibinfo{person}{Liping Zhang},
  \bibinfo{person}{Wei Lin}, {and} \bibinfo{person}{Yu Ding}.}
  \bibinfo{year}{2022}\natexlab{}.
\newblock \showarticletitle{{MLaaS} in the Wild: Workload Analysis and
  Scheduling in {Large-Scale} Heterogeneous {GPU} Clusters}. In
  \bibinfo{booktitle}{\emph{19th USENIX Symposium on Networked Systems Design
  and Implementation}} \emph{(\bibinfo{series}{NSDI '22})}.
\newblock


\bibitem[\protect\citeauthoryear{Wu, Xu, and Wang}{Wu et~al\mbox{.}}{2020}]%
        {wu2020irina}
\bibfield{author}{\bibinfo{person}{Xiaorui Wu}, \bibinfo{person}{Hong Xu},
  {and} \bibinfo{person}{Yi Wang}.} \bibinfo{year}{2020}\natexlab{}.
\newblock \showarticletitle{Irina: Accelerating DNN Inference with Efficient
  Online Scheduling}. In \bibinfo{booktitle}{\emph{4th Asia-Pacific Workshop on
  Networking}} \emph{(\bibinfo{series}{APNet '20})}.
\newblock


\bibitem[\protect\citeauthoryear{Wu, Ma, Yan, Liu, Cai, Huang, Cheng, Yuan, and
  Yu}{Wu et~al\mbox{.}}{2022}]%
        {EDL}
\bibfield{author}{\bibinfo{person}{Yidi Wu}, \bibinfo{person}{Kaihao Ma},
  \bibinfo{person}{Xiao Yan}, \bibinfo{person}{Zhi Liu},
  \bibinfo{person}{Zhenkun Cai}, \bibinfo{person}{Yuzhen Huang},
  \bibinfo{person}{James Cheng}, \bibinfo{person}{Han Yuan}, {and}
  \bibinfo{person}{Fan Yu}.} \bibinfo{year}{2022}\natexlab{}.
\newblock \showarticletitle{Elastic Deep Learning in Multi-Tenant GPU
  Clusters}.
\newblock \bibinfo{journal}{\emph{IEEE Transactions on Parallel and Distributed
  Systems}} (\bibinfo{year}{2022}).
\newblock


\bibitem[\protect\citeauthoryear{Xiao, Bhardwaj, Ramjee, Sivathanu, Kwatra,
  Han, Patel, Peng, Zhao, Zhang, Yang, and Zhou}{Xiao et~al\mbox{.}}{2018}]%
        {Gandiva}
\bibfield{author}{\bibinfo{person}{Wencong Xiao}, \bibinfo{person}{Romil
  Bhardwaj}, \bibinfo{person}{Ramachandran Ramjee}, \bibinfo{person}{Muthian
  Sivathanu}, \bibinfo{person}{Nipun Kwatra}, \bibinfo{person}{Zhenhua Han},
  \bibinfo{person}{Pratyush Patel}, \bibinfo{person}{Xuan Peng},
  \bibinfo{person}{Hanyu Zhao}, \bibinfo{person}{Quanlu Zhang},
  \bibinfo{person}{Fan Yang}, {and} \bibinfo{person}{Lidong Zhou}.}
  \bibinfo{year}{2018}\natexlab{}.
\newblock \showarticletitle{Gandiva: Introspective Cluster Scheduling for Deep
  Learning}. In \bibinfo{booktitle}{\emph{{USENIX} Symposium on Operating
  Systems Design and Implementation}}.
\newblock


\bibitem[\protect\citeauthoryear{Xiao, Ren, Li, Zhang, Hou, Li, Feng, Lin, and
  Jia}{Xiao et~al\mbox{.}}{2020}]%
        {Antman}
\bibfield{author}{\bibinfo{person}{Wencong Xiao}, \bibinfo{person}{Shiru Ren},
  \bibinfo{person}{Yong Li}, \bibinfo{person}{Yang Zhang},
  \bibinfo{person}{Pengyang Hou}, \bibinfo{person}{Zhi Li},
  \bibinfo{person}{Yihui Feng}, \bibinfo{person}{Wei Lin}, {and}
  \bibinfo{person}{Yangqing Jia}.} \bibinfo{year}{2020}\natexlab{}.
\newblock \showarticletitle{{AntMan}: Dynamic Scaling on {GPU} Clusters for
  Deep Learning}. In \bibinfo{booktitle}{\emph{14th USENIX Symposium on
  Operating Systems Design and Implementation}} \emph{(\bibinfo{series}{OSDI
  '20})}.
\newblock


\bibitem[\protect\citeauthoryear{Xie, Zhai, Wu, Wang, Zhang, Sun, and Yan}{Xie
  et~al\mbox{.}}{2020}]%
        {ELAN}
\bibfield{author}{\bibinfo{person}{Lei Xie}, \bibinfo{person}{Jidong Zhai},
  \bibinfo{person}{Baodong Wu}, \bibinfo{person}{Yuanbo Wang},
  \bibinfo{person}{Xingcheng Zhang}, \bibinfo{person}{Peng Sun}, {and}
  \bibinfo{person}{Shengen Yan}.} \bibinfo{year}{2020}\natexlab{}.
\newblock \showarticletitle{Elan: Towards Generic and Efficient Elastic
  Training for Deep Learning}. In \bibinfo{booktitle}{\emph{2020 IEEE 40th
  International Conference on Distributed Computing Systems}}
  \emph{(\bibinfo{series}{ICDCS '20})}.
\newblock


\bibitem[\protect\citeauthoryear{Yadwadkar, Romero, Li, and
  Kozyrakis}{Yadwadkar et~al\mbox{.}}{2019}]%
        {Neeraja2019INFaaSw}
\bibfield{author}{\bibinfo{person}{Neeraja~J. Yadwadkar},
  \bibinfo{person}{Francisco Romero}, \bibinfo{person}{Qian Li}, {and}
  \bibinfo{person}{Christos Kozyrakis}.} \bibinfo{year}{2019}\natexlab{}.
\newblock \showarticletitle{A Case for Managed and Model-less Inference
  Serving}. In \bibinfo{booktitle}{\emph{Proceedings of the Workshop on Hot
  Topics in Operating Systems}} \emph{(\bibinfo{series}{HotOS '19})}.
\newblock


\bibitem[\protect\citeauthoryear{Yan, Ruwase, He, and Smirni}{Yan
  et~al\mbox{.}}{2016}]%
        {yan2016SERF}
\bibfield{author}{\bibinfo{person}{Feng Yan}, \bibinfo{person}{Olatunji
  Ruwase}, \bibinfo{person}{Yuxiong He}, {and} \bibinfo{person}{Evgenia
  Smirni}.} \bibinfo{year}{2016}\natexlab{}.
\newblock \showarticletitle{SERF: Efficient Scheduling for Fast Deep Neural
  Network Serving via Judicious Parallelism}. In \bibinfo{booktitle}{\emph{SC
  '16: Proceedings of the International Conference for High Performance
  Computing, Networking, Storage and Analysis}}.
\newblock


\bibitem[\protect\citeauthoryear{Ye, Sun, Gao, Zhang, Wang, Yan, and Luo}{Ye
  et~al\mbox{.}}{2021}]%
        {ASTRAEA}
\bibfield{author}{\bibinfo{person}{Zhisheng Ye}, \bibinfo{person}{Peng Sun},
  \bibinfo{person}{Wei Gao}, \bibinfo{person}{Tianwei Zhang},
  \bibinfo{person}{Xiaolin Wang}, \bibinfo{person}{Shengen Yan}, {and}
  \bibinfo{person}{Yingwei Luo}.} \bibinfo{year}{2021}\natexlab{}.
\newblock \showarticletitle{ASTRAEA: A Fair Deep Learning Scheduler for
  Multi-tenant GPU Clusters}.
\newblock \bibinfo{journal}{\emph{IEEE Transactions on Parallel and Distributed
  Systems}} (\bibinfo{year}{2021}).
\newblock


\bibitem[\protect\citeauthoryear{Yeh, Chen, and Chou}{Yeh
  et~al\mbox{.}}{2020}]%
        {KubeShare}
\bibfield{author}{\bibinfo{person}{Ting-An Yeh}, \bibinfo{person}{Hung-Hsin
  Chen}, {and} \bibinfo{person}{Jerry Chou}.} \bibinfo{year}{2020}\natexlab{}.
\newblock \showarticletitle{KubeShare: A Framework to Manage GPUs as
  First-Class and Shared Resources in Container Cloud}. In
  \bibinfo{booktitle}{\emph{Proceedings of the 29th International Symposium on
  High-Performance Parallel and Distributed Computing}}
  \emph{(\bibinfo{series}{HPDC '20})}.
\newblock


\bibitem[\protect\citeauthoryear{Yeung, Borowiec, Friday, Harper, and
  Garraghan}{Yeung et~al\mbox{.}}{2020}]%
        {Yeung}
\bibfield{author}{\bibinfo{person}{Gingfung Yeung}, \bibinfo{person}{Damian
  Borowiec}, \bibinfo{person}{Adrian Friday}, \bibinfo{person}{Richard Harper},
  {and} \bibinfo{person}{Peter Garraghan}.} \bibinfo{year}{2020}\natexlab{}.
\newblock \showarticletitle{Towards GPU Utilization Prediction for Cloud Deep
  Learning}. In \bibinfo{booktitle}{\emph{{USENIX} Workshop on Hot Topics in
  Cloud Computing}}.
\newblock


\bibitem[\protect\citeauthoryear{Yeung, Borowiec, Yang, Friday, Harper, and
  Garraghan}{Yeung et~al\mbox{.}}{2022}]%
        {Horus}
\bibfield{author}{\bibinfo{person}{Gingfung Yeung}, \bibinfo{person}{Damian
  Borowiec}, \bibinfo{person}{Renyu Yang}, \bibinfo{person}{Adrian Friday},
  \bibinfo{person}{Richard Harper}, {and} \bibinfo{person}{Peter Garraghan}.}
  \bibinfo{year}{2022}\natexlab{}.
\newblock \showarticletitle{Horus: Interference-Aware and Prediction-Based
  Scheduling in Deep Learning Systems}.
\newblock \bibinfo{journal}{\emph{IEEE Transactions on Parallel and Distributed
  Systems}} (\bibinfo{year}{2022}).
\newblock


\bibitem[\protect\citeauthoryear{Yi, Zhang, Luo, Long, Diao, Wu, Zheng, Yang,
  and Lin}{Yi et~al\mbox{.}}{2020}]%
        {HeteroG}
\bibfield{author}{\bibinfo{person}{Xiaodong Yi}, \bibinfo{person}{Shiwei
  Zhang}, \bibinfo{person}{Ziyue Luo}, \bibinfo{person}{Guoping Long},
  \bibinfo{person}{Lansong Diao}, \bibinfo{person}{Chuan Wu},
  \bibinfo{person}{Zhen Zheng}, \bibinfo{person}{Jun Yang}, {and}
  \bibinfo{person}{Wei Lin}.} \bibinfo{year}{2020}\natexlab{}.
\newblock \showarticletitle{Optimizing distributed training deployment in
  heterogeneous GPU clusters}. In \bibinfo{booktitle}{\emph{Proceedings of the
  16th International Conference on emerging Networking EXperiments and
  Technologies}} \emph{(\bibinfo{series}{CoNext '20})}.
\newblock


\bibitem[\protect\citeauthoryear{Yoo, Jette, and Grondona}{Yoo
  et~al\mbox{.}}{2003}]%
        {SLURM}
\bibfield{author}{\bibinfo{person}{Andy~B. Yoo}, \bibinfo{person}{Morris~A.
  Jette}, {and} \bibinfo{person}{Mark Grondona}.}
  \bibinfo{year}{2003}\natexlab{}.
\newblock \showarticletitle{SLURM: Simple Linux Utility for Resource
  Management}. In \bibinfo{booktitle}{\emph{Job Scheduling Strategies for
  Parallel Processing}}.
\newblock


\bibitem[\protect\citeauthoryear{Yu, Wang, Shangguan, Zhang, Liu, and Chen}{Yu
  et~al\mbox{.}}{2022b}]%
        {yu2022survey}
\bibfield{author}{\bibinfo{person}{Fuxun Yu}, \bibinfo{person}{Di Wang},
  \bibinfo{person}{Longfei Shangguan}, \bibinfo{person}{Minjia Zhang},
  \bibinfo{person}{Chenchen Liu}, {and} \bibinfo{person}{Xiang Chen}.}
  \bibinfo{year}{2022}\natexlab{b}.
\newblock \showarticletitle{A Survey of Multi-Tenant Deep Learning Inference on
  GPU}.
\newblock \bibinfo{journal}{\emph{CoRR}}  \bibinfo{volume}{abs/2203.09040}
  (\bibinfo{year}{2022}).
\newblock


\bibitem[\protect\citeauthoryear{Yu, Wang, Shangguan, Zhang, Tang, Liu, and
  Chen}{Yu et~al\mbox{.}}{2021c}]%
        {SurveyDLServing}
\bibfield{author}{\bibinfo{person}{Fuxun Yu}, \bibinfo{person}{Di Wang},
  \bibinfo{person}{Longfei Shangguan}, \bibinfo{person}{Minjia Zhang},
  \bibinfo{person}{Xulong Tang}, \bibinfo{person}{Chenchen Liu}, {and}
  \bibinfo{person}{Xiang Chen}.} \bibinfo{year}{2021}\natexlab{c}.
\newblock \showarticletitle{A Survey of Large-Scale Deep Learning Serving
  System Optimization: Challenges and Opportunities}.
\newblock \bibinfo{journal}{\emph{CoRR}} (\bibinfo{year}{2021}).
\newblock


\bibitem[\protect\citeauthoryear{Yu, Jiang, Ng, Wang, Chen, and Li}{Yu
  et~al\mbox{.}}{2021a}]%
        {yu2021Gillis}
\bibfield{author}{\bibinfo{person}{Minchen Yu}, \bibinfo{person}{Zhifeng
  Jiang}, \bibinfo{person}{Hok~Chun Ng}, \bibinfo{person}{Wei Wang},
  \bibinfo{person}{Ruichuan Chen}, {and} \bibinfo{person}{Bo Li}.}
  \bibinfo{year}{2021}\natexlab{a}.
\newblock \showarticletitle{Gillis: Serving Large Neural Networks in Serverless
  Functions with Automatic Model Partitioning}. In
  \bibinfo{booktitle}{\emph{2021 IEEE 41st International Conference on
  Distributed Computing Systems (ICDCS)}}.
\newblock


\bibitem[\protect\citeauthoryear{Yu, Tian, Ji, Wu, Rajan, and Liu}{Yu
  et~al\mbox{.}}{2022a}]%
        {GADGET}
\bibfield{author}{\bibinfo{person}{Menglu Yu}, \bibinfo{person}{Ye Tian},
  \bibinfo{person}{Bo Ji}, \bibinfo{person}{Chuan Wu}, \bibinfo{person}{Hridesh
  Rajan}, {and} \bibinfo{person}{Jia Liu}.} \bibinfo{year}{2022}\natexlab{a}.
\newblock \showarticletitle{GADGET: Online Resource Optimization for Scheduling
  Ring-All-Reduce Learning Jobs}.
\newblock \bibinfo{journal}{\emph{arXiv preprint arXiv:2202.01158}}
  (\bibinfo{year}{2022}).
\newblock


\bibitem[\protect\citeauthoryear{Yu, Wu, Ji, and Liu}{Yu
  et~al\mbox{.}}{2021d}]%
        {SMD}
\bibfield{author}{\bibinfo{person}{Menglu Yu}, \bibinfo{person}{Chuan Wu},
  \bibinfo{person}{Bo Ji}, {and} \bibinfo{person}{Jia Liu}.}
  \bibinfo{year}{2021}\natexlab{d}.
\newblock \showarticletitle{A Sum-of-Ratios Multi-Dimensional-Knapsack
  Decomposition for DNN Resource Scheduling}. In \bibinfo{booktitle}{\emph{IEEE
  Conference on Computer Communications}} \emph{(\bibinfo{series}{INFOCOM
  '21})}.
\newblock


\bibitem[\protect\citeauthoryear{Yu and Chowdhury}{Yu and Chowdhury}{2020}]%
        {Salus}
\bibfield{author}{\bibinfo{person}{Peifeng Yu} {and} \bibinfo{person}{Mosharaf
  Chowdhury}.} \bibinfo{year}{2020}\natexlab{}.
\newblock \showarticletitle{Fine-Grained GPU Sharing Primitives for Deep
  Learning Applications}. In \bibinfo{booktitle}{\emph{Proceedings of Machine
  Learning and Systems}} \emph{(\bibinfo{series}{MLSys '20})}.
\newblock


\bibitem[\protect\citeauthoryear{Yu, Liu, and Chowdhury}{Yu
  et~al\mbox{.}}{2021b}]%
        {Fluid}
\bibfield{author}{\bibinfo{person}{Peifeng Yu}, \bibinfo{person}{Jiachen Liu},
  {and} \bibinfo{person}{Mosharaf Chowdhury}.}
  \bibinfo{year}{2021}\natexlab{b}.
\newblock \showarticletitle{Fluid: Resource-aware Hyperparameter Tuning
  Engine}. In \bibinfo{booktitle}{\emph{Proceedings of Machine Learning and
  Systems}} \emph{(\bibinfo{series}{MLSys '21})}.
\newblock


\bibitem[\protect\citeauthoryear{Zaharia, Borthakur, Sen~Sarma, Elmeleegy,
  Shenker, and Stoica}{Zaharia et~al\mbox{.}}{2010}]%
        {zaharia2010delay}
\bibfield{author}{\bibinfo{person}{Matei Zaharia}, \bibinfo{person}{Dhruba
  Borthakur}, \bibinfo{person}{Joydeep Sen~Sarma}, \bibinfo{person}{Khaled
  Elmeleegy}, \bibinfo{person}{Scott Shenker}, {and} \bibinfo{person}{Ion
  Stoica}.} \bibinfo{year}{2010}\natexlab{}.
\newblock \showarticletitle{Delay scheduling: a simple technique for achieving
  locality and fairness in cluster scheduling}. In
  \bibinfo{booktitle}{\emph{Proceedings of the 5th European conference on
  Computer systems}}. \bibinfo{pages}{265--278}.
\newblock


\bibitem[\protect\citeauthoryear{Zhan, Liu, Gong, Zhang, Chung, and Li}{Zhan
  et~al\mbox{.}}{2015}]%
        {SurveyCloudSched}
\bibfield{author}{\bibinfo{person}{Zhi-Hui Zhan}, \bibinfo{person}{Xiao-Fang
  Liu}, \bibinfo{person}{Yue-Jiao Gong}, \bibinfo{person}{Jun Zhang},
  \bibinfo{person}{Henry Shu-Hung Chung}, {and} \bibinfo{person}{Yun Li}.}
  \bibinfo{year}{2015}\natexlab{}.
\newblock \showarticletitle{Cloud Computing Resource Scheduling and a Survey of
  Its Evolutionary Approaches}.
\newblock \bibinfo{journal}{\emph{Comput. Surveys}} (\bibinfo{year}{2015}).
\newblock


\bibitem[\protect\citeauthoryear{Zhang, Yu, Wang, and Yan}{Zhang
  et~al\mbox{.}}{2019}]%
        {zhang2019mark}
\bibfield{author}{\bibinfo{person}{Chengliang Zhang}, \bibinfo{person}{Minchen
  Yu}, \bibinfo{person}{Wei Wang}, {and} \bibinfo{person}{Feng Yan}.}
  \bibinfo{year}{2019}\natexlab{}.
\newblock \showarticletitle{{MArk}: Exploiting Cloud Services for
  {Cost-Effective}, {SLO-Aware} Machine Learning Inference Serving}. In
  \bibinfo{booktitle}{\emph{2019 USENIX Annual Technical Conference (USENIX ATC
  19)}}.
\newblock


\bibitem[\protect\citeauthoryear{Zhang, Yu, Yan, et~al\mbox{.}}{Zhang
  et~al\mbox{.}}{2020d}]%
        {zhang2020mark}
\bibfield{author}{\bibinfo{person}{Chengliang Zhang}, \bibinfo{person}{Minchen
  Yu}, \bibinfo{person}{Feng Yan}, {et~al\mbox{.}}}
  \bibinfo{year}{2020}\natexlab{d}.
\newblock \showarticletitle{Enabling Cost-Effective, SLO-Aware Machine Learning
  Inference Serving on Public Cloud}.
\newblock \bibinfo{journal}{\emph{IEEE Transactions on Cloud Computing}}
  (\bibinfo{year}{2020}).
\newblock


\bibitem[\protect\citeauthoryear{Zhang, Li, Ai, Luo, Wen, Jin, and Ta}{Zhang
  et~al\mbox{.}}{2020b}]%
        {zhang2020hysia}
\bibfield{author}{\bibinfo{person}{Huaizheng Zhang}, \bibinfo{person}{Yuanming
  Li}, \bibinfo{person}{Qiming Ai}, \bibinfo{person}{Yong Luo},
  \bibinfo{person}{Yonggang Wen}, \bibinfo{person}{Yichao Jin}, {and}
  \bibinfo{person}{Nguyen Binh~Duong Ta}.} \bibinfo{year}{2020}\natexlab{b}.
\newblock \showarticletitle{Hysia: Serving DNN-Based Video-to-Retail
  Applications in Cloud}. In \bibinfo{booktitle}{\emph{Proceedings of the 28th
  ACM International Conference on Multimedia}} \emph{(\bibinfo{series}{MM
  '20})}.
\newblock


\bibitem[\protect\citeauthoryear{Zhang, Zheng, Xu, Dai, Ho, Liang, Hu, Wei,
  Xie, and Xing}{Zhang et~al\mbox{.}}{2017}]%
        {Poseidon}
\bibfield{author}{\bibinfo{person}{Hao Zhang}, \bibinfo{person}{Zeyu Zheng},
  \bibinfo{person}{Shizhen Xu}, \bibinfo{person}{Wei Dai},
  \bibinfo{person}{Qirong Ho}, \bibinfo{person}{Xiaodan Liang},
  \bibinfo{person}{Zhiting Hu}, \bibinfo{person}{Jinliang Wei},
  \bibinfo{person}{Pengtao Xie}, {and} \bibinfo{person}{Eric~P Xing}.}
  \bibinfo{year}{2017}\natexlab{}.
\newblock \showarticletitle{Poseidon: An efficient communication architecture
  for distributed deep learning on GPU clusters}. In
  \bibinfo{booktitle}{\emph{2018 {USENIX} Annual Technical Conference}}
  \emph{(\bibinfo{series}{ATC '18})}.
\newblock


\bibitem[\protect\citeauthoryear{Zhang, Elnikety, Zarar, Gupta, and Garg}{Zhang
  et~al\mbox{.}}{2020a}]%
        {jeff2020modelswitching}
\bibfield{author}{\bibinfo{person}{Jeff Zhang}, \bibinfo{person}{Sameh
  Elnikety}, \bibinfo{person}{Shuayb Zarar}, \bibinfo{person}{Atul Gupta},
  {and} \bibinfo{person}{Siddharth Garg}.} \bibinfo{year}{2020}\natexlab{a}.
\newblock \showarticletitle{{Model-Switching}: Dealing with Fluctuating
  Workloads in {Machine-Learning-as-a-Service} Systems}. In
  \bibinfo{booktitle}{\emph{12th USENIX Workshop on Hot Topics in Cloud
  Computing (HotCloud 20)}}.
\newblock


\bibitem[\protect\citeauthoryear{Zhang, Zhang, Pu, and Xu}{Zhang
  et~al\mbox{.}}{2020e}]%
        {zhang2020QoS}
\bibfield{author}{\bibinfo{person}{Jianfeng Zhang}, \bibinfo{person}{Wensheng
  Zhang}, \bibinfo{person}{Lingjun Pu}, {and} \bibinfo{person}{Jingdong Xu}.}
  \bibinfo{year}{2020}\natexlab{e}.
\newblock \showarticletitle{QoS Optimization of DNN Serving Systems Based on
  Per-Request Latency Characteristics}. In
  \bibinfo{booktitle}{\emph{International Conference on Mobility, Sensing and
  Networking (MSN)}}.
\newblock


\bibitem[\protect\citeauthoryear{Zhang, Zhou, Wu, Jiao, and Li}{Zhang
  et~al\mbox{.}}{2020f}]%
        {AOnline}
\bibfield{author}{\bibinfo{person}{Qin Zhang}, \bibinfo{person}{Ruiting Zhou},
  \bibinfo{person}{Chuan Wu}, \bibinfo{person}{Lei Jiao}, {and}
  \bibinfo{person}{Zongpeng Li}.} \bibinfo{year}{2020}\natexlab{f}.
\newblock \showarticletitle{Online scheduling of heterogeneous distributed
  machine learning jobs}. In \bibinfo{booktitle}{\emph{Proceedings of the
  Twenty-First International Symposium on Theory, Algorithmic Foundations, and
  Protocol Design for Mobile Networks and Mobile Computing}}
  \emph{(\bibinfo{series}{MOBIHOC '20})}.
\newblock


\bibitem[\protect\citeauthoryear{Zhang, Li, Wang, Tari, and Zomaya}{Zhang
  et~al\mbox{.}}{2020c}]%
        {zhang2020dybatch}
\bibfield{author}{\bibinfo{person}{Shaojun Zhang}, \bibinfo{person}{Wei Li},
  \bibinfo{person}{Chen Wang}, \bibinfo{person}{Zahir Tari}, {and}
  \bibinfo{person}{Albert~Y. Zomaya}.} \bibinfo{year}{2020}\natexlab{c}.
\newblock \showarticletitle{DyBatch: Efficient Batching and Fair Scheduling for
  Deep Learning Inference on Time-sharing Devices}. In
  \bibinfo{booktitle}{\emph{2020 20th IEEE/ACM International Symposium on
  Cluster, Cloud and Internet Computing (CCGRID)}}.
\newblock


\bibitem[\protect\citeauthoryear{Zhao, Cui, Chen, Leng, Yu, Zeng, Li, and
  Guo}{Zhao et~al\mbox{.}}{2020a}]%
        {CODA}
\bibfield{author}{\bibinfo{person}{Han Zhao}, \bibinfo{person}{Weihao Cui},
  \bibinfo{person}{Quan Chen}, \bibinfo{person}{Jingwen Leng},
  \bibinfo{person}{Kai Yu}, \bibinfo{person}{Deze Zeng}, \bibinfo{person}{Chao
  Li}, {and} \bibinfo{person}{Minyi Guo}.} \bibinfo{year}{2020}\natexlab{a}.
\newblock \showarticletitle{CODA: Improving Resource Utilization by Slimming
  and Co-locating DNN and CPU Jobs}. In \bibinfo{booktitle}{\emph{2020 IEEE
  40th International Conference on Distributed Computing Systems}}
  \emph{(\bibinfo{series}{ICDCS '20})}.
\newblock


\bibitem[\protect\citeauthoryear{Zhao, Han, Yang, Zhang, Yang, Zhou, Yang, Lau,
  Wang, Xiong, and Wang}{Zhao et~al\mbox{.}}{2020b}]%
        {HiveD}
\bibfield{author}{\bibinfo{person}{Hanyu Zhao}, \bibinfo{person}{Zhenhua Han},
  \bibinfo{person}{Zhi Yang}, \bibinfo{person}{Quanlu Zhang},
  \bibinfo{person}{Fan Yang}, \bibinfo{person}{Lidong Zhou},
  \bibinfo{person}{Mao Yang}, \bibinfo{person}{Francis~C.M. Lau},
  \bibinfo{person}{Yuqi Wang}, \bibinfo{person}{Yifan Xiong}, {and}
  \bibinfo{person}{Bin Wang}.} \bibinfo{year}{2020}\natexlab{b}.
\newblock \showarticletitle{HiveD: Sharing a {GPU} Cluster for Deep Learning
  with Guarantees}. In \bibinfo{booktitle}{\emph{14th {USENIX} Symposium on
  Operating Systems Design and Implementation}} \emph{(\bibinfo{series}{OSDI
  '20})}.
\newblock


\bibitem[\protect\citeauthoryear{Zheng, Xu, Chen, Zhou, and Liu}{Zheng
  et~al\mbox{.}}{2019}]%
        {Cynthia}
\bibfield{author}{\bibinfo{person}{Haoyue Zheng}, \bibinfo{person}{Fei Xu},
  \bibinfo{person}{Li Chen}, \bibinfo{person}{Zhi Zhou}, {and}
  \bibinfo{person}{Fangming Liu}.} \bibinfo{year}{2019}\natexlab{}.
\newblock \showarticletitle{Cynthia: Cost-Efficient Cloud Resource Provisioning
  for Predictable Distributed Deep Neural Network Training}. In
  \bibinfo{booktitle}{\emph{Proceedings of the 48th International Conference on
  Parallel Processing}} \emph{(\bibinfo{series}{ICPP' 19})}.
\newblock


\bibitem[\protect\citeauthoryear{Zhou, He, Luo, Yu, and Sun}{Zhou
  et~al\mbox{.}}{2020}]%
        {JPAS}
\bibfield{author}{\bibinfo{person}{Pan Zhou}, \bibinfo{person}{Xinshu He},
  \bibinfo{person}{Shouxi Luo}, \bibinfo{person}{Hongfang Yu}, {and}
  \bibinfo{person}{Gang Sun}.} \bibinfo{year}{2020}\natexlab{}.
\newblock \showarticletitle{JPAS: Job-progress-aware flow scheduling for deep
  learning clusters}.
\newblock \bibinfo{journal}{\emph{Journal of Network and Computer
  Applications}} (\bibinfo{year}{2020}).
\newblock


\bibitem[\protect\citeauthoryear{Zhou, Pang, Zhang, Wu, Jiao, Zhong, and
  Li}{Zhou et~al\mbox{.}}{2022}]%
        {AOnlineTPDS}
\bibfield{author}{\bibinfo{person}{Ruiting Zhou}, \bibinfo{person}{Jinlong
  Pang}, \bibinfo{person}{Qin Zhang}, \bibinfo{person}{Chuan Wu},
  \bibinfo{person}{Lei Jiao}, \bibinfo{person}{Yi Zhong}, {and}
  \bibinfo{person}{Zongpeng Li}.} \bibinfo{year}{2022}\natexlab{}.
\newblock \showarticletitle{Online Scheduling Algorithm for Heterogeneous
  Distributed Machine Learning Jobs}.
\newblock \bibinfo{journal}{\emph{IEEE Transactions on Cloud Computing}}
  (\bibinfo{year}{2022}).
\newblock


\end{thebibliography}

\end{document}